\documentclass{aa}  
\usepackage{graphicx}
\usepackage{physics}
\usepackage{booktabs}
\usepackage{threeparttable} \usepackage{empheq}
\usepackage{siunitx}
\usepackage{txfonts}
\usepackage{xcolor}
\usepackage{placeins}
\definecolor{xlinkcolor}{cmyk}{1,1,0,0}

\PassOptionsToPackage{hyphens}{url}
\usepackage[bookmarks=true,         pdfnewwindow=true,      colorlinks=true,    linkcolor=xlinkcolor,     citecolor=xlinkcolor,     filecolor=xlinkcolor,  urlcolor=xlinkcolor,      final=true,
 ]{hyperref}

\usepackage{mathabx}
\usepackage{wasysym}

\newcommand{\taus}{\tau_\mathrm{s}}
\begin{document}

   \title{Forming Earth-like and low-mass rocky exoplanets through pebble and planetesimal accretion}

   \author{Mitchell Yzer\inst{1}\fnmsep\thanks{Now at: Astrophysics, University of Oxford, Denys Wilkinson Building, Keble Road, Oxford OX1 3RH, UK\\ E-mail: mitchell.yzer@physics.ox.ac.uk}\and
          Ramon Brasser\inst{2}\fnmsep\inst{3}
          \and
          Inge Loes ten Kate\inst{1}\fnmsep\inst{4}
          }

   \institute{Astronomical Institute Anton Pannekoek, University of Amsterdam,
              Science Park 904, PO box 94249, Amsterdam, The Netherlands
              \and
              Konkoly Observatory, HUN-REN CSFK; MTA Centre of Excellence; 15-17 Konkoly Thege Miklos Rd., Budapest, 1121, Hungary
              \and
              Centre for Planetary Habitability, University of Oslo, Sem Saelands vei 2A, Oslo, 0371, Norway
              \and
              Department of Earth Sciences, Utrecht University, Princetonlaan 8A, 3584 CB Utrecht, The Netherlands
             }
\authorrunning{Yzer, Brasser and ten Kate}
   \date{}

  \abstract
{The theory of planet formation through pebble accretion has gained in popularity over the past decade. Recent studies claim that pebble accretion could potentially explain the mass and orbits of the terrestrial planets in the Solar system, the size and water contents of the planets in the TRAPPIST-1 system, and the formation of super-Earth systems at small orbital radii. However, all these studies start with planetary embryos much larger than those expected from the streaming instability.}
{We analyse the formation of terrestrial planets around stars with masses ranging from 0.09 to 1.00 M$_\Sun$ through pebble accretion, starting from small planetesimals with radii between 175 and 450 km.}
{We performed numerical simulations using a modified version of the N-body simulator SyMBA, which includes pebble accretion, type I and II migration, and eccentricity and inclination damping. We analysed two different prescriptions for the pebble accretion rate.}
{We find that Earth-like planets are consistently formed around 0.49, 0.70, and 1.00 M$_\Sun$ stars, irrespective of the pebble accretion model that is used. However, Earth-like planets seldom remain in the habitable zone, for they rapidly migrate to the inner edge of the disc. Furthermore, we find that pebble accretion onto small planetesimals cannot produce Earth-mass planets around 0.09 and 0.20 M$_\Sun$ stars, challenging the proposed narrative of the formation of the TRAPPIST-1 system. }
{Although we have the ability to explain the formation of Earth-mass planets around Sun-like stars, we find a low likelihood of Earth-like planets remaining in the habitable zone. Further research is needed to determine if models with a lower pebble mass flux or with additional migration traps could produce more Solar System-like planetary systems.}


   \maketitle

\section{Introduction} \label{chapter: intro}
Most protoplanetary discs fully dissipate within 3 to 5 Myrs, with hardly any discs surviving past 10 Myrs \citep{Mamajek_2009,Ribas_2015, Li_2016}. Classical theories of planet formation struggle to explain the formation of large planets, especially gas giants such as Jupiter and Saturn, within this short timeframe. For example, classical core formation through runaway accretion of planetesimals (km-sized or larger) takes longer than 10 Myrs beyond 5 au from the star because of the low planetesimal number densities at these distances \citep{Goldreich_2004,Levison_2010}. 

Pebble accretion (PA) is a proposed solution to this problem, whereby planetary cores quickly grow by efficiently accreting mm- to cm-sized solids \citep{Lambrechts_2012}. The theory proposes that planetary seeds\footnote{The terms `planetary seeds or embryos', `planetary cores', `planets' and occasionally even `planetesimals' are used more or less interchangeably in many PA publications. Strictly speaking, these objects differ in mass, though the exact distinction is often arbitrary. The smallest objects in our simulations are planetesimals. Planetary embryos typically have a mass of 0.01 M$_\mathrm{Earth}$, and rapidly grow into planets with masses comparable to Mercury, Mars or higher. The term `planetary core' is often used in the context of gas or ice giant formation and refers to the (hypothesised) solid core of about 10 $M_\mathrm{E}$, around which the thick gas envelope forms. For simplicity, all objects in this study that eventually become planets are referred to as planets throughout their entire evolution. } can efficiently accrete aerodynamically small particles called pebbles, due to an interplay between gravitational and dissipative forces \citep{Ormel_2010}. This leads to high growth rates, even for planets further out in the disc. The theory gained significant traction over the past decade and is supported by the detection of large reservoirs of centimetre-sized particles in protoplanetary discs \citep{Testi2003, Wilner_2005, Ricci_2010}.

In a scenario without gas, pebbles only accrete onto the planet if their trajectory directly collides with the planetary surface. This accretion scenario is called the ballistic regime, characterised by short interaction times and low accretion rates because of the high relative velocities and small collision cross-sections involved \citep{Ormel_2017_PA}. The accretion of large particles (e.g. planetesimals) is always ballistic, since planetesimals are too large to be significantly influenced by gas.

In the presence of a protoplanetary gas disc, however, aerodynamically small pebbles lose energy due to drag from the headwind. As a result, the pebbles settle into the gravitational field of nearby planetesimals or planetary embryos and slowly spiral inwards until they accrete onto the planet. In this settling regime, the pebble accretion rate no longer depends on the physical radius of the planetesimal, but on the size of its gravitational field and thus on its mass \citep{Ormel_2017_PA}. The accretion cross-section, the region from within which material is accreted onto the planet, is significantly enhanced compared to the gas-free scenario \citep{Ormel_2010}, potentially becoming as large as the planet's Hill sphere \citep{Lambrechts_2012}, resulting in rapid growth. Moreover, unlike planetesimals, pebbles are highly mobile, drifting from the outer disc to the inner disc due to drag \citep{Weidenschilling_1977a}. As a result, the pebble reservoir is constantly replenished, allowing for further growth.

Within the Solar System, PA has been used in varying degrees of success, to explain the formation of gas giants \citep{Levison_2015,Matsumura_2017,Matsumura_2021,Raorane_2024,Lau+2024}, as well as the masses and orbits of Venus, Earth, and Mars \citep{Johansen_2021}, the size distribution of asteroids \citep{Johansen_2015b}, and the prograde spin preference of the large bodies in the Solar System \citep{Visser_2019, Takaoka_2023, Yzer_2023}. 

The formation of exoplanetary systems through PA has been studied significantly less. \citet{Schoonenberg_2019} studied the formation of the TRAPPIST-1 planetary system through pebble and planetesimal accretion, proposing a specific solution in which the planets formed at the snowline and migrated inwards sequentially. This model explains the mass and the water contents of the TRAPPIST-1 planets. However, \citet{Schoonenberg_2019} started with large planetary embryos with radii of 1200 km ($\sim$$1.8$$\times$$10^{-3}$ M$_\mathrm{E}$), to limit the number of particles in the numerical simulation. In fact, most PA studies start with large embryos because of computational constraints \citep[see e.g.][]{Morbidelli_2015, Matsumura_2021, Johansen_2021}. Nevertheless, the streaming instability model of planetesimal formation predicts planetesimals start with masses that are one to two orders of magnitude smaller \citep{Simon_2016}. 

The streaming instability is the consequence of the conservation of momentum in pebble accretion models. As the pebbles are slowed down by the aforementioned headwind and drift inwards from the outer disc to the inner disc, they impart their momentum to the gas and speed it up. This reaction, called the back-reaction, locally speeds up the gas and reduces the headwind. This, in turn, reduces the rate at which the pebbles drift inwards, allowing the pebbles to locally pile up. As more pebbles pile up, they impart more momentum onto the gas, further reducing the headwind. Because of this positive feedback loop, a small pebble over-density can grow into a massive cloud of pebbles. Once the mass of the pebble cloud exceeds a critical value, the cloud collapses into planetesimals due to self-gravity \citep{Youdin_and_Goodman_2005, Johansen_and_Youdin_2007, Youdin_and_Johansen_2007, Bai_2010a, Bai_2010c}. Planetesimals formed through the streaming instability have radii between 50 and 450 km \citep{Simon_2016}. This means that the planetesimals must have already significantly grown before they reach the embryo mass used by \citet{Schoonenberg_2019}. 

In this study, we analyse the formation of terrestrial planets through pebble accretion, starting with around 400 planetesimals with radii between 175 and 450 km. Using numerical N-body simulations that include pebble and planetesimal accretion and type I and II migration \citep{Goldreich_and_Tremaine_1979, Tanaka_2002, Paardekooper_2010, Paardekooper_2011}, we aim to answer the question if PA dominated growth can explain the formation of planetary systems close to the star. Aside from the Sun and TRAPPIST-1 (0.09 M$_\Sun$), we tested the theory of PA for an M-dwarf star (0.20 M$_\Sun$), a star at the edge between M-dwarfs and K-dwarfs (0.49 M$_\Sun$), and a K-dwarf star (0.70 M$_\Sun$) \citep{Habets_and_Heintze_1981}. Altogether, these stars represent the most abundant stellar types in the Milky Way.

The structure of this paper is as follows. Section \ref{chapter: models} introduces models describing the disc (Sect. \ref{sec: discmodel}), pebble accretion (Sect. \ref{sec: PAmodel}), and planetary migration (Sect. \ref{sec: migration}). Section \ref{sec: parameters} discusses the simulation set-up and parameters. In Section \ref{sec: PA_afo_r_and_Mini} the results of the analytical model with only PA and planet migration are presented. The results of the full N-body simulations are presented in Sect. \ref{Chapter: Nbody-results} and  further results are given in Sect. \ref{chapter: discussion}. Finally, our main conclusions are summarised in Sect. \ref{chapter: conclusion}.

 \section{Models} \label{chapter: models}
In this study, N-body simulations were performed with around 400 planetesimals with radii between 175 and 450 km and densities of \SI{3}{\g\per\cm}. These planetesimals were represented by physical, gravitating particles in the simulation to account for planetesimal accretion, and to track the dynamics and stability of the planetary systems that form. Pebbles were not included as physical particles. Instead, the pebble flux was calculated based on the disc conditions and the mass accretion rate of pebbles on planetesimals and planets was calculated from analytical equations for the accretion efficiency from \citet{Ida_2016} and \citet{Ormel_and_Liu_2018}. 

\subsection{Disc model} \label{sec: discmodel}
The disc used in this study was assumed to be a steady accretion disc, meaning that the rate at which its material moves inwards (either gas or pebbles) is independent of the distance to the star. This is generally a good approximation in the inner regions of the disc that we are interested in. 

Following \citet{Chambers_2009}, and \citet{Ida_2016}, we assumed that the disc consists of two regimes governed by different types of heating. The inner disc is dominated by viscous heating, while stellar irradiation heats the outer disc \citep{Hueso_and_Guillot_2005, Oka_2011}. The mid-plane temperature, $T$, in these two regimes is given by \citep{Ida_2016}:
\begin{subequations}\label{eq: Tfull}
\begin{equation} \label{eq: Tvisc}
    T_\mathrm{visc} \simeq 
    T_\mathrm{visc,0} 
    M_{*0}^{3/10}
    \alpha_3^{-1/5}
    \dot{M}_{*8}^{2/5}
    \left(\cfrac{r}{\mathrm{au}} \right)^{-9/10}\mathrm{K},
\end{equation}
for the viscous regime, and \citep[see also][]{Chiang_and_Goldreich_1997}:
\begin{equation} \label{eq: Tirr}
    T_\mathrm{irr} \simeq 
    T_\mathrm{irr,0} 
    L_{*0}^{2/7}
    M_{*0}^{-1/7}
    \left(\cfrac{r}{\mathrm{au}} \right)^{-3/7}\mathrm{K},
\end{equation}    
\end{subequations}
for the irradiative regime. In these equations, $M_{*0}$ and $L_{*0}$ are the mass and luminosity of the central star, normalised by the solar values M$_\Sun$ and L$_\Sun$. Moreover, $\dot{M}_{*8}$ and $\alpha_3$ are renormalisations of the gas accretion rate onto the star, $\dot{M}_*$, and the $\alpha$-viscosity parameter, $\alpha_\mathrm{acc}$, from \citet{Shakura_and_Sunyaev_1973}, respectively, using typical values at the beginning of the accretion phase \citep[see e.g.][]{Ida_2016}, such that:
\begin{equation}
    \dot{M}_{*8}\equiv \cfrac{\dot{M}_*}{10^{-8} M_\Sun/\mathrm{yr}} \quad \mathrm{and} \quad 
    \alpha_3 \equiv \cfrac{\alpha_\mathrm{acc}}{10^{-3}}.
\end{equation}
Finally, we assumed that $T_\mathrm{visc,0}=200$ K and $T_\mathrm{irr,0}=150$ K \citep{Chiang_and_Goldreich_1997,Ida_2016}.

The gas scale height, $H\equiv c_\mathrm{s}/\Omega_\mathrm{k}$, is related to these midplane temperatures through the orbital frequency, $\Omega_\mathrm{k}=\sqrt{GM_*/r^3}$, and the sound speed, $c_\mathrm{s}=\sqrt{\gamma k_\mathrm{b}T/(\mu m_\mathrm{p})}$, with $k_\mathrm{b}$ the Boltzmann constant, and $m_\mathrm{p}$ the proton mass. Assuming a heat capacity ratio, $\gamma = 7/5$, and a mean molecular weight, $\mu = 2.34$, the scale height is given by: 
\begin{subequations}\label{eq: hgfull}
\begin{align} \label{eq: hgvisc}
    &H_\mathrm{g,visc} \simeq C_\mathrm{H} T_\mathrm{visc,0}^{1/2}
    M_{*0}^{-7/20}
    \alpha_3^{-1/10}
    \dot{M}_{*8}^{1/5}
    \left(\cfrac{r}{\mathrm{au}} \right)^{21/20} \mathrm{au},\\
    \label{eq: hgirr}
     &H_\mathrm{g,irr} \simeq C_\mathrm{H} T_\mathrm{irr,0}^{1/2}
     L_{*0}^{1/7}
     M_{*0}^{-4/7}
     \left(\cfrac{r}{\mathrm{au}} \right)^{9/7}\mathrm{au},
\end{align}    
\end{subequations}
in which the subscripts visc and irr correspond to the viscous and irradiative regime, respectively, and $C_\mathrm{H}\approx 1.9949\times 10^{-3}$ au K$^{-1/2}$. Finally, following the steady accretion assumption, and the relation between $\dot{M}_*$, $H_\mathrm{g}$, and $\Sigma_\mathrm{g}$,
\begin{equation} \label{eq: Sigmageneral}
    \dot{M}_* = 3\pi \alpha_\mathrm{acc} H^2_\mathrm{g}\Sigma_\mathrm{g}\Omega_\mathrm{k},
\end{equation}
the gas surface density is given by
\begin{subequations}\label{eq: Sigmafull}
\begin{align}
    &\Sigma_\mathrm{g,visc} \simeq C_\Sigma T_\mathrm{visc,0}^{-1}M^{1/5}_{*0}\alpha_3^{-4/5}\dot{M}_{*8}^{3/5}\left(\cfrac{r}{\mathrm{au}} \right)^{-3/5} \mathrm{\ g\ cm}^{-2}, \\
    \label{eq: Sigmairr}
    &\Sigma_\mathrm{g,irr} \simeq C_\Sigma T_\mathrm{irr,0}^{-1}L^{-2/7}_{*0}M_{*0}^{9/14}\alpha_3^{-1}\dot{M}_{*8}\left(\cfrac{r}{\mathrm{au}} \right)^{-15/14} \mathrm{\ g\ cm}^{-2},
\end{align}
\end{subequations}
with $C_\Sigma \approx 3.7708\times 10^5$ g cm$^{-2}$ K.

The boundary between the viscous and radiative regimes lies at the orbital radius, $r$, where $T_\mathrm{visc}(r)=T_\mathrm{irr}(r)$, and is given by \citep[see also][]{Chambers_2009,Ida_2016}:
\begin{equation}\label{eq: r_visc-irr}
    r_\mathrm{visc-irr} \simeq \left( \cfrac{T_\mathrm{visc,0}}{T_\mathrm{irr,0}}\right)^{70/33}
    L_{*0}^{-20/33}
    M_{*0}^{31/33}
    \alpha_3^{-14/33}
    \dot{M}_{*8}^{28/33} \mathrm{\ au}.
\end{equation}
This boundary shifts radially inwards as the disc evolves, due to the decreasing value of $\dot{M}_*$ as the disc is being depleted.

In fact, all of the time evolution of the disc parameters was modelled by the time dependence of $\dot{M}_{*(8)}$.
In the steady accretion state, this stellar gas accretion rate is given by \citep{Hartmann_1998}:
\begin{equation}\label{eq: Mdot*}
\begin{split}
    \dot{M}_*&=\cfrac{M_\mathrm{d,0}}{2t_\mathrm{diff}(2-\xi)}\left(\frac{t}{t_\mathrm{diff}} +1\right)^{-(5/2 - \xi)/(2-\xi)}\\
    &\simeq \frac{7}{13}\frac{M_\mathrm{d,0}}{t_\mathrm{diff}}\left(\frac{t}{t_\mathrm{diff}}+1 \right)^{-20/13},
\end{split}
\end{equation}
where $M_\mathrm{d,0}$ is the initial disc mass, $t_\mathrm{diff}$ is the diffusion time, and $\xi=-\frac{\mathrm{d}\ln \Sigma_\mathrm{g,irr}}{\mathrm{d} \ln r}=15/14$ is the negative slope of the gas surface density power law in the irradiative regime (see Eq. \ref{eq: Sigmairr}). The diffusion time, $t_\mathrm{diff}$, is, in turn, related to the accretion $\alpha$-viscosity parameter, $\alpha_\mathrm{acc}$, which is assumed to be constant in a steady accretion disc,  given by \citep{Hartmann_1998}:
\begin{equation}
    \alpha_\mathrm{acc}=\frac{h_\mathrm{g,D}^{-2}}{6\pi(2-\xi)^2}\frac{t_\mathrm{orb,D}}{t_\mathrm{diff}},
\end{equation}
where $h_\mathrm{g,D} \equiv H_\mathrm{g,D}/r_\mathrm{D}$ is the aspect ratio of the disc and $t_\mathrm{orb,D}=2\pi/\Omega_\mathrm{k,D}$ is the orbital period of a circular orbit, both measured at the outer edge of the disc. 

Following \citet{Matsumura_2021}, we assumed the initial values for the disc mass, $M_\mathrm{d,0}$, the diffusion timescale, $t_\mathrm{diff}$, and the outer radius of the disc, $r_\mathrm{D}$, to fix the disc evolution to a specific model. Especially the latter parameter is important for pebble accretion since the pebble flux rapidly decreases once the pebble formation front reaches the edge of the disc \citep{Sato_2016}. The initial conditions are shown in Table \ref{tab:standard_disc_conditions}, and discussed in Sect. \ref{sec: parameters}.

 \subsection{Pebble accretion model} \label{sec: PAmodel}
The growth rate of a planet undergoing pebble accretion can be parameterised as 
\begin{equation}
    \dot{M}_\mathrm{p}=\epsilon\dot{M}_\mathrm{F},
\end{equation}
where $\epsilon$ is the pebble accretion efficiency and $\dot{M}_\mathrm{F}$ is the pebble mass flux available to the planet. The accretion efficiency strongly depends on the size of the pebbles and the mass of the planet. Meanwhile, the pebble mass flux depends on the location of the pebble formation front, which is the region of the disc in which dust coagulates into pebbles. 

\subsubsection{Pebble radius and mass flux}\label{sec: Rpeb_and_Mdotf}
This study used semi-analytical expressions for the monodisperse (i.e. single-sized) pebble radius, $R_\mathrm{p}$, and Stokes number, $\taus$, based on the work of, among others, \citet{Ida_2016}. The Stokes number, $\taus$, describes how the orbit of a pebble is influenced by the gas of the disc; in particular, by drag from the headwind discussed in the introduction. It depends on both the pebble radius and the local gas conditions. 

We assumed that the pebble radius is drift-limited, which is to say that pebbles grow in situ until their drift timescale, $t_\mathrm{drift}$ becomes shorter than their growth timescale, $t_\mathrm{grow}$. This is a fair approximation when the location at which most pebbles are being formed, the pebble formation front, is in the outer disc \citep{Ida_2016}. The effects of the bouncing and fragmentation barriers were assumed to be negligible, which is valid for the icy grains outside the snowline for discs with $\alpha_\mathrm{turb}\lesssim10^{-2}$ \citep{Ida_2016}. For a discussion about the bouncing or fragmentation barriers for pebbles inside the snowline, we refer to Appendix \ref{App: sublimation}.

In this study, the planetesimals were located at orbital radii smaller than about 3 au. Using the assumption of a drift-limited pebble radius in combination with our disc conditions, we find that the Stokes number of pebbles in these inner regions of the disc typically ranges between 0.1 and 2 (see, e.g., Fig. \ref{fig: rpeb_in_rsnow}). For the full analytical expressions for the pebble radius, we refer to \citet{Ida_2016}.

The pebble mass flux, $\dot{M}_\mathrm{F}$, is determined by the amount of dust that is kicked up by the pebble formation front per unit of time. Following the models of \citet{Lambrechts_2014,Ida_2016,Sato_2016,Ida_2019}, we assumed the pebble mass flux is given by:
\begin{equation}\label{eq: Mdotf_final}
\begin{split}
    \dot{M}_\mathrm{F}=3.&099\ \times
    \frac{\ln 10^{4}}{\ln \frac{R_\mathrm{peb}}{R_0}}
    T_\mathrm{irr,0}^{-1}
    L_{*0}^{-2/7} 
    M_{*0}^{8/7}
    \alpha_3^{-1} \dot{M}_{*8}\left(\frac{\Sigma_\mathrm{pg,0}}{0.01} \right)^{2}\\ \times
    &\left(\frac{r_\mathrm{D}}{\mathrm{au}}\right)^{-4/7}
    \left(\frac{t}{t_\mathrm{pf}}\right)^{-8/21}
    \left(1+\frac{t}{t_\mathrm{pf}}\right)^{-\gamma_\mathrm{pf}} \ \mathrm{M}_\Earth/\mathrm{yr}.
\end{split}
\end{equation}
In this expression, $R_\mathrm{peb}/R_0$ is the typical ratio between pebble and dust particle radius ($\sim$$10^4$), $\Sigma_\mathrm{pg,0}\equiv\Sigma_\mathrm{p,0}/\Sigma_\mathrm{g,0}$ is the initial ratio between the pebble and gas surface density, $r_\mathrm{D}$ is the outer radius of the disc, and $t_\mathrm{pf}$ is the time it takes the pebble formation front to reach this outer radius. When $t>t_\mathrm{pf}$, the pebble mass flux rapidly decreases due to the factor $(1+t/t_\mathrm{pf})^{-\gamma_\mathrm{pf}}$, introduced by \citet{Sato_2016}. Here, $\gamma_\mathrm{pf}=1+\gamma_\mathrm{pf,2}\times(300\ \mathrm{au})/r_\mathrm{D}$ with $\gamma_\mathrm{pf,2} \sim 0.15$ a fit parameter.

\begin{figure}
    \centering
    \includegraphics[width=\linewidth]{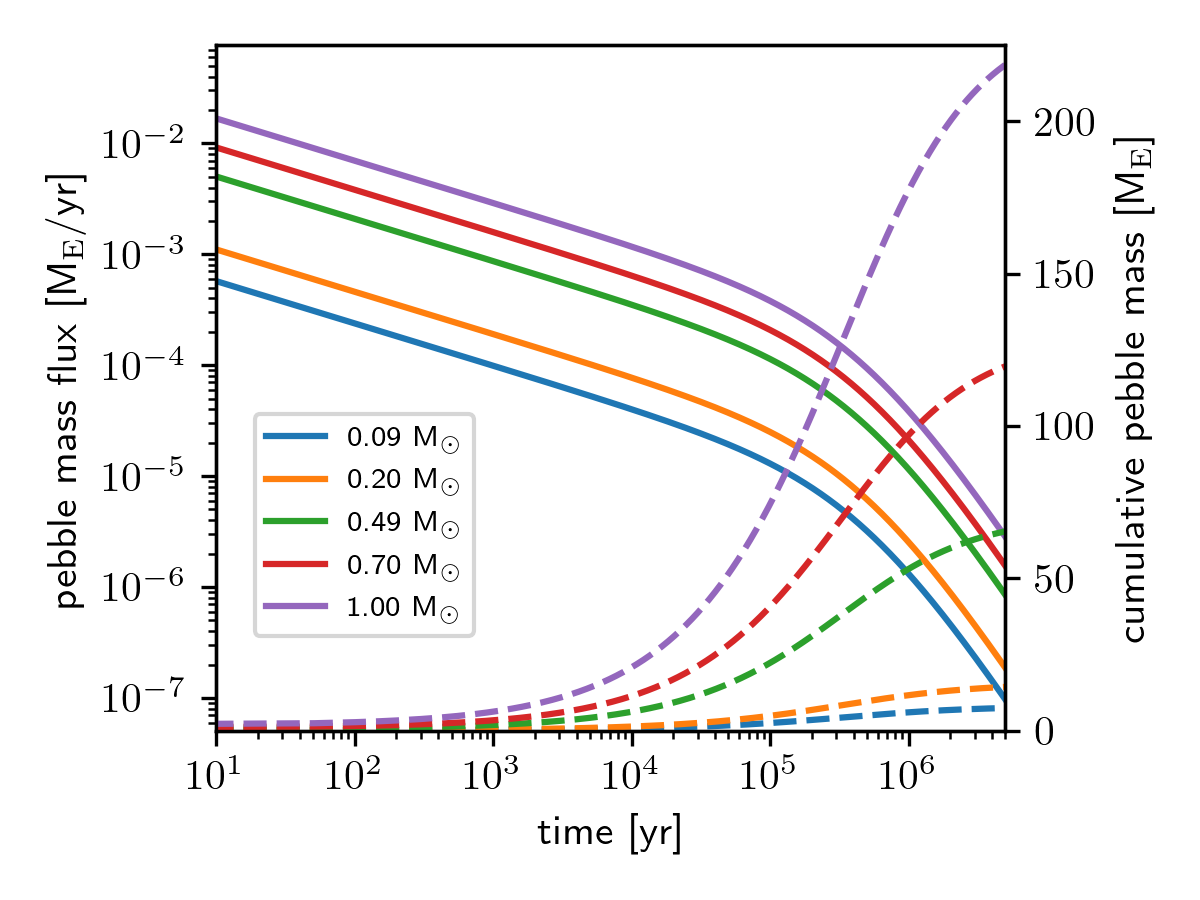}
    \caption[Radial pebble mass flux and cumulative pebble mass as a function of time.]{Radial pebble mass flux (solid lines, left-hand axis) and cumulative pebble mass (dashed lines, right-hand axis) in the inner disc as a function of time for the five stellar masses used in this study. After around 0.1 Myrs, the pebble formation front reaches the outer edge of the disc, after which the slope of the pebble mass flux becomes significantly more negative. The disc around the 0.09 M$_\Sun$ star is weighted as 10\% of the stellar mass, while all others are weighted as 5\% (see the discussions in Sects. \ref{subsec: disc_parameters} and \ref{sec: PA_afo_r_and_Mini}).}
    \label{fig: Mdotf0}
\end{figure}

Figure \ref{fig: Mdotf0} shows the pebble mass flux, and cumulative pebble mass as a function of time for the five stellar masses used in this study. There is a clear change in the slope of the pebble mass flux after the pebble formation front reaches the outer edge of the disc, after approximately $10^5$ yr. 

Since pebbles contain significant amounts of water ice, the snowline, located where $T_\mathrm{disc}=170$ K, is an important boundary in analytical pebble prescriptions. Once the pebbles drift inwards past the snowline, their ice sublimates, changing the radius and density of the pebble, and reducing the total flux of pebble mass by 50\% \citep{Lodders_2003}. The location of the snowline is given by $r_\mathrm{snow} \sim \max(r_\mathrm{snow,visc},r_\mathrm{snow,irr})$, whereby 
\begin{subequations}
\begin{equation}\label{eq: rsnow_visc}
    r_\mathrm{snow,visc} \simeq \left(\cfrac{T_\mathrm{snow}}{T_\mathrm{visc,0}} \right)^{-10/9}
    M_{*0}^{1/3} \alpha_3^{-2/9} \dot{M}_{*8}^{4/9} \mathrm{\ au},
\end{equation}
is the snowline calculated in the viscous regime, and 
\begin{equation}
    r_\mathrm{snow,irr} \simeq \left(\cfrac{T_\mathrm{snow}}{T_\mathrm{irr,0}} \right)^{-7/3}
    L_{*0}^{2/3}M_{*0}^{-1/3}\mathrm{\ au},
\end{equation}
\end{subequations}
the snowline calculated in the irradiative regime. For a detailed discussion of the influence of sublimation on the pebble radius, we refer to Appendix \ref{App: sublimation}.

We did not consider shading effects, which can complicate the behaviour of the snowline in the irradiative regime; nor did we take into account the slow outward movement of the snowline due to a decrease in vapour pressure resulting from a diminishing influx of icy pebbles over time \citep{Schoonenberg_2019}. We only considered the inwards movement of the snowline when it is located in the viscously heated part of the disc, due to the decrease in $\dot{M}_{*8}$ as the disc depletes.
 \subsubsection{Pebble accretion efficiency}\label{subsubsec: PA_efficiency}
In this study, we considered two different prescriptions for the accretion efficiency: $\epsilon_\mathrm{IGM16}$ by \citet{Ida_2016}, and $\epsilon_\mathrm{OL18}$ by \cite{Ormel_and_Liu_2018}. The model of \citet{Ida_2016} is valid in the settling regime for planets on circular orbits, and contains terms for both 2D and 3D accretion. In the settling regime, the relative velocity between the planet and the pebbles is small enough, and therefore the encounter time long enough, for the pebbles to settle in the gravitational field of the planet, and spiral inwards, leading to an accretion cross-section that is orders of magnitude larger than the geometric cross-section of the planet. According to \citet{Ida_2016}, the accretion efficiency in this regime is given by
\begin{equation}\label{eq: eps_IGM16}
    \epsilon_\mathrm{IGM16}= \min \left(1,\frac{C_{\epsilon,\mathrm{i}} C_\epsilon b^2 }{4\sqrt{2\pi} h_\mathrm{p}}\frac{\sqrt{1+4\taus^2}}{\taus}\left(1+\frac{3b}{2\chi\eta} \right)\right).
\end{equation}
The parameters in this equation are
\begin{equation*}
    \chi = \frac{\sqrt{1+4\taus^2}}{1+\taus^2},
\end{equation*}
\begin{equation*}
     b \equiv \frac{B}{r} = \min \left(1, \sqrt{3\taus^{1/3}r_\mathrm{H}/(\chi\eta)} \right) \times 2\kappa \taus^{1/3} r_\mathrm{H},
\end{equation*}
\begin{equation*}
    \kappa = \exp \left(- \left( \frac{\taus}{\min\left(2,\taus^*\right)}\right)^{0.65}\right),
\end{equation*}
\begin{equation*}
    \taus^*=4(M_\mathrm{p}/M_*)/\eta^3
\end{equation*}
\begin{equation*}
    C_\epsilon = \min \left( \sqrt{\frac{8}{\pi}}\frac{h_\mathrm{p}}{b},1\right),
\end{equation*}
\begin{equation*}
    h_\mathrm{p}\simeq\left(1+\frac{\taus}{\alpha_\mathrm{turb}}\right)^{-1/2}h_\mathrm{g}\simeq \left(\frac{\taus}{\alpha_\mathrm{turb}}\right)^{-1/2}h_\mathrm{g},
\end{equation*}
\begin{equation*}
     C_{\epsilon,\mathrm{i}} = \frac{1}{2}\frac{\left(\mathrm{erf}\left(\frac{z+B}{\sqrt{2}H_\mathrm{p}}\right) - \mathrm{erf}\left(\frac{z-B}{\sqrt{2}H_\mathrm{p}} \right)\right)}{\mathrm{erf} \left(\frac{B}{\sqrt{2}H_\mathrm{p}} \right)}.
\end{equation*}

The reduction factor $C_{\epsilon,\mathrm{i}}$ was proposed by \citet{Matsumura_2021} to account for the effect of the orbital inclination of the planet on the amount of pebbles it encounters. Finally, $r_\mathrm{H} \equiv R_\mathrm{H}/r$, where $R_\mathrm{H}=r\left(M_\mathrm{p}/3M_*\right)^{1/3}$ is the Hill radius of the planet with mass, $M_\mathrm{p}$.\\

The second accretion efficiency model, proposed by \citet{Ormel_and_Liu_2018}, is based on 2D \citep{Liu_and_Ormel_2018} and 3D simulations of pebble accretion, and includes the orbit-averaged influence of eccentricity and inclination of the planet's orbit, as well as disc turbulence. In the 2D limit, the accretion efficiency is given by \citep{Liu_and_Ormel_2018}:
\begin{equation}\label{eq: eps_O_2D}
     \epsilon_\mathrm{OL18,2D}=\frac{A_2}{\eta} \sqrt{\frac{M_\mathrm{p}}{M_*}\frac{\Delta v_\mathrm{y}}{\taus v_\mathrm{K}}} f_\mathrm{set},
\end{equation}
and in the 3D limit by \citep{Ormel_2017_PA,Liu_and_Ormel_2018,Ormel_and_Liu_2018}:
\begin{equation}\label{eq: eps_O_3D}
    \epsilon_\mathrm{OL18,3D}=\frac{A_3}{\eta h_\mathrm{p,eff}}\left(\frac{M_\mathrm{p}}{M_*}\right)f_\mathrm{set}^2.
\end{equation}
In these equations, $A_2=0.32$ and $A_3=0.39$ are fitting constants. Furthermore, $\Delta v_\mathrm{y}$ is the azimuthal approach velocity, $h_\mathrm{p,eff}$ is the effective pebble aspect ratio, and $f_\mathrm{set}$ is the settling fraction \citep[see][]{Ormel_and_Liu_2018}. The effective pebble aspect ratio includes a correction for the inclination, $i_\mathrm{p}$, of the planet's orbit. The settling fraction, $f_\mathrm{set}$, dependents on both the inclination, and the eccentricity of the planetary orbit, through the vertical and azimuthal approach velocities, $\Delta v_\mathrm{y}$ and $\Delta v_\mathrm{z}$, respectively. 

If these approach velocities become larger than the critical settling velocity, $v_*$, given by \citep{Ormel_2010,Liu_and_Ormel_2018}:
\begin{equation}\label{eq: vcrit_sett_Ormel}
    v_*=\left(\frac{M_\mathrm{p}}{M_*\taus}\right)^{1/3}v_\mathrm{K},
\end{equation}
the pebbles have too little time to settle and spiral inwards. Accretion then enters the inefficient ballistic regime, in which only pebbles that are on a direct collision course with the planet are accreted. The expressions for the ballistic regime are provided by \citet{Liu_and_Ormel_2018}.

Herein lies the core difference between $\epsilon_\mathrm{IGM16}$ and $\epsilon_\mathrm{OL18}$. Whereas the IGM16 model assumes that all planets are on circular orbits, the OL18 model considers the fact that for planets on significantly excited orbits, the relative velocities between the planet and the pebbles become too large for settling, leading to a rapid reduction in PA efficiency. 

\begin{figure}
    \centering
    {\includegraphics[width=1.0\linewidth]{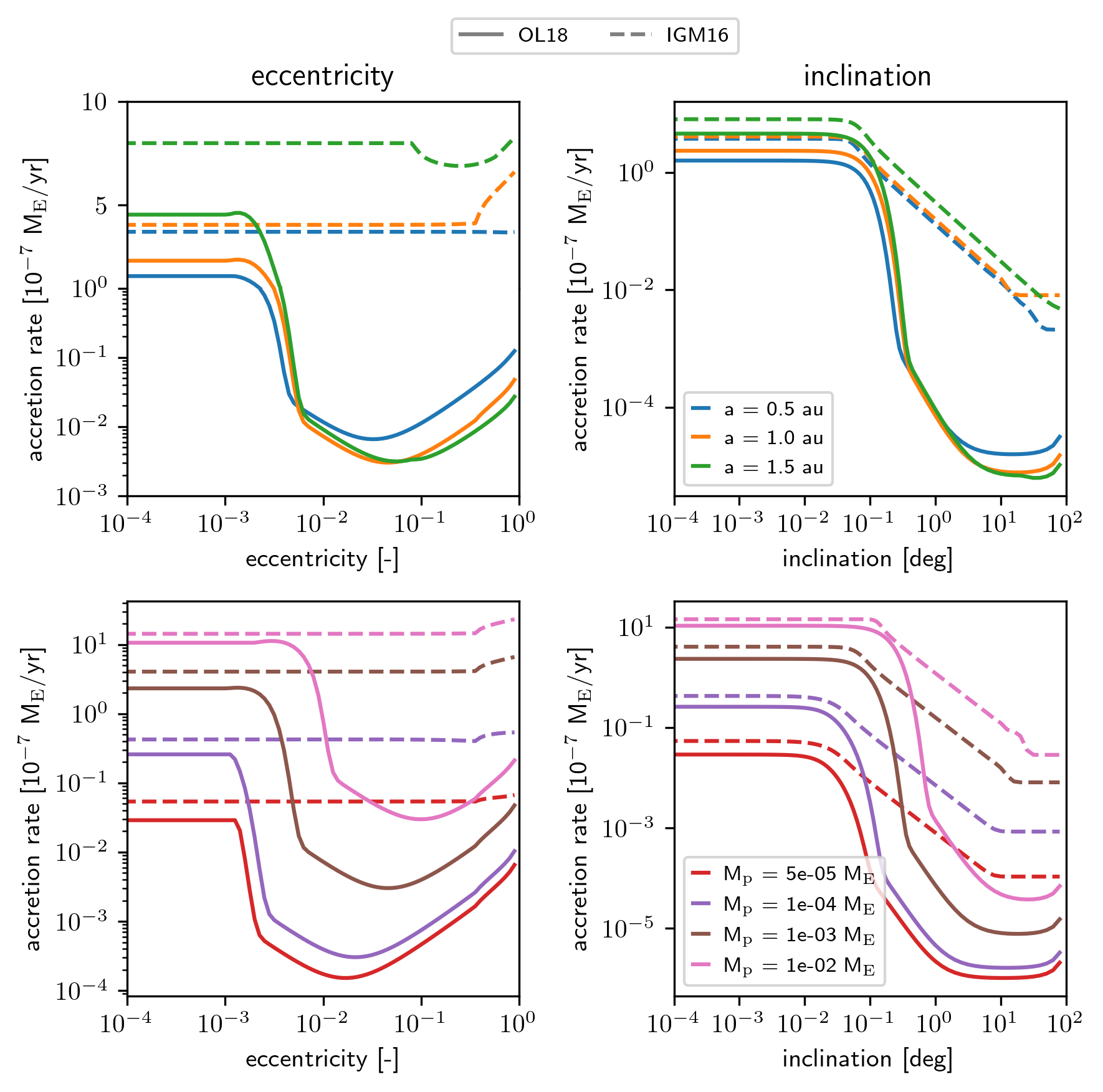}}
        \caption[Orbit-averaged pebble accretion rate as a function of eccentricity and inclination.]{Orbit-averaged pebble accretion rate as a function of eccentricity (left panels) and inclination (right panels) for planets around a 1.0 M$_\Sun$ star, 0.01 Myrs after the formation of the disc. The top plots show the accretion rate for a $10^{-3}$ M$_\mathrm{E}$ planetesimal at different orbital radii, while the bottom plots shows results for planets of different masses at 1 au. The solid lines were calculated with OL18, the dashed lines with IGM16. The OL18 prescription contains explicit expressions for the influence of the eccentricity and inclination on the accretion rate, while the IGM16 model only includes corrections for part of the influence of the inclination. For large $e$, the change in $r_\mathrm{p}$ along a single orbit becomes significant enough for the variations in the encountered disc conditions to influence the accretion rate, both in the OL18 and the IGM16 model. The y-axis of the top-left plot changes from a logarithmic to a linear scale at $1\times10^{-7}$ M$_\mathrm{E}$/yr to highlight the effect.}
    \label{fig: orbavg_ecc_inc}
\end{figure}

This effect is demonstrated in Fig. \ref{fig: orbavg_ecc_inc}. This figure shows the orbit-averaged pebble accretion rate for planets with different semi-major axes, and masses, as a function of eccentricity ($e$), and inclination ($i$). These values have been calculated for our standard disc model (see Table \ref{tab:standard_disc_conditions} for initial conditions) around a solar-mass star, 0.01 Myrs after the formation of the disc. The results for varying $e$ were calculated with $i=0$, and vice versa. 

For the smallest planetesimals, an eccentricity between $10^{-3}$ and $10^{-2}$ is already sufficient to reduce the growth rate by two orders of magnitude in the OL18 model. Meanwhile, in the IGM16 model, only for eccentricities of order 0.1 and above does the orbit-averaged accretion rate change, and by less than a factor of 2, primarily due to the fact that the orbit starts crossing the snowline (1.37 au for these conditions), leading to an increase (for $0.5 r_\mathrm{snow}<a<r_\mathrm{snow}$) or decrease (for $a>r_\mathrm{snow}$) in the orbit-averaged encountered pebble flux.

Similarly, an induced inclination leads to a much steeper decline in accretion rate in OL18, than in IGM16, since the former considers both the reduced pebble density at high altitudes, and the increased relative velocity between the planet and the pebbles, while IGM16 only includes the reduction factor $C_{\epsilon,\mathrm{i}}$ for the encountered pebble density from \citet{Matsumura_2021}.

For the N-body simulations, this means that if the planetesimals are quickly excited, PA might come to a halt in OL18, while the planets in IGM16 continue growing, leading to more and more massive planets in the latter simulations.

\subsubsection{Pebble isolation mass} \label{subsubsec: Miso}
Pebble accretion ceases once the planet's mass exceeds the pebble isolation mass, $M_\mathrm{iso}$ \citep{Lambrechts_2014}. At this point, the planet significantly perturbs the gas disc, creating a local pressure bump just outside its orbit. Since pebbles drift against the pressure gradient, the local maximum traps the pebbles, preventing them from drifting further inwards. This not only halts pebble accretion for the planet in question, but also for all planets interior to it.

There are multiple prescriptions for $M_\mathrm{iso}$. In this study, we used the prescription by \citet{Ataiee_2018}, who used 2D hydrodynamical simulations with gas and dust with turbulence parameters in the range of $\alpha_\mathrm{turb}=\left[5\times10^{-4}, 1\times10^{-2} \right]$ to determine the dependence of $M_\mathrm{iso}$ on the disc aspect ratio, pressure gradient, and (turbulent) viscosity. They proposed
\begin{equation}\label{eq: Miso}
\begin{split}
    M_\mathrm{iso}\simeq h^3_\mathrm{g}&\sqrt{37.3\alpha_\mathrm{turb}+0.01} \\ &\times\left(1+0.2\left(\frac{\sqrt{\alpha_\mathrm{turb}}}{h_\mathrm{g}}\sqrt{\frac{1}{\taus^2}+4} \right)^{0.7} \right) M_*.
\end{split}
\end{equation}

In the 2D low-viscosity limit ($\alpha_\mathrm{turb}\sim 10^{-4}$) used in our study, the prescription of \citet{Ataiee_2018} agrees well with the results of the main competing model by \citet{Bitsch_2018} \citep[see Fig. 8 of][]{Ataiee_2018}. In the high-viscosity limit ($\alpha_\mathrm{turb}\sim 10^{-2}$), the two models do significantly differ, by a factor of 3, but this range is not relevant to this study. 

 \subsection{Planetary migration model} \label{sec: migration}
In this study, the equations of motion of planetesimals were solved numerically in order to analyse the dynamical evolution of the system. Aside from the gravitational force terms from the central star and the other planetesimals in the system, migration through planet-disc interactions enters into a planet's equation of motion. 

We used the equations of motion for migrating planets proposed by \citet{Ida_2020}. This prescription predicts planet-disc interactions well, both in the subsonic \citep{Tanaka_and_Ward_2004} and the supersonic \citep{Muto_2011} regimes. The migration component of the equation of motion is given by
\begin{equation}\label{eq: EOM_migration}
    \left(\frac{\mathrm{d}\textbf{v}}{\mathrm{dt}}\right)_\mathrm{migr} = -\frac{v_\mathrm{K}}{2\tau_\mathrm{a}}\textbf{e}_\theta - \frac{v_\mathrm{r}}{\tau_\mathrm{e}}\textbf{e}_\mathrm{r}-\frac{v_\theta-v_\mathrm{K}}{\tau_\mathrm{e}}-\frac{v_\mathrm{z}}{\tau_\mathrm{i}}\textbf{e}_\mathrm{z},
\end{equation}
where $v_\mathrm{K}$ is the Keplerian orbital speed at radius $r$; $v_\mathrm{r},\ v_\theta$ and $v_\mathrm{z}$ are the velocity components in the radial, azimuthal and vertical direction, respectively; $\textbf{e}_\mathrm{r},\ \textbf{e}_\theta$ and $\textbf{e}_\mathrm{z}$ are the corresponding unit vectors; and $\tau_\mathrm{a},\ \tau_\mathrm{e}\ \mathrm{and\ } \tau_\mathrm{i}$ are the characteristic timescales for the evolution of the semi-major axis, eccentricity and inclination, respectively. The timescales for type I migration are discussed in detail by \citet{Ida_2020}.

As planets grow more massive and transition from type I to type II migration, their inwards motion slows down. According to \citet{Kanagawa_2018}, type II migration is the same as type I migration, but with a reduced surface density due to the planet gap that has formed. They showed that the timescale of semi-major axis decay can be written as
\begin{equation}
    \tau_\mathrm{a}'\simeq (1+0.04K)\tau_\mathrm{a},
\end{equation}
with
\begin{equation}
    K=\left(\frac{M_\mathrm{p}}{M_*} \right)^2 \left(\frac{H_\mathrm{g}}{a}\right)^{-5}\alpha_\mathrm{turb}^{-1},
\end{equation}
which is valid for both type I and type II migration. However, there is no consensus on the evolution of the eccentricity and inclination during type II migration. Following \citet{Matsumura_2021}, we assumed that $\tau_\mathrm{e}$ and $\tau_\mathrm{i}$ evolve similarly to $\tau_\mathrm{a}$, such that
\begin{equation}
    \tau_\mathrm{e}' \simeq (1+0.04K)\tau_\mathrm{e}, \quad \mathrm{and} \quad \tau_\mathrm{i}'\simeq (1+0.04K)\tau_\mathrm{i}.
\end{equation}
These adjusted timescales were used in the equation of motion of Eq. \ref{eq: EOM_migration}. The back reactions of the planets on the gas were ignored.

  \section{Simulation parameters and initial conditions} \label{sec: parameters}

\begin{table}[]
\centering
\begin{threeparttable}
\caption{Standard disc parameters.}
\label{tab:standard_disc_conditions}

\begin{tabular}{l|l}
\toprule \toprule
Parameter              & Standard Value             \\ \midrule
Initial disc mass $M_\mathrm{D,0}$       & 0.05 $M_*$                 \\
Disc outer radius $r_\mathrm{D}$         & 100 au                     \\
Diffusion timescale $t_\mathrm{diff}$      & 0.5 Myrs                   \\
turbulent $\alpha$-viscosity $\alpha_\mathrm{turb}$ & $10^{-4}$ \\
$T_{\rm visc,0}$ & 200 K \\
$T_{\rm irr,0}$ & 150 K \\
$T_{\rm snow}$ & 170 K
\\ \bottomrule
\end{tabular}
\end{threeparttable}
\tablefoot{For the 0.09 M$_\Sun$ stars, the initial disc mass is 0.1~$M_*$.}
\end{table}

\begin{table*}[]
\centering
\begin{threeparttable}
\caption{Simulation parameters for different stars.}
\label{tab:simulation_parameters}
\begin{tabular}{@{}l|ccccc@{}}
\toprule \toprule
Parameter                                                               & 1.00 M$_\Sun$ & 0.70 M$_\Sun$ & 0.49 M$_\Sun$ & 0.20 M$_\Sun$ & 0.09 M$_\Sun$ \\ \midrule
Initial disc mass ($M_\mathrm{D,0}$) {[}M$_\Sun${]}                     & 0.050         & 0.035         & 0.025         & 0.010         & 0.009         \\
Initial planetesimal ring mass ($M_\mathrm{pl,0}$) {[}M$_\mathrm{E}${]} & 0.015         & 0.015         & 0.015         & 0.015         & 0.010         \\
Wide planetesimal ring size ($r_\mathrm{pl,min,w}\-- r_\mathrm{pl,max,w}$) {[}au{]}   & $0.2\--2.2$   & $0.2\--2.0$   & $0.2\--1.0$   & $0.2\--0.8$   & $0.2\--0.8$   \\
Narrow planetesimal ring size ($r_\mathrm{pl,min,n}\-- r_\mathrm{pl,max,n}$) {[}au{]} & $1.18\--1.60$ & $0.92\--1.25$ & $0.72\--0.97$ & $0.38\--0.52$ & $0.30\--0.40$ \\
Disc inner radius ($r_\mathrm{in}$) {[}au{]}                            & 0.097         & 0.086         & 0.076         & 0.057         & 0.021         \\
Inner truncation radius ($r_\mathrm{trunc}$) {[}au{]}                   & 0.046         & 0.041         & 0.037         & 0.027         & 0.013         \\ \midrule
Total pebble mass flux ($\int\dot{M}_\mathrm{F}\mathrm{d}t$) {[}M$_\mathrm{E}${]}  & 217.0         & 118.7         & 65.0          & 14.3          & 7.4           \\ \bottomrule
\end{tabular}
\end{threeparttable}
\tablefoot{The total pebble mass flux is not an input parameter, but is calculated by integrating Eqs. \ref{eq: Mdot*} and \ref{eq: Mdotf_final}.}
\end{table*}

In this study, we used a modified version of the N-body simulator SyMBA \citep{Duncan_1998} to analyse the planetary systems that form around different low-mass stars. This version of SyMBA was parallelised by \citet{Lau_and_Lee_2023}, and has been augmented to include pebble accretion, type I and II migration, and eccentricity and inclination damping \citep{Matsumura_2017,Matsumura_2021}. The pebbles were not included as actual particles in the N-body simulations. Instead, we analysed two different pebble accretion models which describe the mass accretion rate of the planets, given the disc conditions and the Stokes number of the pebbles: the model of \citet{Ida_2016} (IGM16), and the one of \citet{Ormel_and_Liu_2018} (OL18). We performed simulations for five different stellar masses: 0.09 M$_\Sun$ (M-dwarf, TRAPPIST-1), 0.20 M$_\Sun$ (M-dwarf), 0.49 M$_\Sun$ (M-dwarf/K-dwarf), 0.70 M$_\Sun$ (K-dwarf), and 1.00 M$_\Sun$ (G-dwarf) \citep{Habets_and_Heintze_1981}. These stellar types make up the bulk of the main sequence stars in the galaxy. Moreover, because these stars are relatively long-lived, and have a habitable zone at small orbital radii where terrestrial planets are expected to form, planets around these stars are prime candidates in the search for life.

\subsection{Disc parameters} \label{subsec: disc_parameters}
The initial mass of a protoplanetary disc typically ranges between 1\% and 10\% of the mass of the central star \citep{Pascucci_2016}. For the 0.09 M$_\Sun$ stars, we focussed on discs at the high end of this range. These discs allow for the formation of larger planets due to the increased solid mass, which compensates for the fact that in our simulations of these systems, we include fewer planetesimals than the simulations with higher mass stars, because of computational constraints following from the required step size in the 0.09 M$_\Sun$ simulations, which is discussed below. For all other stars, the initial disc mass was assumed to be 5\% of the stellar mass. Furthermore, we assumed $r_\mathrm{D}=100$ au, based on the observed typical size of protoplanetary discs \citep{Andrews_2018, Andrews_and_Williams_2007, Vicente_and_Alves_2005}, and we took $t_\mathrm{diff}$ = 0.5 Myrs, which fits our assumption that the disc has a lifetime of about 5 Myrs. Finally, we assumed $\alpha_\mathrm{turb}=10^{-4}$.

The other disc parameters follow from the equations presented in Sect. \ref{sec: discmodel} and by \citet{Ida_2016}. In these equations, we assumed the mass-luminosity relation for low-mass protostars to be approximately $L_*/L_\Sun \simeq M_*/M_\Sun$, following the fits with power laws of \citet{Baraffe_2015}, which are close to linear.

Two important control parameters that are not directly derived from the equations are the inner radius of the disc, $r_\mathrm{in}$, and the inner truncation radius, $r_\mathrm{trunc}$. The gas of the disc does not extend all the way to the surface of the star, most likely due to interactions with the star's magnetosphere. The star's magnetic field disrupts the disc and creates a cavity out to the radius where the magnetic energy density is equal to the kinetic energy density of the gas \citep{Long_2005}. For a typical T Tauri star, the magnetospheric boundary lies between 0.05 and 0.1 au \citep{Romanova_2006,Romanova_2019}. Interior to this inner radius, the surface density is assumed to rapidly decrease, such that \citep{Brasser_2018}
\begin{equation}
    \Sigma_\mathrm{g}(r<r_\mathrm{in}) = \Sigma_\mathrm{g}(r_\mathrm{in})\times \tanh \left(\frac{r-0.95r_\mathrm{in}}{H_\mathrm{g}(r)} \right).
\end{equation}
The resulting gas cavity stops type I migration, creating a trap for migrating planets between 0.95$\--$1.00 $r_\mathrm{in}$. This is because the decreased surface density prevents planets from forming the density waves responsible for the torques that cause planetary migration. Without an inner edge of the disc, or any other (artificial) pressure bump stopping migration, all planets with a mass comparable to Earth or higher, drift into the central star within a few hundred thousand years of their formation, significantly limiting the odds of finding stars with (exo)planets, which is not in line with observations.  

Particles that drifted even further inwards and crossed the inner truncation radius, $r_\mathrm{trunc}$, were assumed to have accreted onto the star and were removed from the simulation. This $r_\mathrm{trunc}$ is not necessarily a physical parameter, but a computational constraint. Since SyMBA uses a single, fixed time step size, $\Delta t$, for all particles, the total number of steps in the simulation, and therefore the total runtime, is determined by the smallest necessary time step. For symplectic integrators like SyMBA, this time step should typically be around 1/20th of the orbital period at the truncation radius \citep{Wisdom_and_Holman_1991}.    

However, since the formation and evolution region of terrestrial planets lies so close to the star, $r_\mathrm{trunc}$ must be small, compared to simulations focussing on gas giant formation. As a result, these simulations take many months to complete. Given our limited computing time at the Snellius supercomputer ($10^6$ CPU hours), we decided to use slightly longer time steps. For all stellar masses except 0.09 M$_\Sun$, we set $r_\mathrm{trunc}$ to match an orbit with a period of 0.01 year, with the time step equalling 1/15th of this value. To allow for planets to be pushed to stable orbits interior to $r_\mathrm{in}$ by other massive planets, without immediately being removed from the simulation, we set $r_\mathrm{in}$ to an orbit with a period of 0.03 years for most stars, which corresponds to about 0.1 au for a solar-mass star.

These values are not appropriate for the 0.09 M$_\Sun$ star, however. The most famous 0.09 M$_\Sun$-system, TRAPPIST-1, has five planets with periods shorter than 0.03 years, three of which lie within the habitable zone. Two planets have periods shorter than 0.01 year. For this star, we therefore took $r_\mathrm{in}$ at the radius with an orbital period of 0.01 yr, so that all planets in the habitable zone were exterior to it, and $r_\mathrm{trunc}$ at an orbital period of 0.005 yr. Given our computational constraints, we had to employ a step size of 1/10th of the orbital period at $r_\mathrm{trunc}$, and reduce the total number of particles in the simulation.

\subsection{Initial planetesimal distribution} \label{sec: initial_planetesimal_conditions}
As discussed in Sect. \ref{chapter: intro}, a probable source for planetesimals in protoplanetary discs with significant pebble reservoirs is the streaming instability. In short, a positive feedback loop in the back-reaction of the pebbles on the gas can locally eliminate the headwind, as a result of which pebbles stop drifting inwards and start piling up. Once these dense pebble filaments exceed the threshold of $\rho_\mathrm{peb}\sim\rho_\mathrm{g}$, they collapse under their own gravity and form planetesimals with typical radii between about 100 and 400 km \citep{Simon_2016}.  

In this study, we assumed that planetesimals have an initial radius between 175 and 450 km. Since the number distribution of planetesimals is given by a power law, including planetesimals smaller than 175 km rapidly increases the number of particles in the simulation, and therefore the computational load, without significantly influencing the total mass of the planetesimal ring, which, for a shallow slope, is dominated by the more massive particles. To maintain a reasonable planetesimal ring mass, while keeping the number of particles low, we therefore ignored the smallest planetesimals from \citet{Simon_2016} and focussed on slightly more massive ones. Nevertheless, these planetesimals are still significantly smaller than those used by, for instance, \citet{Schoonenberg_2019}.

For the size distribution of the planetesimals, we sampled a truncated Pareto distribution, such that
\begin{equation}\label{eq: Pareto}
    R_\mathrm{pl} = R_\mathrm{pl,min}\zeta^{-1/\beta},
\end{equation}
where $R_\mathrm{min}=175$ km, $\zeta$ is a randomly generated number from a uniform distribution between (0, 1) and $\beta$ is the slope of the Pareto distribution. We assumed $\beta=2.5$, which follows from the collision equilibrium \citep{Dohnanyi_1969}. Only planets with $R_\mathrm{pl}\leq 450$ km were accepted into the simulation. The initial mass of the planetesimals was estimated by assuming they are homogeneous spheres with a bulk density of $\SI{3000}{\kilo\gram\per\cubic\metre}$, and ranges between $10^{-5}$ and $2\times10^{-4}$ M$_\mathrm{E}$. 

For the initial semi-major axis, we tested two different models. In the first model, the planetesimals were uniformly spread over a wide planetesimal ring. These simulations studied a scenario in which pebble accretion dominates over planetesimal accretion because of the relatively low probability of encountering other planetesimals. 

The inner radius of the planetesimal ring in these simulations was set at 0.2 au, so that in every simulation, the most massive planets were able to migrate inwards by some distance before reaching the inner edge of the disc. The outer edge was set at 0.8 au around 0.09 and 0.20 M$_\Sun$ stars, 1.0 au around 0.49 M$_\Sun$ stars, 2.0 au around 0.70 M$_\Sun$ stars, and 2.2 au around 1.00 M$_\Sun$ stars. These outer edges were motivated by initial estimates of pebble accretion efficiencies of planets with different initial masses around the respective stars (see Sect. \ref{sec: PA_afo_r_and_Mini}). Based on these models, planets with radii smaller than 450 km around 0.09 or 0.20 M$_\Sun$ stars do not accrete any pebbles at orbital radii larger than about 0.5 to 1.0 au. Around 0.70 and 1.00 M$_\Sun$ stars, 450 km planetesimals can efficiently accrete pebbles at orbital distances larger than 2.0 au. However, if 
the semi-major axis range for planetesimals around these stars is expanded even further, the distance between them becomes so large that they are virtually isolated and have a very low probability of encountering other planetesimals. Given our limited number of planetesimals, we decided to limit the outer edge of the planetesimal ring to between 2 and 2.2 au for these stars.       

In the second model, the planetesimals were released in a narrow annulus around the snowline. This model was motivated by the fact that the streaming instability requires a locally enhanced pebble density \citep{Carrera_2015,Yang_and_Johansen_2014,Yang_2017,Yang_2018}, which the snowline could provide. Volatiles that evaporate from the pebbles inside the snowline could diffuse back across the snowline where the vapour pressure is low, and re-deposited onto the solids, enhancing the solid-to-gas ratio just outside the snowline \citep{Schoonenberg_and_Ormel_2017,Drazkowska_and_Alibert_2017,Liu_2019}.

The resulting dense filament of pebbles that can collapse into planetesimals has a typical width of $\Delta r \simeq \eta r_\mathrm{snow}$ \citep{Yang_and_Johansen_2016,Li_2016,Liu_2019}, which is approximately 0.0066 au for a solar-mass star using the disc conditions of \citet{Liu_2019}. However, \citet{Liu_2019} assumed a fully irradiated disc with much higher temperatures than we did, which means that both their $\eta$ and $r_\mathrm{snow}$ were larger than ours. Moreover, their simulations allowed for the planetesimals to be injected one by one, so that the simulations have time to create a semi-stable system. SyMBA injects all planets at once, because of which, initialising simulations with such narrow planetesimal rings is highly unstable. Therefore, we started these simulations with a ring that is slightly further expanded. \citet{Liu_2019} found that within a few thousand years, the planetesimal ring has expanded to a width of about 0.1 au, due to gravitational interactions between the planets. We initialised our planetesimals in these simulations with a semi-major axis uniformly taken from between ($0.85r_\mathrm{snow}$,$1.15r_\mathrm{snow}$), with values  shown in Table \ref{tab:simulation_parameters}. 

For both planetesimal distributions, the initial eccentricity and inclination are uniformly generated from the ranges (0, 0.01) and (0$^\degree$, 0.3$^\degree$), respectively. Though these initial ranges are small, the smaller planetesimals are rapidly excited to much larger values due to gravitational interactions with more massive planetesimals. The longitude of the ascending note, argument of periapsis and mean anomaly are all uniformly chosen from the range (0$^\degree$, 360$^\degree$).

Finally, we took an initial planetesimal ring mass of 0.010 M$_\Earth$ for the 0.09 M$_\Sun$ stars, and 0.015 M$_\Earth$ for all other stars. This translates into approximately 275 massive particles in the 0.09 M$_\Sun$ simulations and 400 in the other simulations. \citet{Liu_2019} found that for their disc model, the planetesimal ring around a solar-mass star weighs about 0.039 M$_\Earth$, using the fact that the total solid mass available to build planetesimals is 
\begin{equation*}
    M_\mathrm{avail} = 2\pi r_\mathrm{snow}\Delta r \Sigma_\mathrm{p}(r_\mathrm{snow}) = 2\pi r_\mathrm{snow}\Delta r \Sigma_\mathrm{g}(r_\mathrm{snow})\frac{H_\mathrm{p}(r_\mathrm{snow})}{H_\mathrm{g}(r_\mathrm{snow})},
\end{equation*}
of which approximately 50\% is converted into planetesimals \citep{Simon_2016}. However, using this equation together with our disc conditions results in unreasonably small planetesimal ring masses ($10^{-3}$ M$_\Earth$ for 1.00 M$_\Sun$, 2.8$\times10^{-4}$ M$_\Earth$ for 0.09 M$_\Sun$). As previously mentioned, the core difference is that \citet{Liu_2019} assumed a disc that is significantly hotter and more turbulent ($\alpha_\mathrm{turb}=10^{-3}$), and contains pebbles whose Stokes number is not drift limited, but fixed to 0.1. These three model differences lead to a difference of over an order of magnitude in the estimated initial planetesimal ring mass. 

Assuming our disc's turbulence is enhanced right after the disc's formation, and the pebbles' Stokes number has not yet reached the drift equilibrium and is instead 0.1 at the moment the planetesimals are being formed, our initial ratio $H_\mathrm{p}/H_\mathrm{g}$ is comparable with that of \citet{Liu_2019}. In this case, we find an initial planetesimal ring mass of 0.013 M$_\Earth$ for a solar-mass star, which matches reasonably well with the value of 0.015 M$_\Earth$ we assumed.

 \section{Semi-analytical results: PA as a function of orbital radius and initial planetesimal mass}\label{sec: PA_afo_r_and_Mini}
To gain a preliminary insight into the role of PA in planet formation, and the general trends expected in the full N-body simulation results in Sect. \ref{Chapter: Nbody-results}, the equations of Sect. \ref{chapter: models} were evaluated using a semi-analytical algorithm. This algorithm determined the rate at which solitary planets on idealised orbits accrete pebbles. It did not include planetary migration, nor the influence of planets on each other's orbit or accretion rate. In this section, the influence of a planet's orbital radius and initial mass on its final mass is discussed. In App. \ref{App: sublimation}, the semi-analytical algorithm is used to compare different pebble-sized models inside the snowline.

\begin{figure}
    \centering
    \includegraphics[width=1.\linewidth]{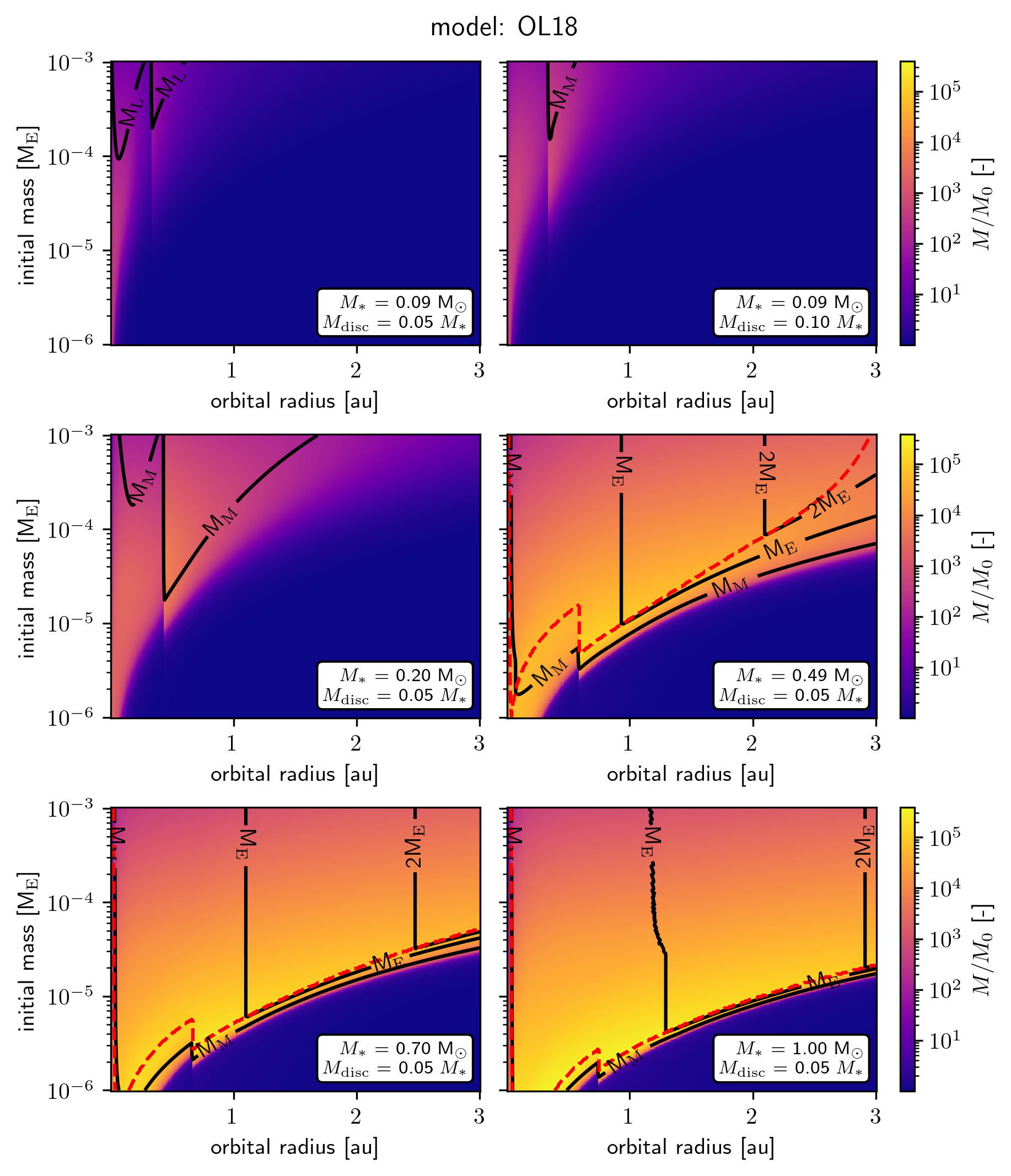}
        \caption[Relative growth of solitary, idealized planets as a function of orbital radius and initial mass, calculated using the OL18 PA model.]{Final mass divided by the initial mass of planets, calculated using only the semi-analytical PA model OL18, shown as a function of orbital radius on the x-axis and initial mass ($M_0$) on the y-axis. The eccentricity and inclination are zero. Black contours indicate final masses higher than a Mars mass ($M_\mathrm{M}$), Earth mass ($M_\mathrm{E}$) and two Earth masses. Planets above the red dashed line have reached the local pebble isolation mass. Around 0.49, 0.70, and 1.00 M$_\Sun$ stars, PA produces Earth-mass planets in a wide range of orbital separations. Around 0.09 and 0.20 M$_\Sun$, PA does not form planets larger than a few Mars masses, even though the pebble isolation mass allows for the formation of Earth-sized planets. For 0.09 M$_\Sun$ stars, planets do not grow larger than about a lunar mass (M$_\mathrm{L}$) in the default disc of 5\% of the stellar mass, which is why a more massive disc of 10\% of the stellar mass is used for this star.}
    \label{fig: Mfinal_Ormel}
\end{figure}

\begin{figure}
    \centering
    \includegraphics[width=1.\linewidth]{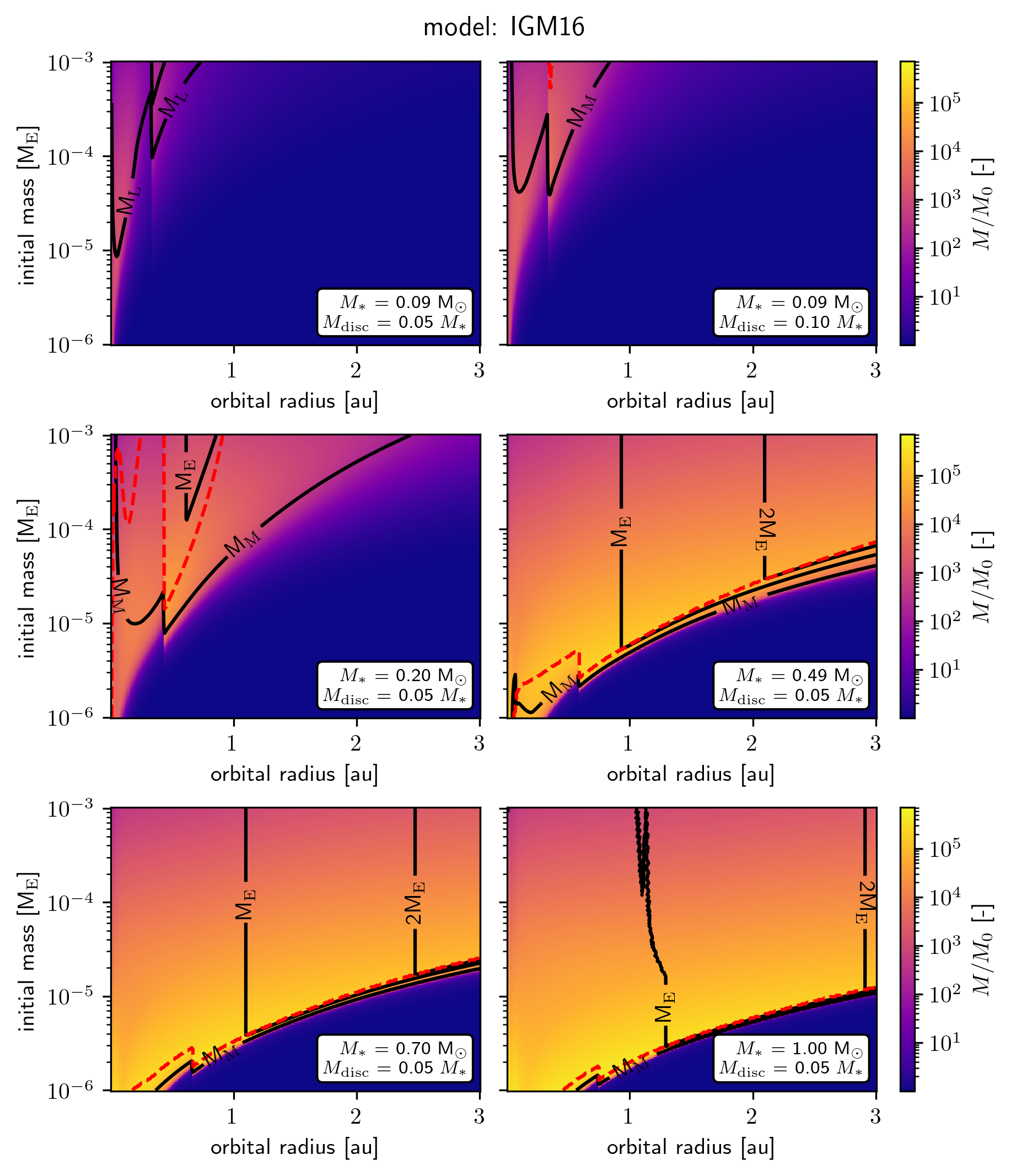}
        \caption[Relative growth of solitary, idealized planets as a function of orbital radius and initial mass, calculated using the IGM16 PA model.]{Same plots as Fig. \ref{fig: Mfinal_Ormel}, but calculated using the semi-analytical PA model IGM16. Pebble accretion using IGM16 is more efficient than using OL18, even for the circular, uninclined orbits assumed here. This is most clearly visible from the contours in the 0.09 and 0.20 M$_\Sun$ subplots.}
    \label{fig: Mfinal_Ida}
\end{figure}

Figure \ref{fig: Mfinal_Ormel} shows the relative growth ($M_\mathrm{final}$/$M_\mathrm{initial}$, $M/M_0$ for short) of these idealised, solitary planets as a function of orbital radius on the x-axis and initial mass on the y-axis. These results were calculated using the PA prescription of \citet{Ormel_and_Liu_2018} (OL18). The results for the prescription of \citet{Ida_2016} (IGM16) are shown in Fig. \ref{fig: Mfinal_Ida}. 

The orbits of the planets in these calculations are circular and uninclined. The black contours indicate the regions of the parameter space within which planets grow more massive than a Mars mass (M$_\mathrm{M}$), an Earth mass (M$_\mathrm{E}$) and two Earth masses. The contour lines for planets more massive than 5 and 10 Earth masses are never reached in these inner regions of the disc due to the pebble isolation mass. The red dashed contour line indicates the boundary above which the planets have reached the pebble isolation mass and cannot grow any further. 

Based on the results in Figs. \ref{fig: Mfinal_Ormel} and \ref{fig: Mfinal_Ida}, neither the OL18 nor the IGM16 PA model seems able to produce Earth-mass planets around 0.09 and 0.20 M$_\Sun$ stars, at least not for the initial mass range of planetesimals in the SyMBA simulations, even though the isolation mass around these stars does allow for the formation of Earth-like planets. The IGM16 prescription allows for slightly more efficient growth, but even in this scenario, the planets do not grow much heavier than a Mars mass. In fact, in calculations with a 0.09 M$_\Sun$ star with a 0.05 M$_*$ disc, planets only reach up to about a lunar mass, being limited primarily by the low pebble mass flux, totalling only a few Earth masses (see Fig. \ref{fig: Mdotf0} for the pebble mass flux in a 0.10 M$_*$ disc for the 0.09 M$_\Sun$ star). We predict that it is highly improbable for Earth-mass planets to form in an N-body simulation with such conditions. Therefore, we focussed on a more massive disc of 0.10 M$_*$ around 0.09 M$_\Sun$ stars, which contains twice the pebble flux a 0.05 M$_*$ disc does. Nevertheless, we predict planetesimal accretion and mergers between several high mass embryos must play an important role for Earth-like planets and TRAPPIST-1-like systems to emerge from the N-body simulations with these low-mass stars.

A final, noteworthy feature in the 0.09 and 0.20 M$_\Sun$ results of Figs. \ref{fig: Mfinal_Ormel} and \ref{fig: Mfinal_Ida} is the discontinuity in $M/M_0$ at an orbital radius shorter than 1 au, the exact location of which depends on the mass of the central star. This feature is caused by the snowline, interior of which the pebble mass flux is halved. It is most visible in the 0.09 and 0.20 M$_\Sun$ stars calculations. For the more massive stars, the feature is less apparent, since most of the planets in these calculations are not limited by the available pebble flux and accretion efficiency, but by the pebble isolation mass.

Around the 0.49, 0.70, and 1.00 M$_\Sun$ stars, planetesimals systematically grow into Earth-mass planets for a wide range of orbital separations, for both OL18 and IGM16. However, unlike what Figs. \ref{fig: Mfinal_Ormel} and \ref{fig: Mfinal_Ida} might suggest, only a few planetesimals are expected to actually grow to Earth-like masses in the full SyMBA simulations of Sect. \ref{Chapter: Nbody-results}, because the planets are not solitary, and need to share the total available pebble flux. Moreover, if a planet relatively far out in the disc reaches the isolation mass early on, it blocks the pebble flux to all planets interior to it, halting their growth.

\begin{figure}
    \centering
    {\includegraphics[width=1.\linewidth]{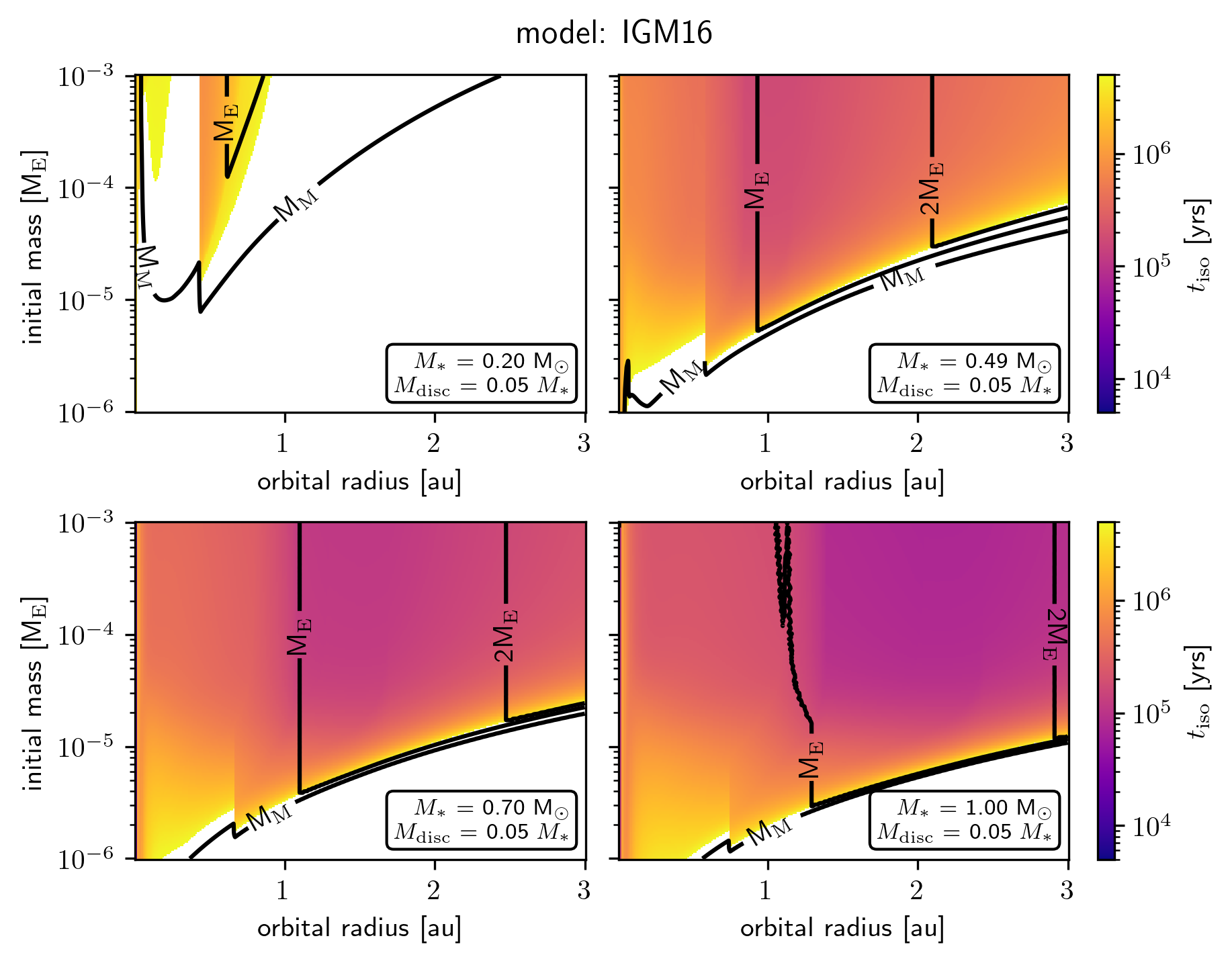}}
        \caption[Time after which idealized planets reach the pebble isolation mass, as a function of orbital radius and initial mass.]{Time after which planets reach the pebble isolation mass $t_\mathrm{iso}\equiv t(M$$=$$M_\mathrm{iso})$, as a function of orbital radius and initial mass, using IGM16. White indicates the isolation mass has not been reached. The contours showing the final mass are the same as in Fig. \ref{fig: Mfinal_Ida}. Around 0.49, 0.70, and 1.00 M$_\Sun$ stars, the first planets reach $M_\mathrm{iso}$ within the first 50,000 to 100,000 years of the simulation, which could halt pebble accretion for planets close to the star prematurely.}\label{fig: tiso_Ida}
\end{figure}

The time at which the planets in Fig. \ref{fig: Mfinal_Ida} reach their isolation mass ($M_\mathrm{iso}$) is shown in Fig. \ref{fig: tiso_Ida} for IGM16. The situation for OL18 is similar, except for the fact that the isolation mass is never reached around 0.20 M$_\Sun$ stars. The OL18 results are therefore not separately shown. Around 0.49, 0.70, and 1.00 M$_\Sun$ stars, the first planets reach $M_\mathrm{iso}$ within the first 50,000 to 100,000 years of the simulation. The range of orbital radii for which $M_\mathrm{iso}$ is rapidly reached, is wider, and the time in which the planets do so is shorter, the more massive the central star is. This could limit the growth of planets closer to the star prematurely, especially for planets inside the snowline, where growth is slower due to the reduced pebble mass flux. This also suggests that the initial disc conditions are more important than how the conditions evolve over time, at least for the largest planets in the system.

 \section{Full N-body simulation results}\label{Chapter: Nbody-results}
In this section, the results from the SyMBA N-body simulations are presented. These simulations include pebble accretion, planetesimal accretion, and type I and type II migration. Simulations were performed for 0.09, 0.20, 0.49, 0.70, and 1.00 M$_\Sun$ stars, and for four different models: two models based on the PA prescription of \citet{Ormel_and_Liu_2018} (OL18 and OL18-Ring), and two for the prescription of \citet{Ida_2016} (IGM16 and IGM16-Ring). In the 'Ring' simulations, the planetesimals were initialised on a narrow ring around the snowline. In the other simulations, the planetesimals were released over a wider range of orbital radii. All other disc conditions were identical for the different models (see Sect. \ref{sec: parameters} for more details). 

For each combination of stellar mass and model, eight simulations were performed with randomly generated planetesimals. The simulations were run for 5 Myrs of evolution. During the first $\sim$2.5 $\--$ 3 Myrs of growth, all bodies in the simulation were self-gravitating. During the final stages of the simulations, only planetesimals more massive than 10$^{-3}$ M$_\mathrm{E}$ were considered self-gravitating, in order to reduce the computational load caused by small planetesimals. As will be shown in the sections below, nearly all planets form within the first million years. The influence of the smallest planetesimals during the final stages of the simulations is negligible.

Each simulation\footnote{The simulations around the 0.09 M$_\Sun$ stars contained ${\sim}275$ planetesimals, had a truncation radius with an orbital period of 0.005 yr, and used a time step of $5\times10^{-4}$ yr.} consisted of ${\sim}400$ planetesimals, with radii between 175 and 450 km, following a truncated Pareto distribution (see Eq. \ref{eq: Pareto}). Planets that moved interior to the truncation radius, defined as the location around the star with an orbital period of 0.01 yr, were removed from the simulation. The size of the time step was $6.7\times10^{-4}$ yr. In total, the simulations took approximately 6 months to complete, using 4 CPU cores per simulation.

In Sect. \ref{sec: 1.00Msun_case_study}, the simulation results for a solar-mass star are presented, and two specific simulations are discussed in detail. The results for the other stars are provided in Sect. \ref{sec: other_star_results}. Finally, the long-term stability of the systems, and the influence of gas accretion, which was not included in the standard results, are analysed in Sects. \ref{sec: system_stability} and \ref{sec: gas_accretion}.

\subsection{Overview of the simulation results for a solar-mass star} \label{sec: 1.00Msun_case_study}

\begin{figure}
    \centering
    {\includegraphics[width=1.0\linewidth]{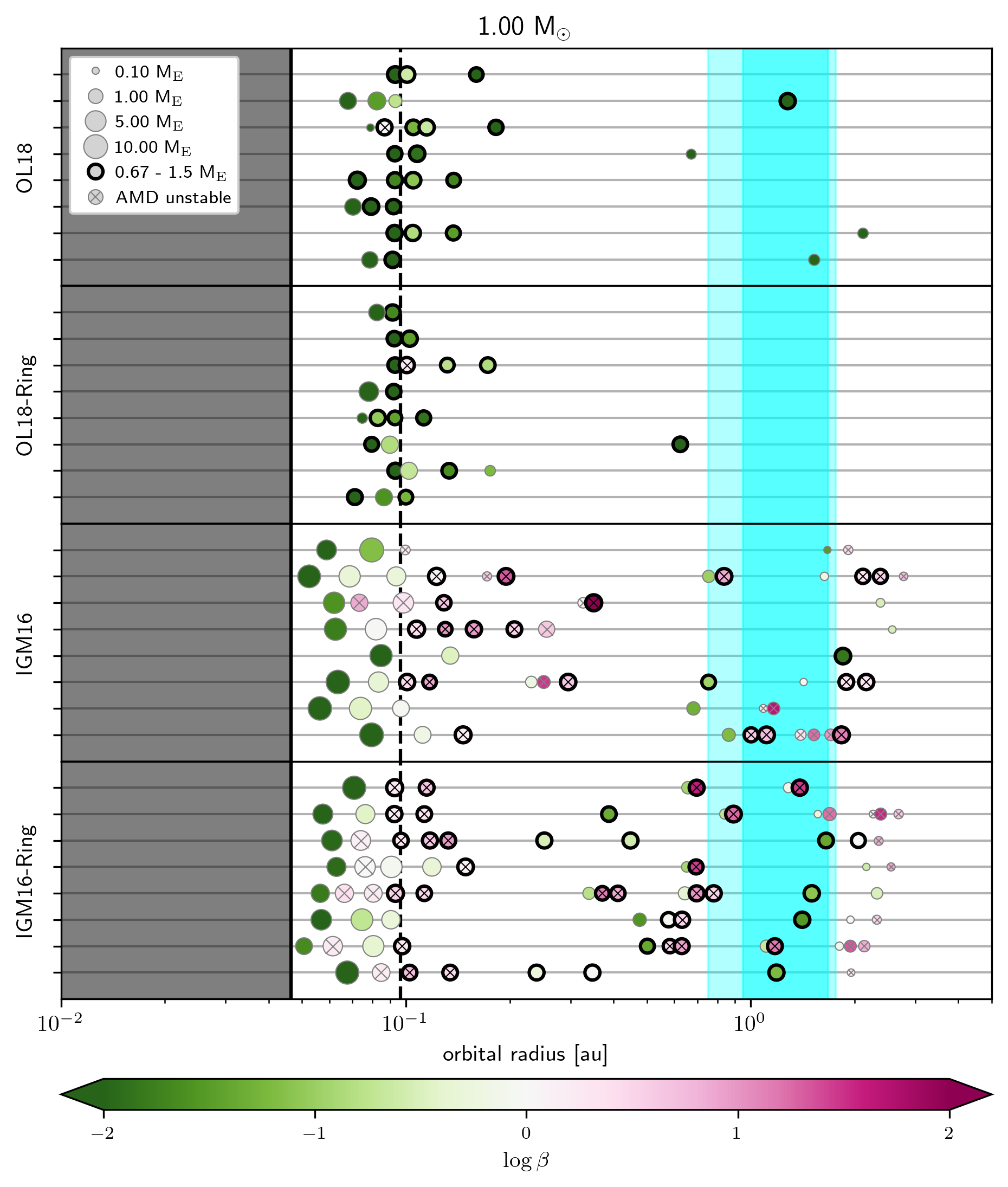}}
        \caption[Simulated planetary systems around a 1.00 M$_\Sun$ star for the four PA models.]{Simulated planetary systems around a 1.00 M$_\Sun$ star for the four PA models. The size of the markers represents the mass of the planet. The simulations are grouped by their model, and each horizontal line represents a simulation. The vertical dashed line at $\sim$$10^{-1}$ au is the inner radius of the gas disc, within which the gas density rapidly drops and type I and type II migration halts. Any planet that enters the gray region, interior to the inner truncation radius, is removed from the simulation. The cyan shaded regions represent the conservative (darker shaded) and optimistic (lighter shaded) habitable zone. The colour of the planets indicates their AMD stability, discussed in Sect. \ref{sec: system_stability}. Only planets more massive than Mars are shown.}
    \label{fig: 1.00Msun_summary}
\end{figure}

The resulting planetary systems after 5 Myrs of simulation are shown in Fig. \ref{fig: 1.00Msun_summary}. This figure shows all simulations for all four models around a solar-mass star. The first thing that stands out is that, as predicted in Sect. \ref{sec: PA_afo_r_and_Mini}, the IGM16 PA-prescription generates planets in greater number, and of significantly higher mass than the OL18 prescription. Nevertheless, regardless of the PA model, every system produces one or more Earth-like planets, which we define as planets with masses between 0.67 and 1.5 M$_\mathrm{E}$ (indicated by the black edge around the markers). Earth-like planets appear in comparable quantities in the general, wide planetesimal disc, spanning from 0.2 to 2.2 au, as in the narrow planetesimal ring around the snowline, spanning from about 1.2 to 1.6 au, even though in the latter, the probability of planetesimals coming together is larger. It is therefore unlikely that in our model, significant planetesimal accretion is a requirement for the formation of Earth-like planets, and even for these relatively small planetesimals of up to 450 km in radius, PA is sufficient for the formation of large planets.

Nearly all of these large planets have migrated to the innermost regions of the disc, especially in the OL18 simulation. The inner edge of the gas disc $r_\mathrm{in}$ is indicated by the vertical dashed line at about 0.1 au. This inner edge acts as a trap, preventing the planets from migrating further inwards because of the rapidly decaying gas surface density, which prevents the planets from producing the density waves that generate the torques required for planet migration. Without this boundary, all planets would have continued drifting into the central star. 

The current understanding of the mechanisms preventing planets from migrating too close to the star is still incomplete, mainly because of uncertainties in the influence of torques near the disc's inner edge \citep[see e.g.][]{Brasser_2018}. The use of a gas cavity within $r_\mathrm{in}$ is a simple, yet general solution, motivated by the star's magnetosphere disrupting the innermost regions of the disc, causing the gas surface density to rapidly drop at around 0.05 to 0.1 au \citep{Long_2005,Romanova_2006,Romanova_2019}.

Nevertheless, many planets have migrated significantly further inwards than $r_\mathrm{in}$. This is due to mean motion resonances \citep[MMRs;][]{Terquem_and_Papaloizou_2007}. Migrating protoplanets often get captured in MMRs, forming chains of low-mass planets with orbital periods that are in resonance, migrating through the disc together. As the first planet reaches the inner edge of the disc, the entire chain slows down and stalls close to $r_\mathrm{in}$. However, the resonance chain also transfers part of the torque that is experienced by the planets that are still in the gas disc, to the planets that are in the gas-free region, pushing them further inwards. 

If the chain becomes too long and massive, it become dynamically unstable, leading to orbital crossings, particles being ejected, and giant impacts \citep{Izidoro_2017,Izidoro_2021}, which break the resonances. This typically happens when the gas disc disperses, or shortly thereafter. Since the disc in this study is exponentially drained with a diffusion time of 0.5 Myrs, there is not a specific moment at which the gas of the disc has fully dissipated, but generally, most collisions happen within the first ${\sim}2$ Myrs \citep{Ogihara_2015,Zawadzki_2021,Hatalova_2023}, though late dynamic instabilities can occur for up to a 100 Myrs after the formation of the disc (not modelled in this study). 

Evidence of giant impacts is seen in the IGM16 results in Fig. \ref{fig: 1.00Msun_summary}. Many of the planets in the innermost regions of these discs have masses exceeding 5 or even 10 M$_\mathrm{E}$, far above the pebble isolation mass, which is closer to 1 or 2 M$_\mathrm{E}$ (see Figs. \ref{fig: Mfinal_Ormel} and \ref{fig: Mfinal_Ida}). The IGM16 models produce so many large planets, that the systems become unstable, causing the many protoplanets to merge into giant planets when the MMR chain reaches $r_\mathrm{in}$.

\subsection{Planets in the habitable zone around solar-mass stars}
A consequence of the rapid planetary migration is that very few planets remain in the habitable zone (HZ), at least in simulations with the OL18 model. The location of the conservative and optimistic HZ are shown using the darker and lighter cyan shading in Fig. \ref{fig: 1.00Msun_summary}. The HZ was calculated using the algorithms from \citet{Kopparapu_2013, Kopparapu_2014}, and assumed solar values for the effective temperature and luminosity, and a planet mass of 1 M$_\mathrm{E}$. The planet mass influences the expected thickness of the atmospheres, and therefore the maximum strength of the greenhouse effect. 

The normal OL18 simulations produced only one Earth-like planet in the HZ, and one Mars-like planet, in the eight different realisations of the system, with two other simulations with Mars-like planets close to the HZ. The OL18-Ring simulations produced even fewer planets in or close to the HZ, even though all planetesimals started in the HZ. This shows that forming an Earth-like planet in or around the HZ using PA is not as challenging as keeping it there, given the rapid migration.\\ 

The IGM16 simulations produce far more planets in the habitable zone, even more in the 'Ring' configuration than in the normal planetesimal distribution. This is because the IGM16 model is more efficient than the OL18 model at accreting pebbles onto small planetesimals (see Sect. \ref{sec: PA_afo_r_and_Mini}), especially onto those that have been excited (see Fig. \ref{fig: orbavg_ecc_inc}), since the reduction in accretion efficiency for planets on eccentric orbits is ignored. The IGM16 prescription is therefore far more likely to produce planets at late stages of the disc evolution than the OL18 prescription. The gas of the disc dissipates before these late-forming planets have time to migrate significantly inwards, allowing them to remain in the HZ. 

\subsection{Dynamical evolution of solar-like systems} \label{sec: dynamic_evolution}

\begin{figure*}
    \centering
    {\includegraphics[width=0.8\linewidth]{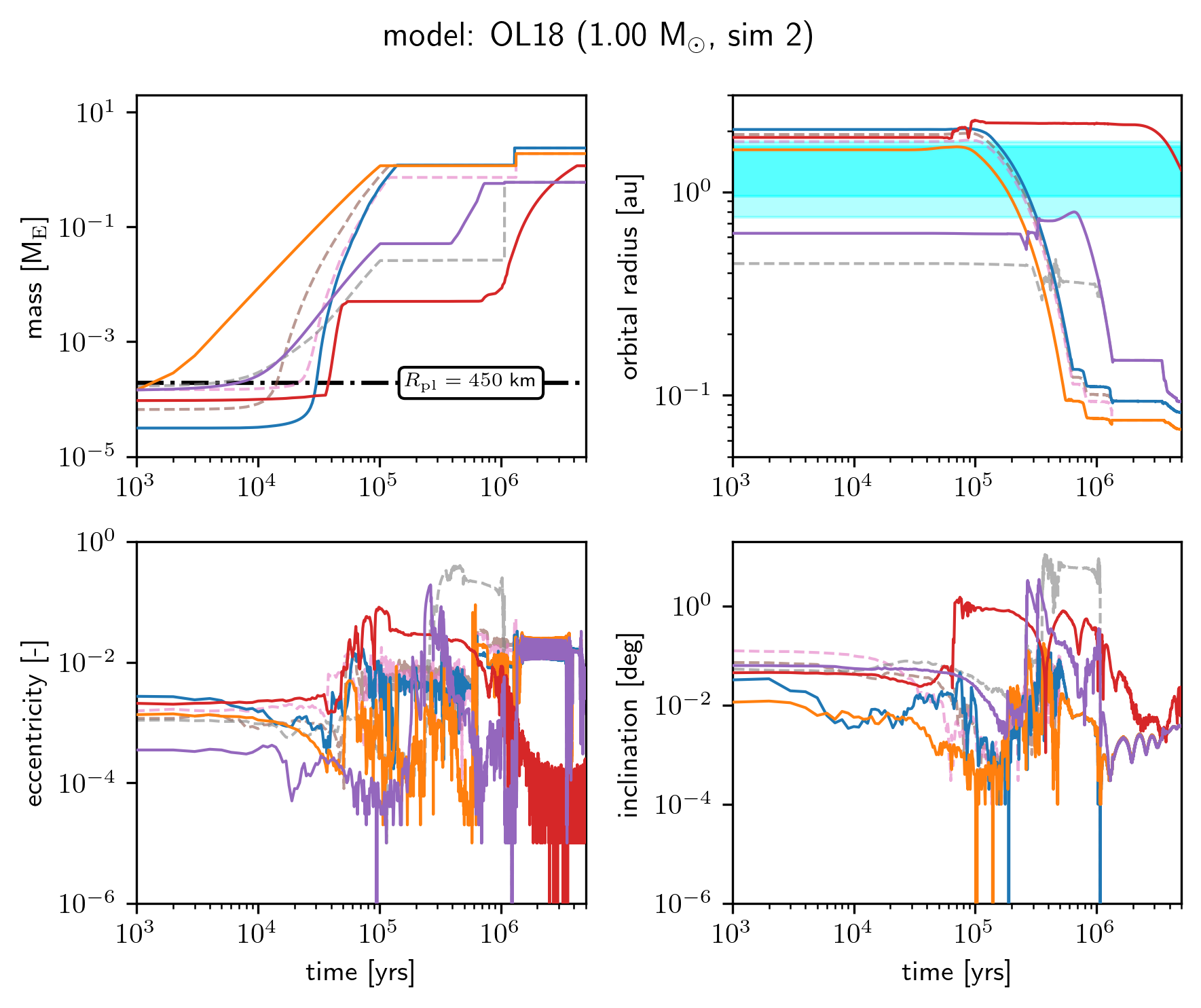}}
        \caption[The dynamical evolution tracks ($M_\mathrm{p},a,e$ and $i$) of all large planets in a single OL18 simulation for a solar-mass star.]{Dynamical evolution tracks ($M_\mathrm{p},a,e$ and $i$) of all large planets in OL18 simulation 7 around a solar-mass star. The different coloured solid lines represent different planets, the red one being the planet that is currently in the habitable zone (cyan shaded region). The transparent dashed lines represent large planetesimals that merged with the planets. Sudden stepwise increases in planet masses signify these mergers. Only planets with masses $> 0.1$ M$_\mathrm{E}$ are included; however, for this simulation, there are no other objects with masses of $\gtrsim 10^{-3}$ M$_\mathrm{E}$.}
    \label{fig: 1.00Msun_case_OL18}
\end{figure*}

To get a closer look into the general growth track of planets, and the formation of planets in the HZ in particular, we present an analysis of the dynamic evolution of OL18 simulation 2, which produced the Earth-like planet in the HZ, in Fig. \ref{fig: 1.00Msun_case_OL18}. This simulation shows both the general trends observed in OL18 simulations, and the specific sequence of events that led to the formation of an Earth-like planet in the HZ.  

\subsubsection{The dynamical evolution of OL18 systems}
As can be seen in Fig. \ref{fig: 1.00Msun_case_OL18}, the first massive planets form within the first 0.1 Myrs of the simulation, as predicted by Fig. \ref{fig: tiso_Ida}. Most of these planets formed from planetesimals at the high end of the size distribution. Higher mass seeds have an advantage over lower-mass seeds, since they can efficiently accrete pebbles for higher values of $e$ and $i$, and have a higher chance of starting their growth early on, when there is still little competition for the pebble flux. However, the initial mass is not the primary limiting factor to the growth of planets.

This is demonstrated by the blue planet, which grew to be the biggest in the system, despite its seed being only $3.125\times 10^{-5}$  M$_\mathrm{E}$ (245 km in radius). One of the reasons this planet was able to grow so large is that it was initialised close to the outer edge of the planetesimal disc, at 2.02 au from the star. With no large seeds exterior to it, and with the interior seeds being far enough not to perturb its orbit, the blue planet was allowed to develop undisturbed. 

The blue planet's exponential growth started equally abruptly as that of the brown and pink protoplanets (shown dashed and transparent, for they later merged with other planets), which were initialised around the same region as the blue planet, but with significantly higher mass. The blue planet did not need to first gradually grow to a specific mass threshold before entering the rapid accretion regime. Instead, it had to rid itself of its initial eccentricity and inclination through disc interactions. 

With a starting eccentricity and inclination of $e=2.75\times10^{-3}$ and $i=3.25\times10^{-2}$, the blue planetesimal was initialised in the inefficient ballistic regime (see Fig. \ref{fig: orbavg_ecc_inc} for reference). Only when its eccentricity dropped below 10$^{-3}$, at around 20,000 yrs, did the planet enter the settling regime, and start runaway pebble accretion. By the time its eccentricity increased again, as a result of a close encounter with the red planet at around 50,000 yrs, the blue planet was sufficiently massive for it to remain in the settling regime, despite its high eccentricity. 

After about 100,000 yrs, the blue planet reached the pebble isolation mass. The orange planet reached the isolation mass earlier, which promptly stopped the growth of the purple and gray (dashed) planets, but since the blue planet was exterior to the orange one, the blue planet could continue to grow. After the blue planet reached its isolation mass, it started migrating inwards, joining in an MMR chain with the brown (dashed), pink (dashed) and orange planet. Together, the chain migrated to the inner edge of the gas disc at 0.09 au in about 0.5 Myrs. 

Several dynamic instabilities, one of which being caused by the purple planet migrating inwards and joining the chain, led the brown and pink planets to collide with the blue and orange planets, respectively, and forced the latter two into the gas cavity. For a more detailed discussion about different mean motion resonance structures and their dynamic evolution, we refer to, for example, \citet{Brasser_2022}, and \citet{Hatalova_2023}.

\begin{figure*}[h]
    \centering
    {\includegraphics[width=0.8\linewidth]{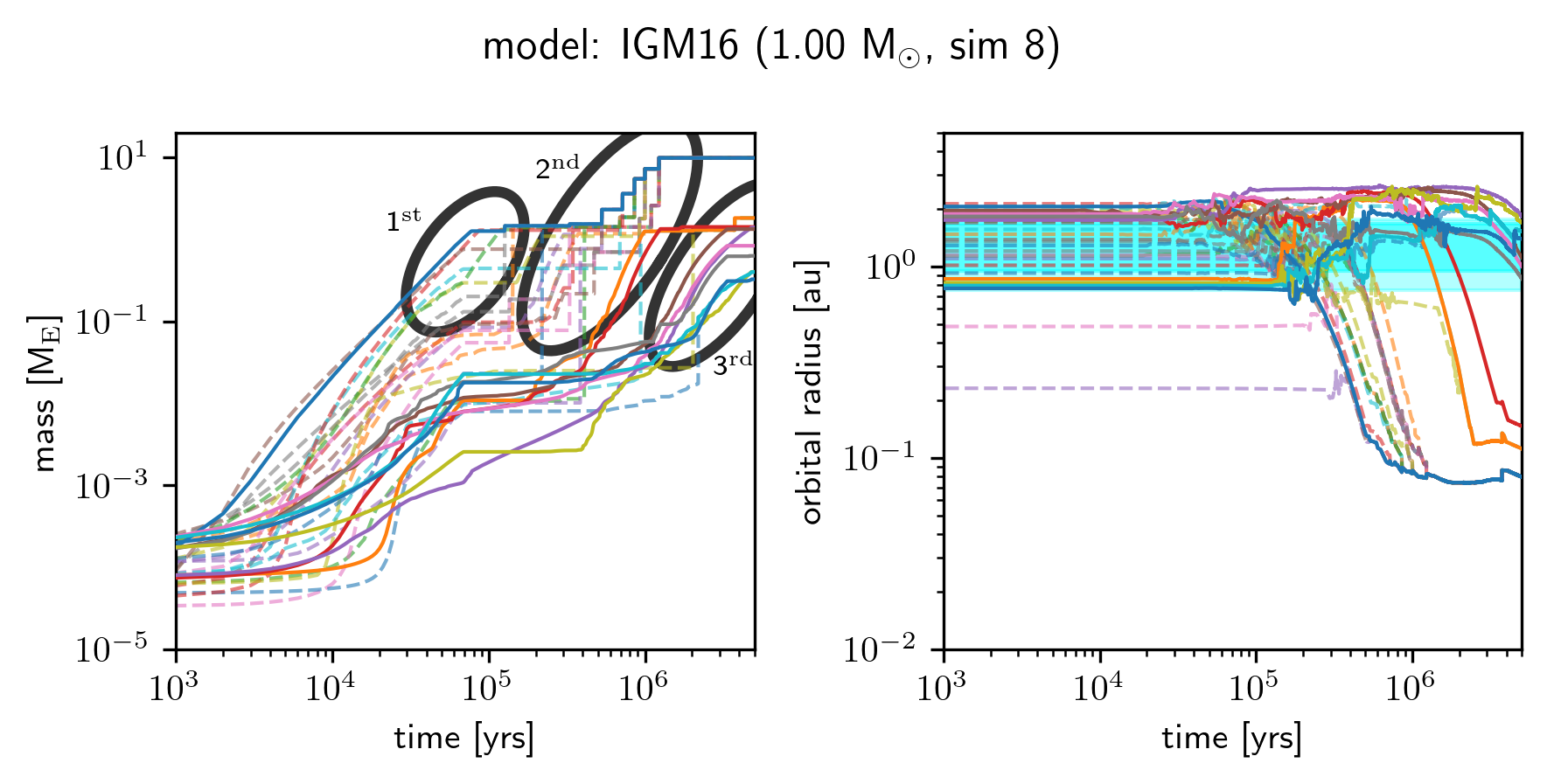}}
        \caption[The dynamical evolution tracks ($M_\mathrm{p}$ and $a$) of all large planets in a single IGM16 simulation for a solar-mass star.]{Dynamical evolution tracks ($M_\mathrm{p}$ and $a$) of all large planets in IGM16 simulation 1 around a solar-mass star. The different coloured solid lines represent different planets. The transparent dashed lines represent large planetesimals that merged with the planets. Sudden stepwise increases in planet masses signify these mergers. Only planets with masses $> 0.1$ M$_\mathrm{E}$ are included. The cyan shaded region represents the habitable zone. The thick black ellipses in the lefthand panel highlight the first, second, and third generation of planets. IGM16 produces far more planets than OL18. Those formed as part of the third generation, after 1 $\--$ 2 Myrs, remain in the HZ because they have insufficient time for migration.}
    \label{fig: 1.00Msun_case_IGM16}
\end{figure*}
The evolution of the planets discussed above is very typical for all OL18(-Ring) systems. The first planets start growing within 1,000 to 10,000 yrs of the planetesimal formation. After about 50,000 to 100,000 yrs, the first planets in the outer disc reach the isolation mass, halting the growth of the protoplanets interior to them. Within about 0.5 Myrs, the most massive planets in the outer disc migrate to the inner edge of the disc, forming a resonance chain with the other large planets they drag with them. When these large planets cross the orbits of the smaller protoplanets closer to the star, they excite the orbits of the smaller protoplanets. If the mass of these smaller protoplanets is great enough for them to quickly lose the induced eccentricity and inclination through disc torques \citep[e.g.][]{Matsumura_2021}, they resume their growth. Within the next ${\sim}1$ Myrs, this second group of planets grows to the isolation mass, and migrates to the inner edge of the disc, joining the MMR chain. This often leads to multiple dynamic instabilities, which push the innermost planets into the gas cavity, and cause other planets to merge or be ejected.

\subsubsection{The dynamical evolution of OL18 planets in the habitable zone}
The formation of Earth-like planets in the habitable zone (HZ) in OL18 simulations is less trivial, and therefore far less common, than the general formation of Earth-like planets in these systems. Since most Earth-mass planets are formed in just a few ten thousand years, but rapidly migrate to the inner edge of the disc within the next million years, a very specific sequence of events is required for an Earth-like planet to remain in the HZ. The only realisation for this case is the red planet in Fig. \ref{fig: 1.00Msun_case_OL18}. 

This planetesimal was initialised relatively far out in the disc at 1.85 au, which is a prerequisite for any planet in our simulations to end up in the HZ, since planets in our model generally do not migrate outwards. Like the other planets, the red planet entered runaway pebble accretion once its eccentricity and inclination dipped far enough for pebbles to settle into its gravitational field, at about 35,000 yrs into the simulation. However, its growth was stopped at a very specific moment in its evolution, which allowed it to end up in the habitable zone. This happened due to two consecutive close encounters, first with the brown (dashed) planet and then with the blue planet. These encounters kicked the red planet even further out in the disc and increased its eccentricity and inclination by about two orders of magnitude, which halted its growth. The key element of these events is that the planet had exactly the right mass to eventually end up in the HZ. Had it been more massive, then it would have dampened its eccentricity and inclination faster, reached an Earth-like mass sooner, and it would have had time to migrate to the inner disc, as the purple planet did. Had it been less massive, then it might not have lost its eccentricity in time, and would have remained a sub-lunar object, like all the other planetesimals in its vicinity. 

In fact, the mass of the red planet at the time of its excitation might have already been slightly too high for it to remain in the HZ. After all, at the end of the simulation, the planet is still migrating inwards. This migration might quickly cease due to the gas disc becoming too thin, especially if we assume photoevaporation kicks in, and blows away the remainder of the gas. Nevertheless, it might also be that this planet drifts out of the HZ because of its mass, just as all the others, if the simulations were continued for another million years.

On the other hand, the planet required 4.5 Myrs to reach its final mass, which means its mass should not have been much lower either at the time of its excitation, for it would not have had enough time to grow then. This only demonstrates how specific the conditions need to be for an Earth-like planet to remain in the HZ when using a simple migration model.

Nevertheless, since the planet formed outside the snowline and accreted all of its mass from pebbles there, it consists for up to 50\% of water. Even if a large fraction of this water is lost to evaporation due to the heat from the planet's formation, there should still be more than enough left for the planet to possibly be suitable for life.

\subsubsection{The dynamical evolution of IGM16 systems}\label{sec: dynamic_evo_IGM16}
Figure \ref{fig: 1.00Msun_case_IGM16} shows the dynamic evolution of a single system with the IGM16 PA-prescription. The evolution of the largest planets follows the same patterns as of those in the OL18 systems. The main difference is that IGM16(-Ring) produces far more planets than OL18(-Ring). In fact, the IGM16 PA-prescription is so efficient that it is able to create a third generation of planets. This is most likely due to the fact that the influence of the eccentricity and the ballistic regime are not included in IGM16, which were the main limiting factors for growth in the OL18 simulations. As the number of massive planets in the OL18 simulation grows, the other planetesimals become increasingly excited, making it less and less likely that additional planets form. By ignoring this negative feedback loop, the IGM16 model likely significantly overestimates the probability of planets arising from the planetesimal disc.  

Either way, similarly to the OL18 simulations, the first generation of planets forms far out in the disc, reaches the isolation mass within the first 50,000 to 100,000 yrs, and then migrates inwards. The second generation of planets forms when the first generation moves interior to it, and reaches the isolation mass at about 1 Myrs into the simulation. However, unlike in the OL18 model, as the second generation migrates inwards, a third generation of planets has time to grow, starting its final growth phase between 1 and 2 Myrs into the simulation. These planets do have time to grow to Earth-like masses, but not to migrate significantly inwards, which is why IGM16 simulations produce significantly more planets further out in the disc, in particular in the habitable zone. 

Furthermore, since the system is oversaturated with Earth-mass planets, it is highly unstable, leading to a lot of mergers between the innermost massive planet, and the other planets joining the MMR chain. As a result, the innermost planet absorbs many of its neighbours, growing to over 10 M$_\mathrm{E}$. Gas accretion was not included in these simulations, and because of the low pebble isolation mass in these regions of the disc, planets generally do not grow massive enough to start gas accretion, for which cores of masses between 5 and 10 M$_\mathrm{E}$ are required \citep{Mizuno_1978, Stevenson_1982, Bodenheimer_and_Pollack_1986, Hubickyj_2005}. However, due to these many mergers between massive planets, the IGM16 model is capable of creating gas giants close to the star. These are discussed in Sect. \ref{sec: gas_accretion}. 

On a final note, both Fig. \ref{fig: 1.00Msun_case_OL18} and Fig. \ref{fig: 1.00Msun_case_IGM16} suggest a clear preference for planets to form in the outer regions of the planetesimal disc ($r>\sim 0.8$ au), even though the planetesimals in these simulations are evenly distributed between 0.2 to 2.2 au. An important factor is likely that the pebble mass flux inside the snowline, which starts at about 1.4 au, is reduced by 50\%, giving planets outside the snowline a significant advantage. Another reason could be that the Stokes number in the inner disc is too high for the small planetesimals to efficiently accrete pebbles due to the high relative velocity (see Fig. \ref{fig: rpeb_in_rsnow}).

 \subsection{Planet formation for different stellar masses} \label{sec: other_star_results}

\begin{figure*}[]
    \includegraphics[width=\textwidth]{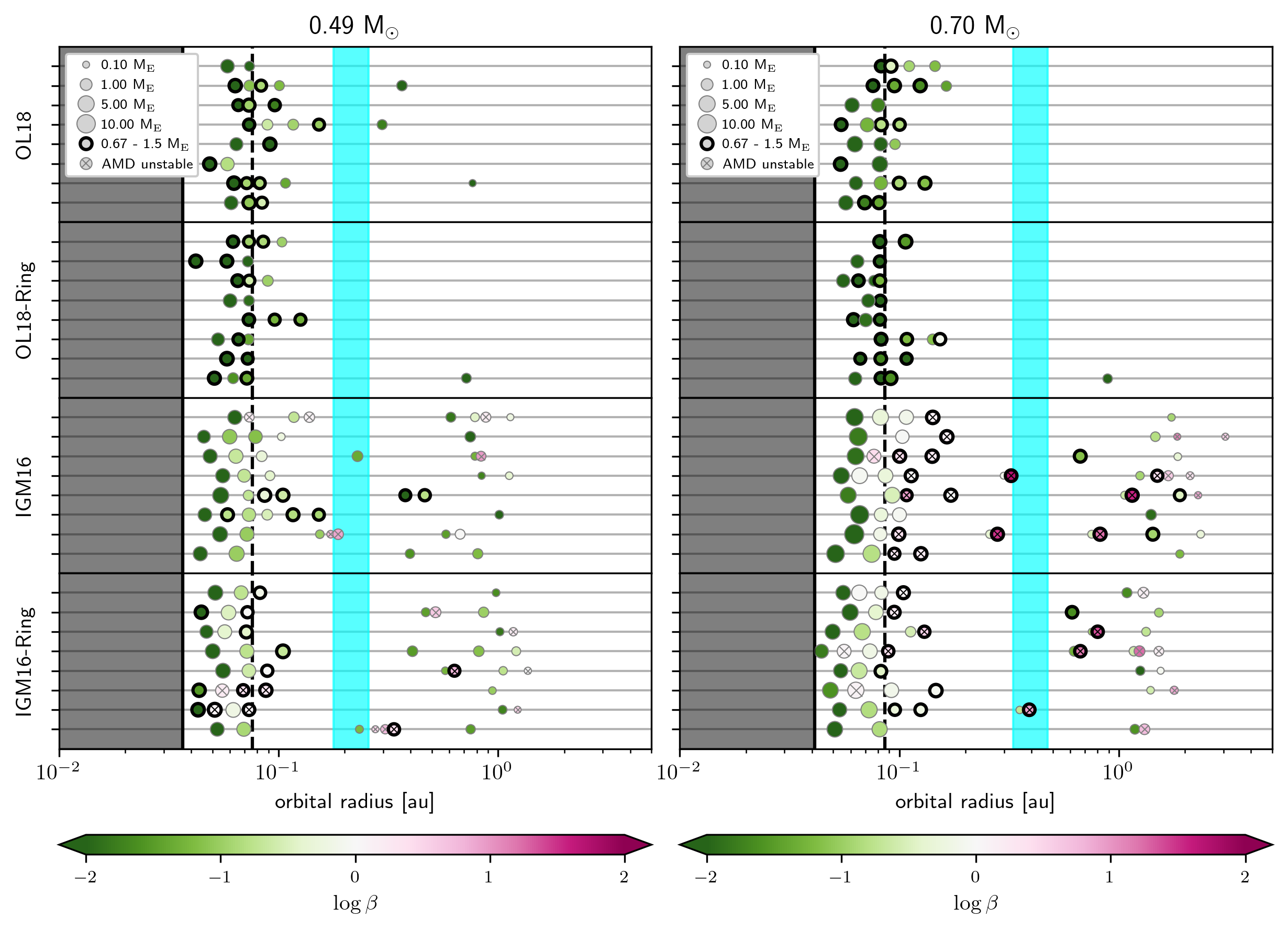}
    \caption[Simulated planetary systems around the 0.49 and 0.70 M$_\Sun$ stars for the four PA models.]{Simulated planetary systems around the 0.49 (left) and 0.70 (right) M$_\Sun$ stars for the four PA models. The size of the markers represents the mass of the planet, while their colour represents the planet's AMD stability (see Sect. \ref{sec: system_stability}). The layout of the figure is explained in the caption of Fig. \ref{fig: 1.00Msun_summary}.}
    \label{fig: 0.49+0.70_summary}
\end{figure*}
\subsubsection{0.49 and 0.70 M$_\Sun$ stars}
The results of the simulations around 0.49 and 0.70 M$_\Sun$ stars are shown in Fig. \ref{fig: 0.49+0.70_summary}. The results are very similar to those around a solar-mass star. Earth-like planets are systematically formed, generally more than one per system. This is due to the fact that the pebble isolation mass to which nearly all planets grow in our simulations lies at around an Earth-mass for our assumed disc conditions, limiting the formation of larger cores.

Similarly to in the solar-mass simulations, nearly all planets in the OL18(-Ring) simulations migrate to the inner edge of the disc, where they form mean motion resonance (MMR) chains, which push some planets even further in. Ultimately, no planet remains in the habitable zone. To avoid making assumptions about the effective temperature of the stars, the inner and outer edge of the habitable zone in these simulations were calculated using the simpler relationships: 
\begin{equation*}
    r_\mathrm{i}=\sqrt{\frac{L/L_\Sun}{1.1}} \quad \mathrm{and} \quad r_\mathrm{o}=\sqrt{\frac{L/L_\Sun}{0.53}},
\end{equation*}
in which 1.1 and 0.53 are constants representing the stellar flux at the inner and outer edge of the HZ, respectively \citep{Kasting_1993,Whitmire_1996}. Furthermore, $L$ is the absolute bolometric luminosity of the star on the main sequence, which for M and K dwarfs is approximately given by \citep{Cuntz_2018}:
\begin{equation*}
    \frac{L}{L_\Sun} = \left(\frac{M_*}{M_\Sun} \right)^{n(M)}\quad \left(0.20\ \mathrm{M}_\Sun \leq M \leq 0.85\ \mathrm{M}_\Sun\right), 
\end{equation*}
with the fitted exponent
\begin{equation*}
    n(M)=-141.7M^4+232.4M^3-129.1M^2+33.29M+0.215.
\end{equation*}
This estimate of the HZ is generally much more narrow than both the conservative and optimistic HZ of the more sophisticated model of \citet{Kopparapu_2013,Kopparapu_2014}, which was used to calculate the HZ around the 0.09 and 1.00 M$_\Sun$ stars, which have been modelled to TRAPPIST-1 and the Sun, respectively. The HZ around the 0.20, 0.49, and 0.70 M$_\Sun$ might therefore be underestimated.

For the IGM16(-Ring) simulations, relatively fewer Earth-like planets remain in the HZ than in the 1.00 M$_\Sun$ simulations. This is because the HZ lies significantly closer inwards than the snowline, which means that the third generation no longer forms inside the HZ, but needs to migrate to it. Moreover, since growth around these stars is slightly slower, the third generation has less time to reach Earth-like masses, and especially for the 0.49 M$_\Sun$ star, most late-forming planets grow only to a few Mars masses an example of which can be seen in Fig. \ref{fig: 0.49Msun_case_IGM16} in Appendix \ref{app:supplement}.

\subsubsection{0.09 and 0.20 M$_\Sun$ stars} \label{subsec: 0.09_and_0.20Msun_stars}
\begin{figure*}[]
    \includegraphics[width=\textwidth]{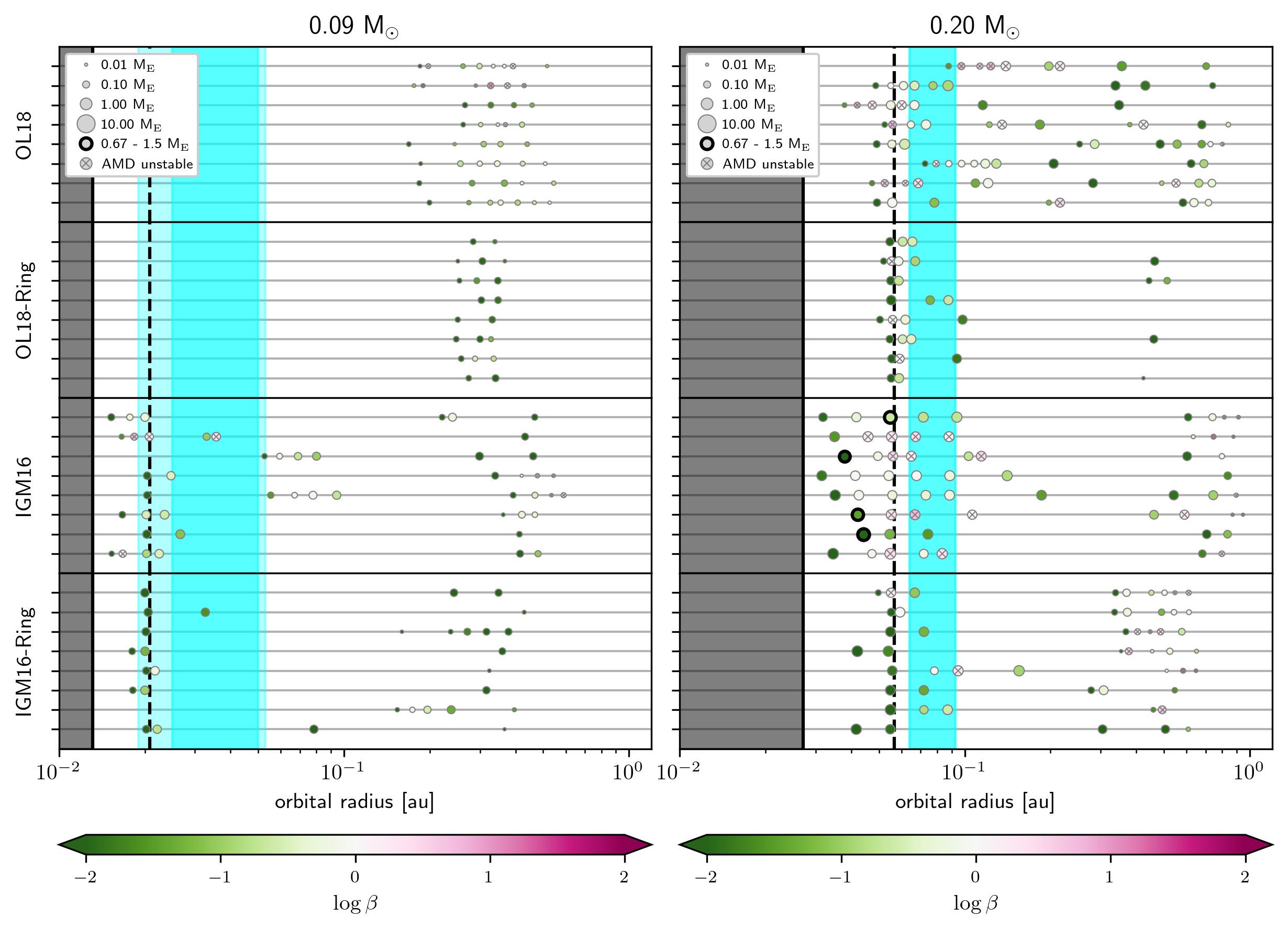}
    \caption[Simulated planetary systems around the 0.09 and 0.20 M$_\Sun$ stars for the four PA models.]{Simulated planetary systems around the 0.09 (left) and 0.20 (right) M$_\Sun$ stars for the four PA models. The size of the markers represents the mass of the planet, while their colour represents the planet's AMD stability (see Sect. \ref{sec: system_stability}). The layout of the figure is explained in the caption of Fig. \ref{fig: 1.00Msun_summary}.}
    \label{fig: 0.09+0.20_summary}
\end{figure*}

The results of the 0.09 and 0.20 M$_\Sun$ simulations are shown in Fig. \ref{fig: 0.09+0.20_summary}. The most important observation is that no Earth-like planets form around 0.09 stars, irrespective of the model that is used, according to the predictions in Figs. \ref{fig: Mfinal_Ormel} and \ref{fig: Mfinal_Ida}. In half of the IGM16 simulations with a broad planetesimal disc around a 0.20 M$_\Sun$ star, a single Earth-like planet managed to form through the merger of the many sub-Earth planets that arise in this model. However, the highest mass planets produced by the OL18(-Ring) model around 0.20 M$_\Sun$ stars are a few Mars masses, suggesting that the OL18 PA model is unable to form Earth-like planets around these low-mass stars. This is in stark contrast with the results of \citet{Ormel_2017}, and \citet{Schoonenberg_2019}. The latter authors find that for each combination of their parameters, multiple Earth-mass planets form around a 0.09 M$_\Sun$ star, even though they use the same OL18 PA-prescription that in our 0.09 M$_\Sun$ simulations cannot produce planets more massive than a few lunar masses, regardless of the initial width of the planetesimal disc.

There are two core differences between our models and those of \citet{Schoonenberg_2019}. They assumed that a very massive disc of planetesimals forms from the streaming instability, weighing around 1.4 M$_\mathrm{E}$ for most of their models, with each planetesimal having a mass of 3.6$\times 10^{-3}$ M$_\mathrm{E}$ (1200 km) in most simulations. For their fiducial model, the total mass in planetesimals formed by the streaming instability was about 7.5\% of the total solid mass in their disc. We, on the other hand, used a planetesimal disc of only 0.010 M$_\mathrm{E}$, which is comparable to the mass predicted by \citet{Liu_2019}.

Secondly, \citet{Schoonenberg_2019} did not consider the evolution of the disc itself. In our model, the gas accretion rate onto the star decreases over time, as a result of which many of the other parameters, such as the location of the snowline, and the gas surface density, change as well. Moreover, in our model, once the pebble formation front reaches the outer edge of the disc, after about 0.2 Myrs, the pebble flux rapidly declines \citep{Sato_2016}. Growth during the first few hundred thousand years is therefore essential, and around 0.09 and 0.20 M$_\Sun$ stars, this growth is too slow, at least for the small planetesimals that are expected from the streaming instability \citep{Simon_2016}. Figure \ref{fig: Mfinal_Ormel} suggests that for embryos larger than 10$^{-3}$ M$_\mathrm{E}$, planets of at least a few Mars masses should be able to form. 

However, since these large planetesimals are not expected to form during the streaming instability, our results suggest that it is not possible to explain the formation of TRAPPIST-1-like systems using a simple evolving disc model with pebble accretion as the primary mechanism for growth. Further studies are needed to determine if an increased pebble mass flux could produce Earth-mass planets around these stars, and if such an increase can be justified.

\subsection{General overview and statistics}\label{subsec: general_results_overview}
Figure \ref{fig: M,e,i_vs_a_all_planets_normal} provides an overview of all planets that have formed over the different OL18 and IGM16 simulations, showing their mass, eccentricity, and inclination as a function of orbital radius. The results for the OL18-Ring and IGM16-Ring simulations are shown in Fig. \ref{fig: M,e,i_vs_a_all_planets_Ring} in App. \ref{app:supplement}, and are generally very similar. 

In the OL18 simulations, nearly all planets have masses below 3 M$_\mathrm{E}$, and no planet has a mass higher than 5 M$_\mathrm{E}$, meaning that the planets are too small for gas accretion to play any significant role in their formation \citep{Mizuno_1978, Stevenson_1982, Bodenheimer_and_Pollack_1986, Hubickyj_2005}. 

The IGM16 model generates planets in greater number, and of significantly higher mass, with several exceeding 10 M$_\mathrm{E}$. Barring the planets around 0.09 and 0.20 M$_\Sun$ stars (shown in blue and orange), whose mass was limited by their own growth rate, rather than the pebble isolation mass or the interference from more massive planets, most of the planets in the OL18(-Ring) simulations are clustered in a narrow region of the parameter space, with masses between 0.5$\--$2 M$_\mathrm{E}$, and orbital radii between 0.05$\--$0.2 au. Meanwhile, the IGM16(-Ring) model produces planetary objects over the entire range from $10^{-5}$ to 10 M$_\mathrm{E}$, and from 0.05 to 3 au. 
 
The core difference between the models is seen in the eccentricity and inclination results, which points to the problem in the IGM16 model, which was already pointed at in Sects. \ref{subsubsec: PA_efficiency}, and \ref{sec: dynamic_evo_IGM16}. By ignoring the rapid decrease in accretion efficiency due to planets on eccentric or inclined orbits entering the ballistic regime, excited planetesimals in the IGM16 model continue growing.

The result is a clear correlation in $e$ and $i$ as a function of $M$ and $a$ in the IGM16(-Ring) model. After the disc is stirred up by the first generation of planets reaching their isolation mass and migrating inwards within the first ten to a hundred thousand years, a disc of small, heavily excited planetesimals remains. What follows is a continuous stream of small planets with high $e$ and $i$, starting their growth at orbital radii between ${\sim}$0.8 and 2.2 au, growing massive enough to dampen their $e$ and $i$, and to migrate inwards, and finally joining the MMR chain in the inner disc, where they are excited again by dynamic instabilities. If the excitation-induced ballistic regime were included, most of these planets would not have formed, and the mass of the most massive planets would be significantly less, since they would have fewer other planets to merge with.

\begin{figure*}
    \centering
    {\includegraphics[width=\linewidth]{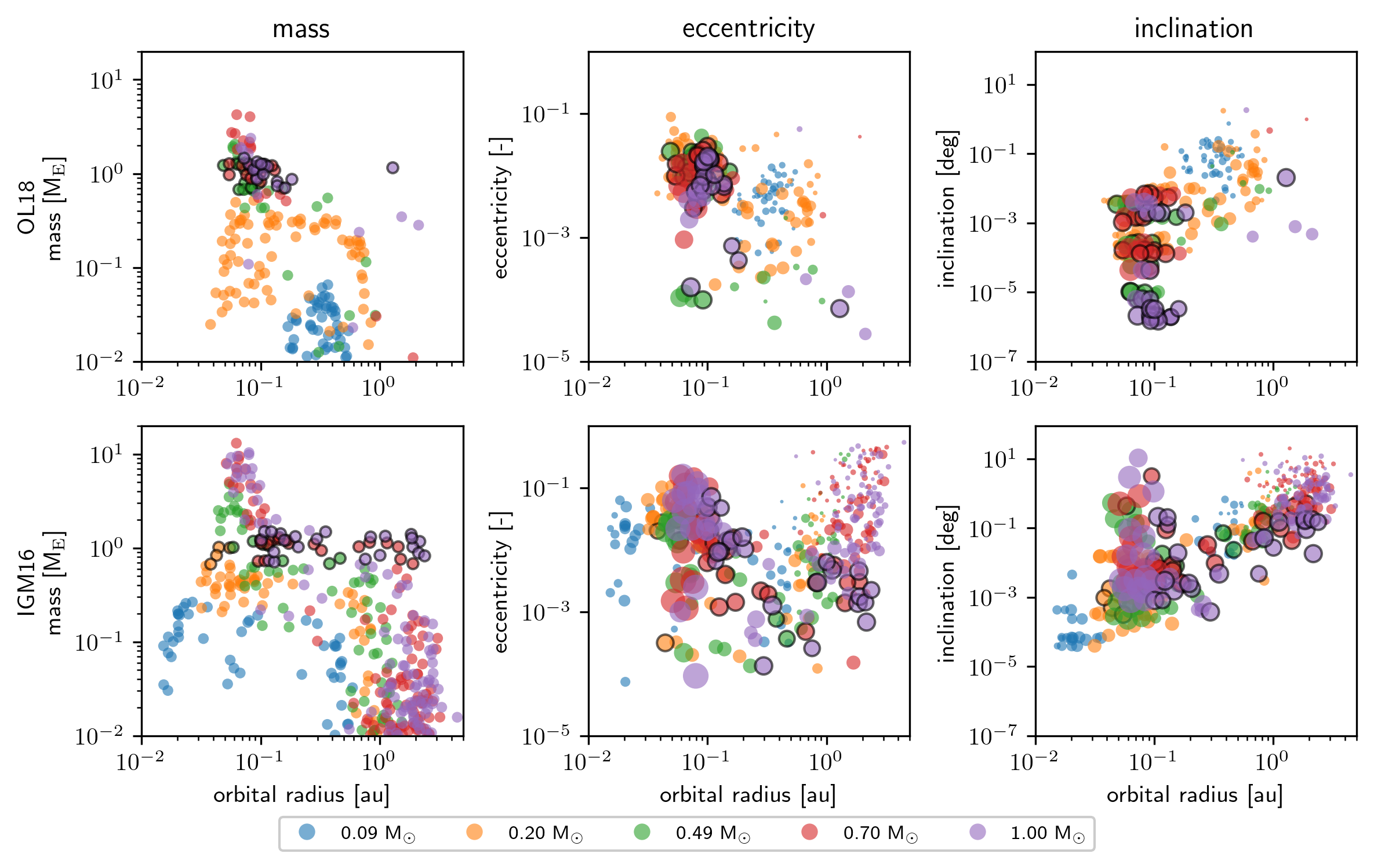}}
        \caption[Mass, eccentricity, and inclination of all planets formed in the OL18 and IGM16 simulations.]{Overview of the mass, eccentricity, and inclination of all planets formed in the OL18 and IGM16 simulations. Earth-like planets have been highlighted using a black edge around the marker. In the $e$ and $i$ plots, the size of the marker is proportional to its mass.}
    \label{fig: M,e,i_vs_a_all_planets_normal}
\end{figure*}

\begin{figure*}
     \centering
     {\includegraphics[width=\linewidth]{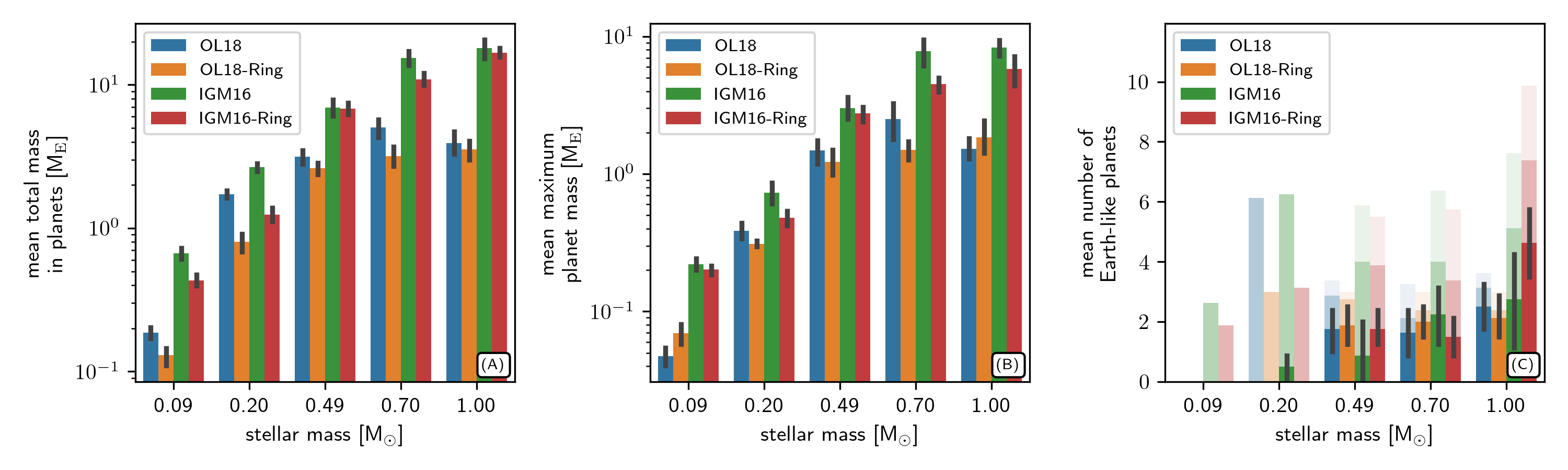}}   
        \caption[Core mean quantities of all simulations, grouped by stellar mass and model.]{Core mean quantities of all simulations, grouped by stellar mass and model. All quantities are calculated per simulation, and then averaged over the 8 simulations per group. (A) Mean total mass of the planetesimals and planets in the disc. (B) Mean maximum planet mass. (C) Mean number of Earth-like (0.67$\--$1.5 M$_\mathrm{E}$) planets. The number of Mars-like planets (0.1$\--$0.67 M$_\mathrm{E}$), and massive planets ($>$1.5 M$_\mathrm{E}$) are stacked on top with increasing transparency.}
        \label{fig: statistics}
\end{figure*}

Figure \ref{fig: statistics} provides a few core statistics describing the systems that have formed, grouped by stellar mass and model. All quantities have been averaged over the eight simulations per model. The mean total mass in planetesimals and maximum planet mass follow very similar trends, with a strong increase in value from the 0.09 to the 0.20 to the 0.49 M$_\Sun$ star, after which the values for the OL18 model flatten out, while they continue growing for the IGM16 model. Even if the entire planetesimal disc around the 0.09 M$_\Sun$ star merged together, the resulting planet would not be larger than a few Mars masses in the OL18 model, and would be a little short of 1 M$_\mathrm{E}$ for the IGM16 model. This makes it highly unlikely that additional simulations, with slightly different initial conditions and minor tweaks to the model, could suddenly produce TRAPPIST-1-like systems with multiple Earth-mass planets. Significant model changes, such as a drastic increase in the pebble mass flux, a further disc edge, or other modifications that lead to far more rapid growth in the initial disc phases, are required to explain the formation of TRAPPIST-1. However, these changes are currently not supported by the existing literature. 

The total mass in planetesimals remains fairly constant between the 0.49, 0.70, and 1.00 M$_\Sun$ OL18 results, as does the maximum planet mass and number of planets, even though more massive stars have more massive discs and therefore more solids available for growth. This constant trend arises because, in all OL18(-Ring) simulations, the dynamical growth profile has the same shape, and when the first planets have reached the isolation mass, these planetesimal discs have all accreted more or less the same amount of mass. Most of the planets can no longer grow after this point, due to their excited orbits. This is irrespective of the time at which the point was reached and the amount of solids that are left in the disc. Since the IGM16 models do not have this limitation, the trend of increased planetesimal disc mass, maximum planet mass, and number of planets continues here.

 \subsection{System structure and long-term stability} \label{sec: system_stability}
As mentioned at the start of Sect. \ref{Chapter: Nbody-results}, the simulations in this study were terminated after 5 Myrs. This end time does not necessarily mark the end of the dynamic evolution of the systems. Since there is no sharp cut-off time after which we assume all gas of the disc has dissipated, there is still a little gas present at the end of these simulations. As the rest of the gas slowly disappears, or is rapidly blown away by photoevaporation from the igniting star, further dynamic instabilities are expected to occur. 

To make a first-order assessment of the long-term orbital stability of the planets in our systems, without resorting to computationally expensive simulations, we make use of two metrics: the Hill stability criterion \citep{Chambers_1996}, and the angular momentum deficit \citep[AMD;][]{Laskar_1997, Laskar_2000, Laskar_and_Petit_2017, Petit_2017}. 

The Hill criterion predicts that orbit crossings will occur if the separation between two planets is less than the critical value of $2\sqrt{3}$ times their mutual Hill radius. These orbit crossings result in system instabilities and collisions. The mutual Hill radius is given by
\begin{equation*}
    R_\mathrm{H} = \frac{a_\mathrm{p1}+a_\mathrm{p2}}{2}\left(\frac{M_\mathrm{p1}+M_\mathrm{p2}}{M_*} \right)^{1/3},
\end{equation*}
where the subscripts p1 and p2 indicate the two planets. The separation of the planets is often represented by the dynamic spacing, $\Delta$, which is the ratio between the separation and the mutual Hill radius.

\begin{figure}
    \centering
    \includegraphics[width=\linewidth]{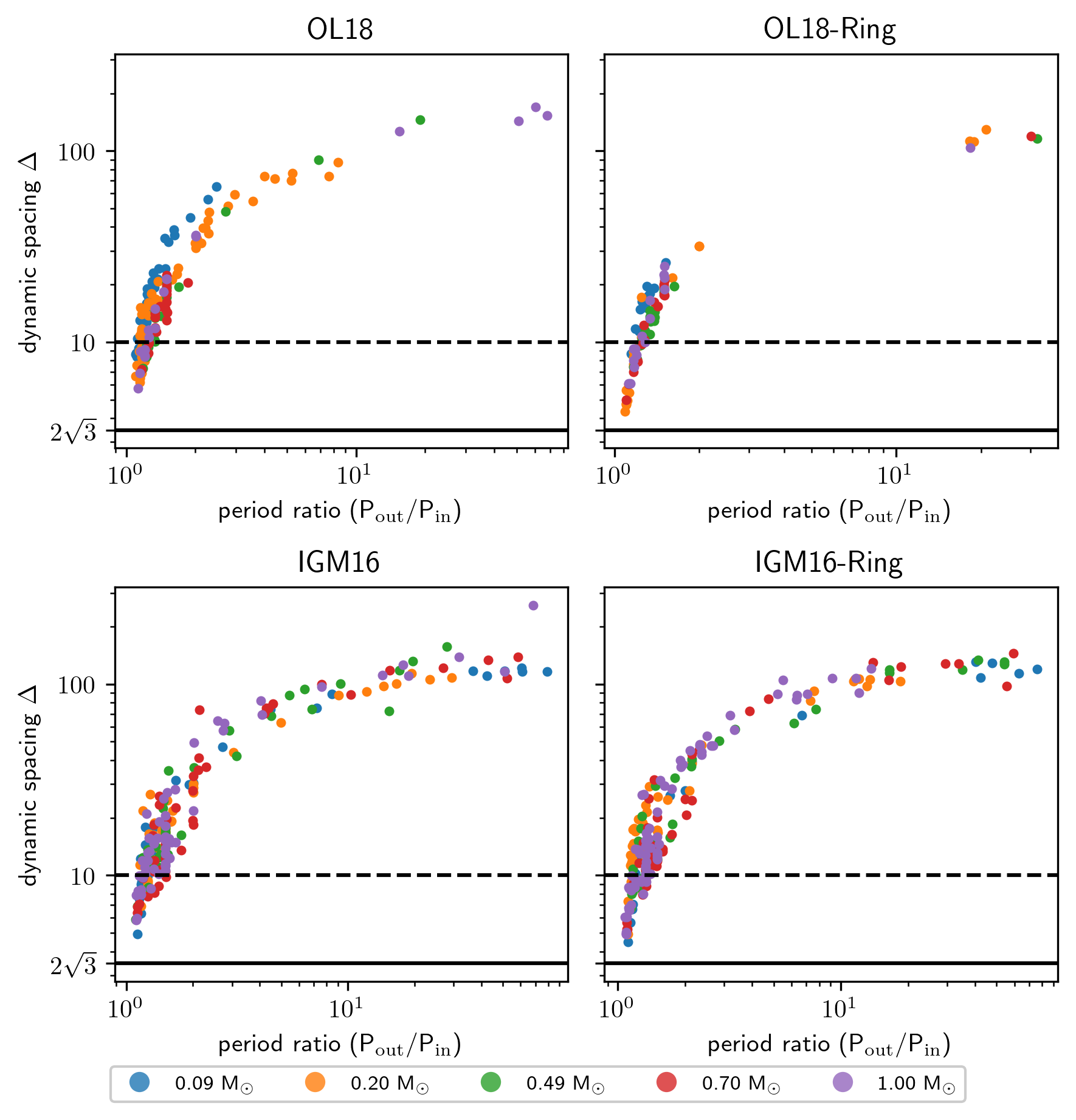}
    \caption{Dynamic spacing $\Delta$ of all planet pairs in all simulations, plotted against their period ratio $P_\mathrm{out}/P_\mathrm{in}$. The planet pairs are grouped per PA model, with their colour indicating the stellar mass. The horizontal lines indicate two important values of the dynamic spacing: the critical value $\Delta=2\sqrt{3}$ (solid), and the value $\Delta=10$ (dashed).}
    \label{fig:Hill_stability}
\end{figure}

For the evaluation of the system stability through the Hill criterion, we only consider the likelihood of orbit crossings between the larger planets in the system, since collisions of large planets with small planetesimals or planetary embryos do not significantly alter the system's architecture. For the 0.49, 0.70, and 1.00 M$_\Sun$ simulations, everything more massive than 0.1 M$_\mathrm{E}$ is considered a planet. Since the 0.09 and 0.20 M$_\Sun$ simulations only contain a few planets (or none) in this mass range, the definition of a planet here is any object more massive than 0.01 M$_\mathrm{E}$, which is the typical mass of a planetary embryo \citep[e.g.][]{Woo_2021,Woo_2022,Voelkel_2021}. The results are shown in Fig. \ref{fig:Hill_stability}. 

None of the simulated planet pairs are Hill-unstable. Still, about 30\% of the planet pairs in the OL18 simulation have a dynamic spacing $\Delta < 10$, which suggests they are less likely to remain stable in the long-term (gigayear-scale) \citep{Pu_and_Wu_2015}. For the IGM16 simulations, this is 19\% of the planet pairs.   

However, the Hill stability should be interpreted with caution, especially when trying to estimate long-term stability. Planets with dynamic spacings $\Delta < 10$ can still be very stable, especially if they are in orbital resonance with each other, as is the case for most of our simulated planets. Moreover, the Hill stability criterion is intended for two-planet scenarios, since it does not consider mutual inclinations between orbits, which are known to influence the evolution and long-term stability of planetary systems significantly. 

For the multi-planet systems we observe, the AMD is a more appropriate measure of the stability, since it accounts for both eccentricities and inclinations. The AMD is a measure of the excitation of a system, indicating its deviation from a perfectly coplanar and circular system. If the total AMD is below a critical value, collisions between planets cannot occur, and the system is stable. The total AMD is given by \citep{Laskar_1997,Laskar_2000}
\begin{equation}\label{eq: total_AMD}
    C = \sum_{k=1}^{n}\Lambda_k\left(1-\sqrt{1-e_k^2}\cos i_k \right),
\end{equation}
with $\Lambda_k=m_k\sqrt{\mu a_k}$ the angular momentum of planet $k$ if its orbit were circular. The AMD stability coefficient is then given by
\begin{equation}
    \beta = \frac{C}{\Lambda'C_\mathrm{c}},
\end{equation}
in which $\Lambda'$ is the circular momentum of the outer planet of the pair, and $C_\mathrm{c}=C_\mathrm{c}(\alpha,\gamma)$ is the critical AMD of the pair, which depends on the ratio of the semimajor axes, $\alpha=a_\mathrm{in}/a_\mathrm{out}$, and the ratio of the masses, $\gamma=m_\mathrm{in}/m_\mathrm{out}$ \citep[see][]{Laskar_and_Petit_2017}. A planet pair is stable if $\beta < 1$, or equivalently, if $\log \beta < 0$. A system is stable if all planet pairs are stable.  

The AMD stability of the planets in the simulations is presented through the colour scales in Figs. \ref{fig: 1.00Msun_summary}, \ref{fig: 0.49+0.70_summary}, and \ref{fig: 0.09+0.20_summary}. For the calculation of the total AMD of the system, all planetesimals, and planets were included in the sum of Eq. \ref{eq: total_AMD}. However, similarly to in the Hill stability calculations, we only consider the stability between orbits of planets more massive than 0.1 $M_\mathrm{Earth}$ (0.01 $M_\mathrm{Earth}$ in the 0.09 and 0.20 M$_\Sun$ simulations). These are the planets shown in Figs. \ref{fig: 1.00Msun_summary}, \ref{fig: 0.49+0.70_summary}, and \ref{fig: 0.09+0.20_summary}. For the innermost planet, the stability is calculated with respect to the central star, in which case $C_\mathrm{c}=1$. 

For the OL18(-Ring) simulations, the systems around the 0.49 and 0.70 M$_\mathrm{sun}$ stars have all reached stability, and the 2 out of 53 planets in the 1.00 M$_\Sun$ simulations that are still unstable, are close to stability. This means that for the OL18(-Ring) simulations, planets settle into a long-term stable system very quickly, in less than 5 Myrs. An exception to this is seen in the OL18 systems around 0.20 M$_\Sun$ stars, most of which have not yet reached stability. Approximately 25\% of the planet pairs in these OL18 simulations are unstable, and 10\% in the OL18-Ring simulations, meaning that these systems would likely experience mergers if the simulations were extended beyond the 5 Myrs, possibly leading to larger planets. However, since most of these unstable planet pairs involve a Moon and a Mars mass object, or in the most extreme case two Mars mass planets, it is highly unlikely that Earth-mass objects would suddenly arise if the simulations were continued for another few million years.

The IGM16(-Ring) simulations are a different story. At the end of our simulations, these systems are still highly unstable, especially for more massive stars. These systems are dynamically active for far longer than the OL18 systems, still forming massive planets after 2 Myrs of evolution. Moreover, since these systems are packed with massive planets, there have been more close encounters, leading to greater average eccentricity and inclination (see Figs. \ref{fig: M,e,i_vs_a_all_planets_normal} and \ref{fig: M,e,i_vs_a_all_planets_Ring}), and the MMR chains are rendered more unstable. To find the long-term architecture of these systems, the simulations must be extended for at least a few more Myrs, but likely until 40 Myrs or even beyond 100 Myrs \citep{Hatalova_2023}.

 \subsection{The influence of gas accretion on the simulated systems} \label{sec: gas_accretion}
\begin{figure*}[]
    \includegraphics[width=\textwidth]{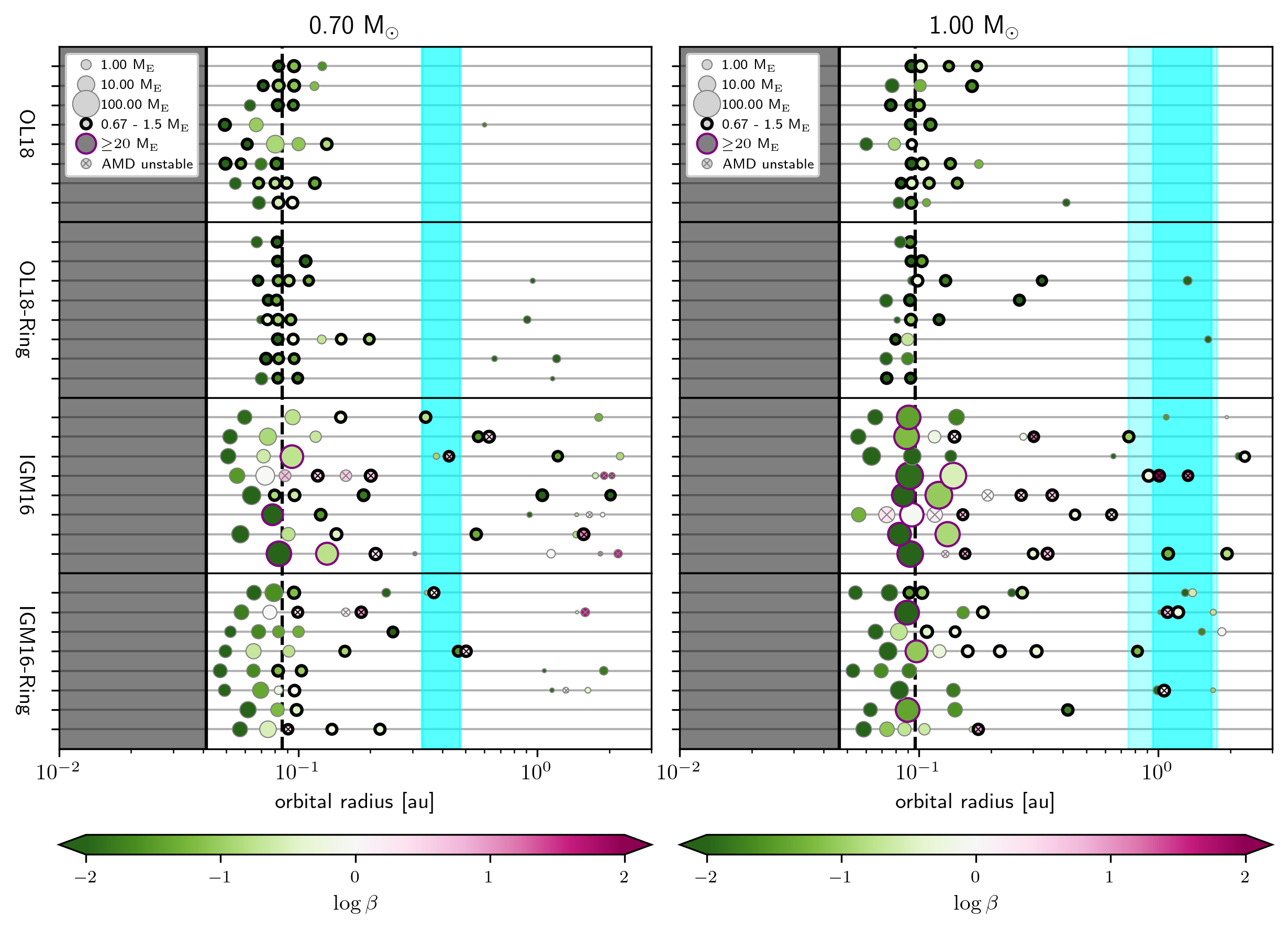}
    \caption[Simulated planetary systems around the 0.70 and 1.00 M$_\Sun$ stars for the four PA models including gas accretion.]{Simulated planetary systems around the 0.70 and 1.00 M$_\Sun$ stars for the four PA models including gas accretion. The mass threshold for particles to become self-gravitating was increased compared to the simulations without gas accretion to conserve computing resources, meaning that these simulations are less accurate than the others. The purpose of this figure is therefore only to indicate whether gas accretion could significantly influence the results from the normal simulations. For the OL18 simulations, gas accretion does not play a significant part. In the IGM16 simulations, on the other hand, multiple gas giants form. The layout of the figure is explained in the caption of Fig. \ref{fig: 1.00Msun_summary}.}
    \label{fig: 0.70+1.00_gasacc}
\end{figure*}
As previously mentioned, gas accretion was not included in the simulations. For most simulations, gas accretion should not play any significant role, since cores with masses between 5 and 10 M$_\mathrm{E}$ are required to start runaway gas accretion \citep{Mizuno_1978, Stevenson_1982, Bodenheimer_and_Pollack_1986, Ikoma_2000, Hubickyj_2005}. Only in the IGM16 simulations around 0.70 and 1.00 M$_\Sun$ stars are there occasionally planets within this mass range. 

To validate that gas accretion did not play a significant role in the OL18 simulations, and to get a first-order idea of the influence of gas accretion on the IGM16 results, the simulations were repeated, starting from the snapshots in which the first generation of planets have just formed. The gas accretion rate in these simulations is given by the equations provided in \citet{Matsumura_2021}. To conserve computing resources, the minimum mass for self-gravitating particles was set at 0.005 and 0.01 M$_\mathrm{E}$ for the OL18 and IGM16 models, respectively. This threshold is too high for all the dynamics of the system to be included, since the gravitational pull of objects slightly smaller than the Moon can influence the orbits of other objects, which is not included in this model. Moreover, the threshold was imposed early in the disc, at a time when sub-lunar objects could still grow through pebble accretion, although in the case of the OL18 model, they hardly ever do. These results therefore do not supersede the results presented in the previous sections. They merely serve to provide an indication of whether gas accretion could significantly alter the conclusions of this study.

The results for a 0.70 and 1.00 M$_\Sun$ star are presented in Fig. \ref{fig: 0.70+1.00_gasacc}. The other stars are not included in the figure, since gas accretion did not play a significant part in these cases. As expected, gas accretion has an insignificant influence on planet formation in the inner disc for the OL18 model. The pebble isolation mass is too low for planets to grow to the required mass for runaway gas accretion through PA, and too few Earth-like planets form for the MMR chain to collapse into massive planetary cores. 

In the IGM16 simulations, however, multiple gas giants form through the mergers of the many planetary embryos that are created. The IGM16 simulations produce more gas giants than the IGM16-Ring simulations, with some systems around the solar-mass star producing two ${\sim}100$ M$_\mathrm{E}$ gas giants in the same simulation. However, since the number of planets formed in the IGM16(-Ring) simulations is most likely a gross overestimation, so is the number of gas giants. Future studies should confirm or deny this.

  \section{Discussion} \label{chapter: discussion}
This study further cements the idea that pebble accretion could explain the formation of Earth-mass planets around low-mass stars of ${\gtrsim}0.5$ M$_\Sun$. Around ${\lesssim}0.20$ M$_\Sun$ stars, no Earth-like planets were observed, except for in the 0.20 M$_\Sun$ IGM16 model, which managed to produce a single Earth-like planet in half of the simulations.

\subsection{OL18 vs IGM16} \label{sec: OL18_vs_IGM16}

\begin{table*}[]
\renewcommand{\arraystretch}{1.5}
\centering
\caption{Quantitative comparison between the OL18 and IGM16 model results.}
\label{tab:OL18_vs_IGM16}
\resizebox{\textwidth}{!}{
\begin{tabular}{l|lll|lll}
\cline{2-7}
 &
  \multicolumn{3}{l|}{OL18} &
  \multicolumn{3}{l}{IGM16} \\ \hline
\begin{tabular}[c]{@{}l@{}}Missing physics compared to \\ the other PA prescription\end{tabular} &
  \multicolumn{3}{l|}{} &
  \multicolumn{3}{l}{\begin{tabular}[c]{@{}l@{}}$\--$   Influence of planet's eccentricity and inclination \\ \hspace{3mm} on pebbles relative velocity\\ $\--$   Reduction of accretion efficiency due to pebbles with \\\hspace{3mm} high relative velocity entering the ballistic regime\end{tabular}} \\ \hline
 &
  0.09 M$_\Sun$ &
  1.00 M$_\Sun$ &
  Overall &
  0.09 M$_\Sun$ &
  1.00 M$_\Sun$ &
  Overall \\ \hline
Mean total accreted mass [M$_\mathrm{E}$] &
  $0.15^{+0.04}_{-0.04}$ &
  $3.7^{+0.7}_{-0.8}$ &
  $2.4^{+1.7}_{-2.2}$ &
  $0.54^{+0.14}_{-0.16}$ &
  $17.3^{+2.4}_{-2.5}$ &
  $7.9^{+8.4}_{-7.2}$ \\
Mean number of planets ($M>M_\mathrm{M}$) &
  $0.00^{+0}_{-0}$ &
  $3.3^{+0.7}_{-0.9}$ &
  $2.8^{+1.2}_{-2.8}$ &
  $2.25^{+0.75}_{-0.25}$ &
  $8.75^{+3.25}_{-2.35}$ &
  $5.5^{+2.5}_{-2.8}$ \\
Mean number of Earth-like planets &
  $0.00^{+0}_{-0}$ &
  $2.3^{+1.3}_{-1.3}$ &
  $1.2^{+1.8}_{-1.2}$ &
  $0.00^{+0}_{-0}$ &
  $3.7^{+1.9}_{-2.3}$ &
  $1.4^{+1.9}_{-1.4}$ \\
Fraction of all planets that remains in the HZ &
  N/A &
  0.038 &
  0.066 &
  0.64 &
  0.19 &
  0.16 \\
Mean maximum planet mass [M$_\mathrm{E}$] &
  $0.06^{+0.18}_{-0.18}$ &
  $1.7^{+0.3}_{-0.5}$ &
  $1.0^{+0.8}_{-1.0}$ &
  $0.21^{+0.23}_{-0.33}$ &
  $7.0^{+2.3}_{-1.9}$ &
  $3.4^{+3.6}_{-3.1}$ \\
Mean $\log_\mathrm{10}\left( \mathrm{AMD}_\mathrm{total}\right)$ &
  $34.9^{+0.4}_{-0.6}$ &
  $36.3^{+0.4}_{-0.5}$ &
  $35.8^{+0.9}_{-0.9}$ &
  $35.2^{+0.5}_{-0.6}$ &
  $38.5^{+0.2}_{-0.2}$ &
  $37.2^{+1.3}_{-1.6}$ \\
Fraction of all planet pairs that is unstable &
  N/A &
  0.038 &
  0.026 &
  0.028 &
  0.52 &
  0.31 \\ \hline
\end{tabular}
}
\tablefoot{The mean values have been calculated by averaging over all simulations with the indicated PA prescription and stellar mass. The fractions have been calculated from the total sample of planets of these simulations combined. Since no planets form around 0.09 M$_\Sun$ stars in the OL18 simulations, these fractions are not applicable to this model. In all calculations, the results from the normal and Ring simulations have been taken together.}
\end{table*}

In this study, two different prescriptions for the pebble accretion efficiency $\epsilon$ were tested, one by \citet[][IGM16]{Ida_2016}, and one by \citet[][OL18]{Ormel_and_Liu_2018}. A few key measurements that show the strong difference between these models are presented in Table \ref{tab:OL18_vs_IGM16}. In general, the IGM16 model produces many more planets, with a significantly higher maximum mass, than the OL18 model. The IGM16 models are also much more excited than the OL18 models, with total angular momentum deficits (AMD$_\mathrm{total}$; a measure of the deviation from a completely circular system) that are orders of magnitude higher, especially for the most massive stars in the simulation, as a result of which most IGM16 simulations are still highly unstable after the 5 Myrs of simulations. 

Both models have their strengths and weaknesses. For example, the OL18 model uses orbit-averaged corrections for the eccentricity and inclination dependence of the accretion rate, but these do not take into account the influence of the argument of periapsis, nor the variations in encountered disc conditions along eccentric orbits. Nevertheless, these small inaccuracies will likely not have a significant influence on the overall accretion of planets, since these parameters only become relevant for highly eccentric and inclined orbits, in which virtually no accretion takes place.  

The shortcomings of IGM16, on the other hand, are very significant. As has extensively been discussed in Sects. \ref{subsubsec: PA_efficiency}, \ref{sec: dynamic_evo_IGM16}, and \ref{subsec: general_results_overview}, the core issue of IGM16 is that it assumes that all planets are on circular orbits and that PA is always in the settling regime. The OL18 simulations show, however, that the main limitations to the growth of planetesimals are their eccentricity and inclination, which can cause them to enter the highly inefficient ballistic pebble accretion regime. This effect is especially of importance given that, as the first planets reach around an Earth-mass and migrate to the inner edge of the disc within a few hundred thousand years of initialisation, they stir up the rest of the disc, inciting high eccentricities and inclinations in all other planetesimals. This increases the relative velocity, $\Delta v$, between the planetesimals and the pebbles so much that the pebbles no longer have time to settle in the gravitational field of the planetesimals, which significantly decreases the accretion efficiency. 

In a realistic scenario, only the planetesimals that had already significantly grown before their excitation, are able to rid themselves of their eccentricity and inclination through interactions with the disc, and resume their growth in the later stages of the disc. This is what is observed in the OL18 simulations. However, by ignoring the effects of the excitation and assuming planets can continue growing uninteruptedly, IGM16 grossly overestimates the accretion rate of planetesimals after the formation of the first planets, and produces a large second, and even a third generation of planets. The many Earth-mass planets in the second generation migrate to the inner disc edge, where they cause dynamic instabilities in the mean motion resonance chains and merge with the first generation of planets, producing multiple massive super-Earths and even gas giants. The third generation, on the other hand, does not have time to migrate inwards and fills the entire disc with Earth-like planets. 

The IGM16 model most likely overestimates the number of planets that are formed, the final mass of the planets at the disc's inner edge, and the probability of encountering Earth-like planets in the habitable zone. A comparison with observations of exoplanetary systems should be performed to fully validate this. The IGM16 model could be very powerful and useful, since its analytical approach and general intuitiveness allow for a very detailed analysis of the influence of different disc conditions on planet formation. However, until its expressions for $\Delta v$ are augmented to include the influence of the eccentricity and inclination, and equations for the accretion rate in the ballistic regime are added, the PA prescription from \citet{Ida_2016} is not appropriate for N-body simulations. In the comparison with other works in the following section,  therefore, we only considered the OL18 results.

\subsection{Comparison with other studies} \label{sec: paper_comparison}
In Sect. \ref{subsec: 0.09_and_0.20Msun_stars}, we compare our results for a 0.09 and 0.20 M$_\Sun$ star to those of \citet{Schoonenberg_2019}. We find that without their assumption of a massive planetesimal disc and with the inclusion of disc evolution, Earth-like planets cannot form around these low-mass stars.

In this section, we  compare our results to two other recent studies: the analysis from \citet{Lambrechts_2019} on the influence of the pebble mass flux on the type of planetary system that forms and that of \citet{Johansen_2021}, who used more specific PA simulation set-ups in an attempt to explain the mass, orbits, and compositions of the terrestrial planets in the Solar System. 

The claims of \citet{Johansen_2021} are still very much disputed (see e.g. the recent work of \citet{Morbidelli+2024} and the response by \citet{Johansen+2024}). The analysis of currently available meteorite samples shows a dichotomy in the isotopic composition between the inner and outer Solar System \citep{Warren_2011}, which is inconsistent with our current understanding of PA. After all, most of the accreted pebble mass originates in the outer disc, so if PA played a significant part in the formation of the terrestrial planets around the Sun, their isotopic composition should be equal to that in the outer Solar System \citep{Mah_2022}. In this section, we, therefore, do not compare our results to those of \citet{Johansen_2021} to try to explain the formation of the Earth specifically, but to analyse the formation of Earth-mass planets at significant orbital radius (in particular within the habitable zone) around Sun-like stars through PA in general. We refer to these systems as `Solar System-like', neglecting the isotopic dichotomy of the actual Solar System.

We show that the artificial introduction of a few 10$^{-3}$ M$_\mathrm{E}$ planetary embryos, as done by \citet{Johansen_2021}, is not necessary for PA to form Earth-mass planets and that the planetesimal size distribution following the streaming instability contains sufficiently massive objects for rapid growth. However, with our general model, we are unable to explain the formation of the aforementioned Solar System-like planetary systems. Because of the rapid growth and migration in our model, there is a low probability of planets remaining in the habitable zone. More importantly, in every simulation, there is a first generation of Earth-mass planets that has migrated to the inner edge of the disc. These planets are not present in the Solar System.

For Solar System-like systems to form, some force must be present to prevent planetary embryos from growing during the first 0.5${\--}$1 Myrs or there must be pressure bumps in the disc preventing migration to the inner edge of the disc \citep{Chambers2023}. \citet{Johansen_2021} focussed on preventing early growth by assuming all initial planetesimals have a radius of 100 km (which is too small for them to growth through pebble accretion) and by introducing the actual protoplanets of 10$^{-3}$ M$_\mathrm{E}$ representing Venus, Earth, Theia, and Mars after $t$ = 0.66, 0.93, 1.50, and 2.67 Myrs, respectively. Though these starting times were based on the expected growth time to 10$^{-3}$ M$_\mathrm{E}$ from an initial mass function of planetesimals that peaks at $M \sim 10^{-7} \-- 10^{-6}$ M$_\mathrm{E}$, they seem inconsistent with our systems that also include a few planetesimals at the higher end of the expected mass range \citep{Simon_2016}. The ${\sim}400$ km planets in our simulations grow into massive planets much earlier. Even if relatively more very low-mass planetesimals were included in our model (which could coagulate over a longer period of time to form embryos later on), the first generation of planets that grew directly from the most massive planetesimals from the streaming instability would still be present in the innermost regions of the systems \citep[see e.g. also][]{Mah_2022}.

An alternative explanation for our observed systems and why they differ from the Solar System is presented by \citet{Lambrechts_2019}. They propose that there are two modes of growth, one that produces super-Earths $\--$ planets with masses between those of Earth and Neptune $\--$ at the inner edge of the disc, and one which produces Earth-like planets spread throughout the disc. The growth mode the system experiences is determined by the radial pebble mass flux. For a low pebble flux, such that the total pebble mass that reaches the inner disc is less than $\approx 110$ M$_\mathrm{E}$ around a solar-mass star, the embryos within the snowline grow slowly without significant migration. The resulting widely spaced population of approximately Mars-mass embryos becomes unstable when the gas disc fully dissipates, which they assume is after 3 Myrs. Collisions between these Mars-mass embryos create a small number of terrestrial planets, with masses of at most five Earth-masses. For high pebble fluxes, with a total pebble mass of more than $\approx 190$ M$_\mathrm{E}$ around a solar-mass star, the embryos within the snowline rapidly grow sufficiently massive to migrate to the inner edge of the gas disc, where they continue accreting pebbles, and merge as a result of dynamic instabilities, forming a system of closely spaced super-Earths with masses between five and twenty Earth-masses. 

This latter growth mode is in some aspects very akin to what we see in our 0.49, 0.70, and 1.00 M$_\Sun$ systems. This would be consistent given that the total pebble mass in the 1.00 M$_\Sun$ simulation is $\approx 215$ M$_\mathrm{E}$ (see Fig. \ref{fig: Mdotf0}), which corresponds to 1.3\% of the total disc mass, placing the simulation in the super-Earth growth mode. However, in many aspects, our simulations are also significantly different. For instance, we find that most planets form outside the snowline. Moreover, we find that the pebble isolation mass is closer to 1 M$_\mathrm{E}$ \citep{Ataiee_2018}, instead of 10 M$_\mathrm{E}$, as assumed by \citet{Lambrechts_2019}. This, combined with the fact that only few planets form, means that the objects at the inner edge of the disc are not in the 5 $\--$ 20 M$_\mathrm{E}$ range, but in the 1 $\--$ 3 M$_\mathrm{E}$ range.

Moreover, \citet{Lambrechts_2019} filled their discs with 25 embryos of $10^{-2}$ M$_\mathrm{E}$ each, which could grow into Mars-mass protoplanets for low pebble fluxes. It is not yet clear if the small planetesimals used in this study would be able to grow fast enough and in sufficiently large numbers under these conditions. 

That being said, the 0.20 M$_\Sun$ simulations in Fig. \ref{fig: 0.09+0.20_summary}, for which the total pebble mass (calculated by integrating Eqs. \ref{eq: Mdot*} and \ref{eq: Mdotf_final}) is approximately 0.4\% of the total disc mass, do somewhat resemble the situation sketched in the terrestrial growth mode of \citet{Lambrechts_2019}, suggesting that there might be a sweet spot in our simulations as well, for the formation of many Mars-like protoplanets which can form Earth-like planets with large orbital radii after the complete dissipation of the disc. However, if there is such a pebble flux sweet spot, it will most likely be extremely narrow, given the fact that the planetesimals in the 0.20 M$_\Sun$ OL18 simulations have not yet accreted enough material to form a single Earth-like planet if they were all brought together, while the protoplanets in the 0.20 M$_\Sun$ OL18-Ring simulations have already migrated to the inner edge of the disc, despite not having accreted sufficient material either. Further research, using different pebble mass fluxes and disc dissipation conditions, is needed to determine whether Solar System-like planetary systems can be created using our general PA model.

\subsection{Limitations of this study} \label{sec: caveats}
Most of the caveats and assumptions of this study have been discussed in Sects. \ref{chapter: models} through \ref{sec: PA_afo_r_and_Mini}. These include the approximations that go into the OL18 and IGM16 PA prescriptions in Sect. \ref{sec: PAmodel}, the planetesimal number and distribution in Sect. \ref{sec: initial_planetesimal_conditions}, and the consequence of the Stokes number being $\gg 1$ in the innermost regions of the disc, discussed in App. \ref{App: sublimation}.

Other aspects of the disc and planets that might influence PA and migration, but that have been ignored in this study, include gas accretion, as discussed in Sect. \ref{sec: gas_accretion}, disc winds, which could limit type I migration in the inner disc \citep{Ogihara_2018, Chambers_2019}, the formation of a primordial atmosphere around the planetary embryos \citep{Brouwers_2020,Takaoka_2023,Yzer_2023}, which could alter the PA efficiency and the planetary composition, and the early formation of gas giants further out in the disc, which could limit the pebble flux to the inner disc. However, these are all higher-order effects that are beyond the scope of this research.

There are three model caveats based on recent developments in the field of PA that we think warrant a slightly more detailed discussion than the list above. These concepts have not yet been widely adopted in numerical PA simulations, but they are worth considering in future work. 

Firstly, the migration model used in this study did not consider thermal torques. Recent work shows that in non-isothermal discs with low viscosities, such as the discs used in this study, thermal torques can slow down, or even reverse planet migration \citep{Masset2017,Guilera+2019,Guilera+2021}, possibly providing a means to limiting the rapid inwards migration seen in most of our simulations. The thermal torque consists of a cold component, originating from two cold and dense lobes that appear on either side of the planet's orbit due to thermal diffusion of the disc, and a heating component, which has the opposite sign of the cold component, and is caused by the heat released by the planet from accreting solids, which creates hot and low-density lobes in its surroundings \citep{Masset2017, Guilera+2019}.

However, as noted by \citet{Guilera+2019}, the thermal torques of \citet{Masset2017} are incompatible with the torques from \citet{Paardekooper_2011} used in this study. While \citet{Masset2017} expanded upon the work of \citet{Jimenez_Masset2017}, deriving the thermal torque by considering the effects of thermal diffusion in the disc which had previously been ignored, \citet{Paardekooper_2011} derived their migration torques based on 2D radiative hydrodynamics simulations which already included thermal diffusion. The torques from \citet{Paardekooper_2011} might, therefore, at least partially incorporate the effects of the cold torque. The heating component of the thermal torque has not been included in the models of \citet{Paardekooper_2011}, since they do not consider the planet's luminosity. However, adding only this heating component without the full cold component is not appropriate either, since it is the balance between the cold and heating components that determine the influence of the thermal torque. The full migration model used in this study should therefore be replaced by that of \citet{Masset2017} for the thermal torque to be considered. We leave this for our future studies.

Secondly, like most PA studies, our models considered monodisperse (single-sized) pebble accretion. The pebble size does depend on the local disc conditions, and is physically motivated by the drift barrier, but it does not include a size distribution. \citet{Lyra+2023} show that in polydisperse (multi-species) PA, the mass of onset of PA lies 1$\--$2 orders of magnitude lower, and the PA efficiency of small planets in the Bondi regime lies 1$\--$2 orders of magnitude higher than in monodisperse PA. In these scenarios, the most efficiently accreted pebble size can be different from the size that dominates the pebble mass flux. Polydisperse PA could promote the formation of larger planets in the 0.09 and 0.20 M$_\Sun$ simulations and might therefore help explain the formation of systems like TRAPPIST-1. We therefore considered employing polydisperse PA models in future studies. 

The final model caveat is related to the pebble isolation mass (PIM) of \citet{Ataiee_2018} used in this study. This semi-analytical expression for the PIM depends only on the disc conditions at the location of the planet. However, using numerical simulations, \citet{Chametla+2022} show that the PIM also depends on the eccentricity of the planetary orbit. For eccentricities lower than the disc's aspect ratio $h_\mathrm{g}$, an increase in eccentricity leads to a decrease in PIM. For even higher eccentricities, the PIM increases monotonically. \citet{Chametla+2022} find that for low-viscosity discs ($\alpha_\mathrm{turb} \leq 10^{-3}$), like the ones used in our study, the PIM is well-defined, and its eccentricity dependence is best reproduced by the expression
\begin{equation}
\begin{split}
    M_\mathrm{iso,2}(\hat{e}) & = \frac{M_\mathrm{iso}(\hat{e}=0)}{M_\mathrm{E}}\times \\ &\left\{1+(400\alpha_\mathrm{turb}+0.2)\hat{e}-\frac{5\hat{e}}{\left[ 8+\hat{e}^3 \right]} \right\} M_\mathrm{E},
\end{split}
\end{equation}
in which $M_\mathrm{iso}(\hat{e}=0)$ is the PIM of \citet{Ataiee_2018} in Eq. \ref{eq: Miso}, and $\hat{e}=e/h_\mathrm{g}$. 

\begin{figure}
    \centering
    \includegraphics[width=\linewidth]{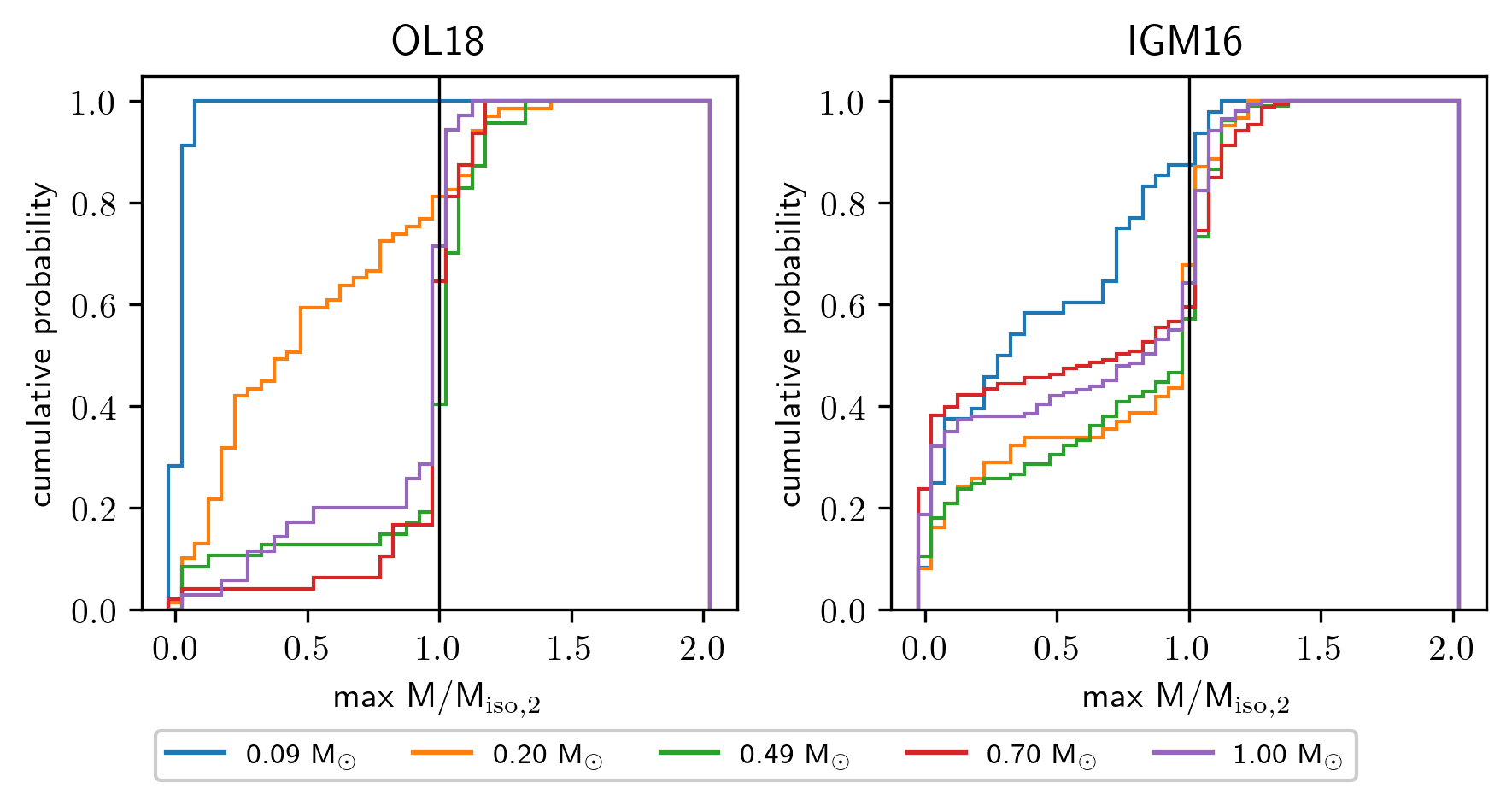}
    \caption{Cumulative distribution function of the maximum ratio between planet mass and eccentricity-dependent PIM for the planets in the different simulations. Only growth phases with non-zero pebble accretion rates for planets with a final mass above 0.01 M$_\mathrm{E}$ are considered in the calculation of the maximum.}
    \label{fig: CDF_Miso}
\end{figure}

To get a first idea of how much this new pebble isolation mass would have influenced the planets in our simulation, we used the planets evolutionary data, shown in e.g. Fig. \ref{fig: 1.00Msun_case_OL18}, to trace the ratio between the mass ($M_\mathrm{p}$) and PIM ($M_\mathrm{iso,2}$) of each planet throughout their pebble-accreting growth phase. The maximum values are reported in Fig. \ref{fig: CDF_Miso}. 

The influence of the eccentricity dependence of the PIM is relatively small for most planets. None of the planets in the OL18 - 0.09 M$_\Sun$ simulations have reached $M_\mathrm{iso}$ or $M_\mathrm{iso,2}$. For the OL18 - 1.00 M$_\Sun$ simulations, 80\% (94\%) of the planets do not exceed $M_\mathrm{iso,2}$ by more than 5\% (7.5\%), with the highest excess being 15\%. The planets around 0.49 M$_\Sun$ stars seem to be influenced most, with 42\% of the planets exceeding $M_\mathrm{iso,2}$ by more than 5\%, and a maximum excess of 36\%. Similar values are found for the IGM16 simulations, though these results are slightly skewed by the large sample of third generation planets that never come close to $M_\mathrm{iso}$.

It is important to note, however, that these results give an upper limit to the effects using the PIM of \citet{Chametla+2022} in our simulations could have had. A maximum $M/M_\mathrm{iso,2}$ of 1.1 does not necessarily mean an overestimation of the planet's mass by 10\%. There is a balance between the growth time scale and the eccentricity demping timescale. If a planet reaches $M_\mathrm{iso,2}$ and stops growing, it could continue to circularise its orbit, assuming that it is not close to another large planet. This in turn brings $M_\mathrm{iso,2}$ closer to $M_\mathrm{iso}$, an effect which is not considered in the calculation of Fig. \ref{fig: CDF_Miso}. Moreover, though only parts of the planetary evolution with a non-zero pebble accretion rate are included in the calculation of the maximum $M/M_\mathrm{iso,2}$, the magnitude of this accretion rate is not considered in the estimations. For most slow-growing planets, such as those around 0.20 M$_\Sun$ stars, the planets do not grow bigger than $M_\mathrm{iso,2}$, as much as $M_\mathrm{iso,2}$ shrinks below their mass due to orbital excitations at later stages of the disc evolution.  

The effects of an increase in isolation mass due to high eccentricity cannot be traced from our data, because there are too many scenarios other than $M_\mathrm{iso,2}>M_\mathrm{iso}$ that lead to a maximum $M/M_\mathrm{iso,2}<1$, such as low PA rates, late formation, or exterior planets reaching the isolation mass. However, especially for the OL18 simulations, it is unlikely that planets would grow bigger when using $M_\mathrm{iso,2}$ instead of $M_\mathrm{iso}$, since planets with high eccentricities cannot efficiently accrete pebbles.

All in all, it is unlikely that the implementation of the pebble isolation mass of \citet{Chametla+2022} in our simulation would influence the final planet masses by more than a few percent. Nevertheless, it is worth including in all future studies, since it is a more complete model of the pebble isolation mass. 

Aside from the three modelling limitations discussed above, there are a few caveats following from computational constraints. The available computing time governed the number of planetesimals, step size, and inner truncation radius. As explained in Sect. \ref{subsec: disc_parameters}, the step size used in this study is slightly larger than advised by \citet{Wisdom_and_Holman_1991}. However, \citet{Hatalova_2023}, who used an even larger step size, showed that decreasing their step size by a factor of four had no significant influence on the outcome of the simulations. Moreover, whereas they studied planetary growth through planetesimal accretion, for which the precise dynamics between all planetesimals is essential, our main growth method is pebble accretion, which is much less sensitive to the step size. Naturally, a sufficiently small step size is important for our simulations as well, in order to properly detect close encounters\footnote{During close encounters, the overall step size is broken up in many smaller steps, so the precise motion during the close encounters is not influenced by using a larger step size, only the detection of the close encounter is.}, and to model the dynamics of the systems that form at the inner edge of the disc. However, if the planetesimal accretion simulations of \citet{Hatalova_2023} are not significantly influenced by the chosen step size, then it is highly unlikely that our pebble accretion simulations would be.

Another important caveat comes in the form of the aforementioned inner truncation radius, interior to which planets are assumed to have merged with the star and are removed from the simulation. This is a necessary computational constraint that follows directly from the step size, as explained in Sect. \ref{subsec: disc_parameters}. However, this radius is not a physical boundary, and there are stars with planets on stable orbits interior to this boundary. As can be seen in Figs. \ref{fig: 1.00Msun_summary} and \ref{fig: 0.49+0.70_summary}, there are several systems with planets that have been pushed close to $r_\mathrm{trunc}$. In fact, several massive particles have been removed from the simulations because they crossed the boundary, primarily in the IGM16 runs. A smaller value of $r_\mathrm{trunc}$ is required to determine the actual stability of these ejected planets. However, this requires a much smaller step size, which was not possible given the available computing resources.

\subsection{Recommendations for future research} \label{sec: future research}

A useful next step for this research would be to perform an in-depth comparison between the simulated systems, and the observed exoplanetary systems, to determine which types of systems are most similar to the ones we predict. Other important steps include improving the IGM16 model to include expressions for the ballistic regime, and the influence of the eccentricity, inclination, and argument of periapsis on the relative velocity between the planetesimals and the pebbles. Furthermore, chemical models for the evolution of the pebble size inside the snowline should be performed, to test our proposed sublimation model.

However, the most relevant next step in our opinion would be to develop a general model that can explain the formation of Earth-like planets in the habitable zone systematically. This means either limiting the initial growth rate of the planets, or decreasing their migration speed. As discussed in Sect. \ref{sec: paper_comparison}, the pebble mass flux and the disc dissipation conditions are the most promising parameters for limiting the planetary growth rate, based on the work of \citet{Lambrechts_2019}. Performing a parameter study of these two variables is, therefore, a very important continuation of this research. Limiting migration could be done by introducing a pressure maximum close to the habitable zone, for example created by disc winds in the inner regions of the disc, or sublimation of volatiles from pebbles at the snowline. Moreover, as discussed in sec \ref{sec: caveats}, the inclusion of thermal torques from \citet{Masset2017} might reduce the rate of inwards migration. Further research is required to determine how realistic these scenarios are, and how they are best combined.  

\section{Conclusion} \label{chapter: conclusion}
We studied the formation of terrestrial planets around low-mass stars using a version of the N-body integrator SyMBA \citep{Duncan_1998}, modified to include pebble accretion (PA), type I and II migration, and eccentricity and inclination damping \citep{Matsumura_2017,Matsumura_2021}. 
The main findings from these simulations are as follows:
\begin{itemize}
    \item Earth-like planets are consistently formed around 0.49, 0.70, and 1.00 M$_\Sun$ stars, irrespective of the model that is used. 
    
    \item Around 0.09 and 0.20 M$_\Sun$ stars, no Earth-mass planets are formed. Even if the final mass of all planetesimals in the disc would somehow concentrate onto one planet, the planet would not be more massive than a few Mars masses. 

    \item In the 0.49, 0.70, and 1.00 M$_\Sun$ simulations, the first planets reach the pebble isolation mass within a few hundred thousand years. As these planets rapidly migrate inwards, they excite the orbits of the other planetesimals so much that PA becomes highly inefficent, limiting planetary growth at later times.

    \item The IGM16 model does not consider the fact that planetesimals with high eccentricities and inclinations enter the ballistic regime. Therefore, this model unrealistically generates far more Earth-like and higher mass planets than the OL18 model. 

\end{itemize}

Overall, PA has a high tendency to create Earth-like planets around low-mass stars of about 0.5 M$_\Sun$ and higher. However, for a planet to remain in the HZ, a series of very specific events must occur. The planet must start its formation relatively slowly and far out in the disc. Its growth must then be stopped at precisely the right moment by interference from a massive planet around it. If its mass is too high, it will rid itself of the induced eccentricity and inclination too soon, grow to the isolation mass too fast, and migrate to the inner edge of the disc like all other planets. If its mass is too low, it holds on to its orbital excitations for too long and will not have time to grow to an Earth-like size, due to the decaying pebble flux. 

If other processes in the disc, such as the sublimation of volatiles at the snowline (or disc winds) could create a pressure maximum within the habitable zone, which would act as a sufficiently strong migration trap, the probability of finding Earth-like planets in the HZ would significantly increase. A lower initial growth rate due to a reduced pebble flux could also increase this likelihood. Further research is needed to determine whether these scenarios are realistic.   

\begin{acknowledgements}
We would like to thank Prof. dr. C. Dominik for the insights he provided during discussions about this work. We also thank Prof. dr. W. Lyra for his comment after reading a pre-print version of our paper. Finally, we thank the anonymous referee for their valuable commentary, making this paper more concise and clear. This work used the Dutch national e-infrastructure with the support of the SURF Cooperative and the Dutch Research Council (NWO) using grant no. EINF-7075. M.J. Yzer further acknowledges funding from the European Research Council (ERC) under the European Union's Horizon 2020 research and innovation program under grant agreement NO 805445.
\end{acknowledgements}
 \bibliographystyle{aa}
\bibliography{references}

\begin{appendix}

\section{The pebble radius within the snowline} \label{App: sublimation}
\begin{figure}[t]
    \centering
    {\includegraphics[width=1.\linewidth]{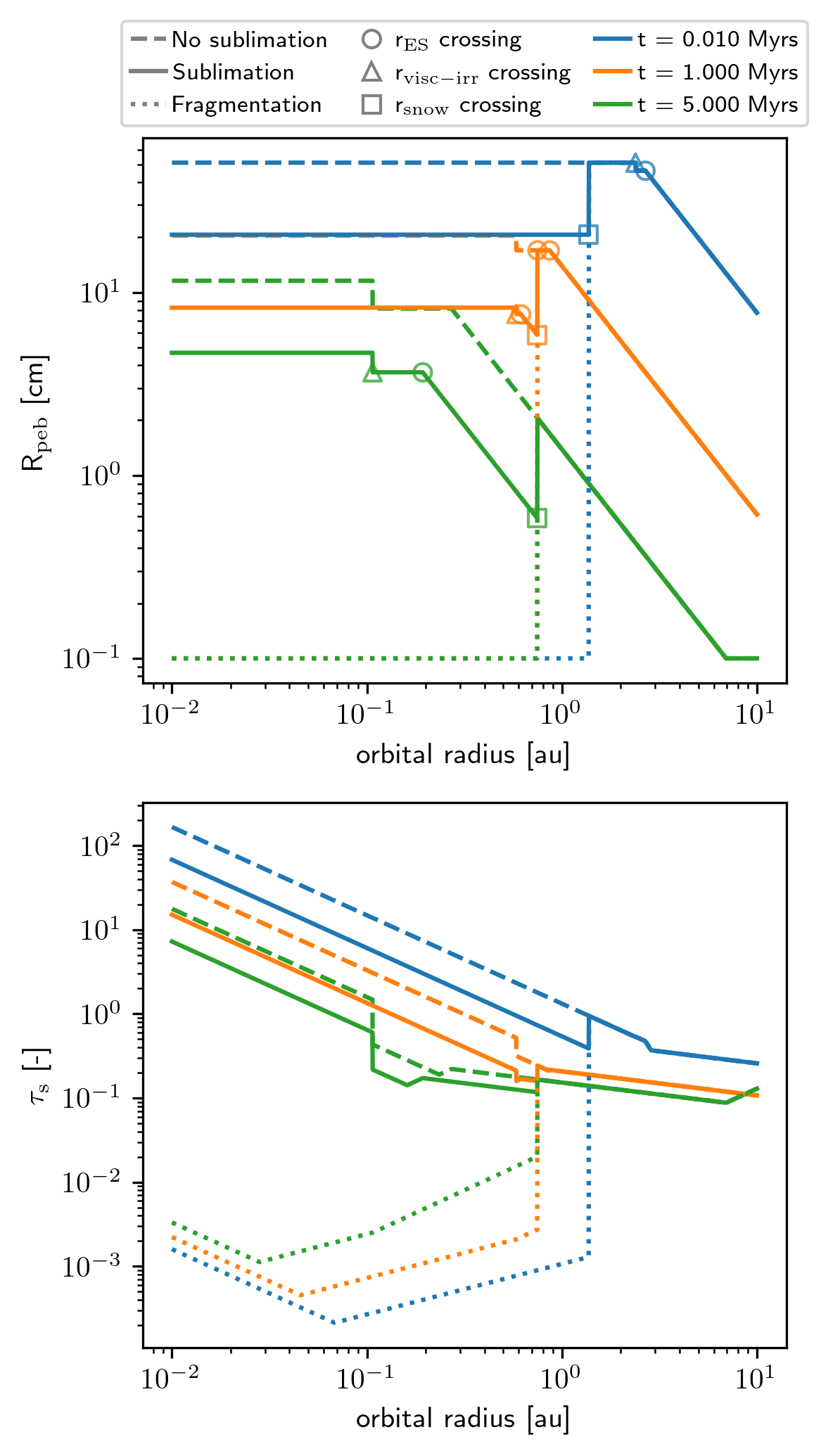}}
        \caption[Influence of sublimation and fragmentation on $R_\mathrm{peb}$ and $\taus$ within the snowline.]{Influence of sublimation and fragmentation on the pebble radius (top) and Stokes number (bottom) within the snowline of a 1.00 M$_\Sun$ star. The solid lines represent the sublimation model, in which pebbles fragment and immediately recoagulate with a solid-dominated internal density. The dotted lines represent the fragmentation model without recoagulation. The dashed lines show the default model, in which the pebbles' size and composition remain unaltered as they cross the snowline. The colours of the lines signify different times of evaluation. The circles, triangles, and squares represent crossings of the Epstein-Stokes boundary, the viscous-irradiative boundary and the snowline respectively. At the snowline, $R_\mathrm{peb}$ instantaneously decreases due to sublimation, and even more so due to fragmentation, which leads to a local reduction in $\taus$.} \label{fig: rpeb_in_rsnow}
\end{figure}

Figure \ref{fig: rpeb_in_rsnow} shows the pebble radius $R_\mathrm{peb}$ (top panel) and Stokes number $\taus$ (bottom panel) as a function of orbital radius for three different times. As discussed in sec \ref{sec: Rpeb_and_Mdotf}, the pebble radius in this study is drift-limited, which is to say, the pebble grows until it reaches the critical Stokes number $\tau_\mathrm{crit,1}$ for which $t_\mathrm{grow}=t_\mathrm{drift}$. The focus of Fig. \ref{fig: rpeb_in_rsnow} is on what happens once the pebbles cross the snowline. In the original model, here referred to as `no sublimation', the equations of \citet{Ida_2016} are smoothly extrapolated to within the snowline, so that the pebble radius and composition are unaltered. However, this assumption is highly unrealistic and inconsistent with the assumption that sublimation of volatiles causes the pebble mass flux to be halved \citep[e.g.][]{Morbidelli_2015}. 

An alternative model is that of fragmentation. This model assumes that due to the sublimation of volatiles, pebbles break apart at the snowline into mm-sized silicate grains, and remain at this size due to the bouncing and fragmentation barrier \citep{Morbidelli_2015,Ida_2016,Matsumura_2017}. However, the mechanics behind fragmentation, and whether it occurs in the first place, are still uncertain. Moreover, because in the fragmentation model $\taus \lll 1$, other model assumptions, such as the assumption that the radial viscous diffusion velocity term $u_\nu$ is negligible, might become invalid.

We therefore propose a third model, which we refer to as the `sublimation' model. Similarly to the fragmentation model, we assume that pebbles disintegrate into dust at the snowline due to sublimation of their icy contents. However, this dust then recoagulates to a new radius, with a new density dominated by rock. We assume $\rho_\mathrm{s}\sim \SI{1.0}{\gram\per\cubic\centi\meter}$ outside the snowline, and $\rho_\mathrm{s}\sim \SI{2.5}{\gram\per\cubic\centi\meter}$ inside the snowline. This, combined with the 50\% reduction in the pebble mass flux $\dot{M}_\mathrm{F}$, results in a change in pebble radius in both the Epstein and the Stokes regime, as well as a shift in the location of the of boundary $r_\mathrm{ES}$ between the two regimes. 

In fact, sublimation could introduce a second and third Epstein-Stokes boundary, as can be seen in the 1.0 Myrs line (shown in orange) of $R_\mathrm{peb}$ in Fig. \ref{fig: rpeb_in_rsnow}. Initially in the outer disc, the pebble is in the Epstein regime and grows as it drifts inwards. Just outside the snowline, the pebble transitions from the Epstein regime to the Stokes regime, signified by the circle marker, after which it no longer grows. However, once the pebble crosses the snowline and falls apart, it grows back to the new value of $\tau_\mathrm{s,crit1}$ \citep[see][for the full expressions]{Ida_2016}, which is again in the Epstein regime. As the pebble drifts further inwards, it continues growing, until it transitions to the Stokes regime one final time. 

The pebble radius in the Stokes regime still has a few caveats, though. For instance, there is a discontinuity in the pebble radius of the green solid and dashed lines (5 Myrs) of Fig. \ref{fig: rpeb_in_rsnow}, which is caused by a transition from the irradiative to the viscous regime, marked by the green triangle at 0.1 au. This discontinuity follows directly from the analytical expressions, which contain a change in parameter dependencies and slopes at the viscous-irradiative boundary. 

When implementing this change in parameters in the analytical expressions for the pebble size of \citet{Ida_2016}, the pebble radius increases once the pebbles cross $r_\mathrm{visc-irr}$. However,  $\tau_\mathrm{s,crit1}$ inside $r_\mathrm{visc-irr}$ remains smaller than $\taus$ just outside the boundary, meaning that sudden in-situ growth is not expected. Much is still unclear about what happens with $R_\mathrm{peb}$ in the Stokes regime within the snowline, and further studies with dust coagulation models within the snowline are required to create a more exact model. For simplicity, we assume the analytical expressions from \citet{Ida_2016} remain valid within the snowline and within the viscous-irradiative boundary, albeit with a different value for $\rho_\mathrm{s}$ and $\dot{M}_\mathrm{F}$. We acknowledge, however, that a model in which $R_\mathrm{peb}$ can only grow if $\taus(R_\mathrm{peb})<\tau_\mathrm{s,crit1}$, might be more self-consistent.

\begin{figure}[t]
    \centering
    \includegraphics[width=1.\linewidth]{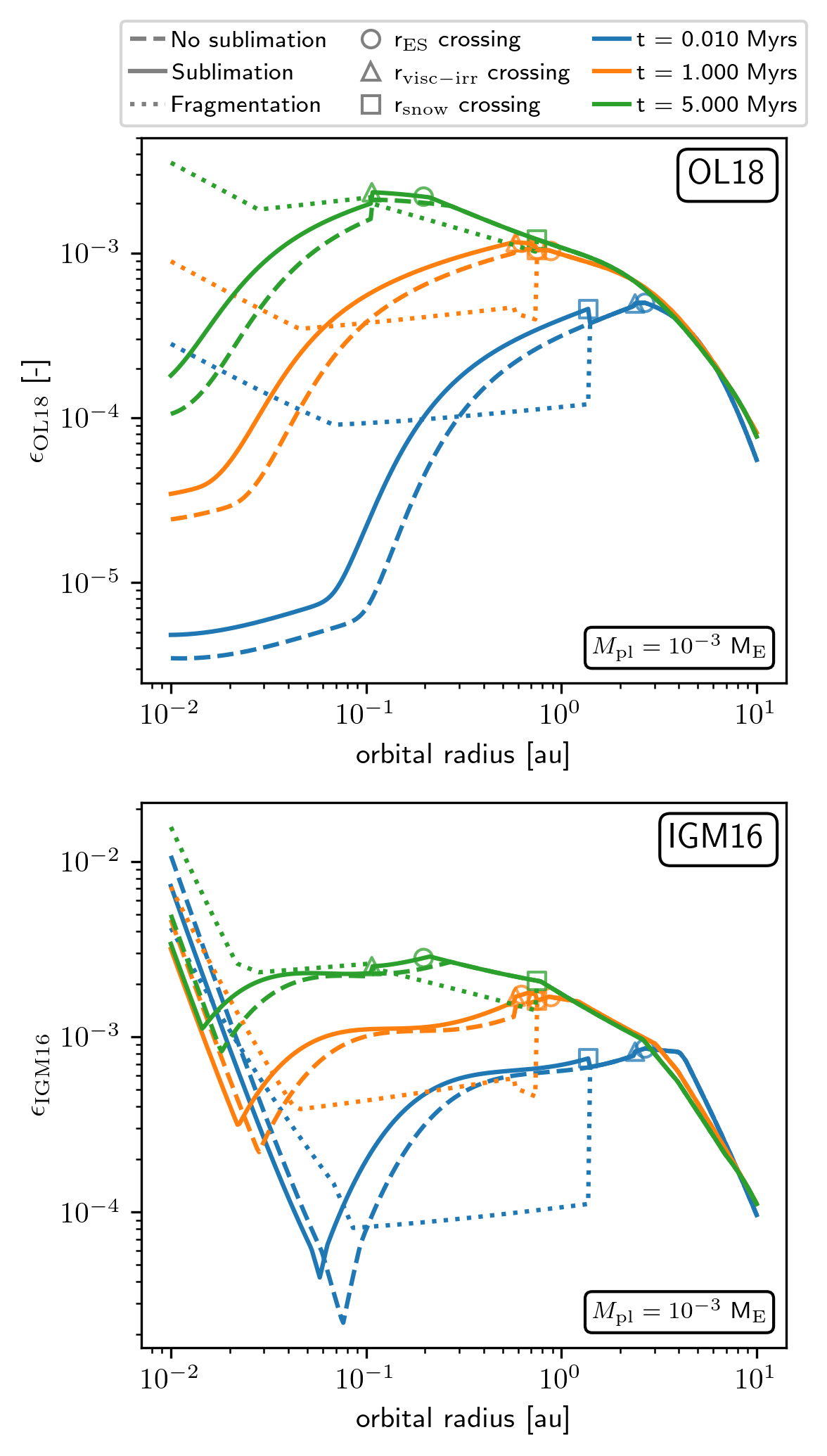}
        \caption[Influence of sublimation and fragmentation on the accretion efficiency within the snowline.]{Influence of sublimation and fragmentation on the accretion efficiency from OL18 (top) and IGM16 (bottom) within the snowline of a 1.00 M$_\Sun$ star. The accretion efficiencies are calculated for a 10$^{-3}$ M$_\mathrm{E}$ planetesimal. The line styles match those in Fig. \ref{fig: rpeb_in_rsnow}. Fragmentation causes a sudden decrease in accretion efficiency at the snowline, especially after 0.01 and 1.0 Myrs, due to the $\taus$ becoming so small that pebbles couple to the gas. Without fragmentation, the efficiency rapidly decreases in the inner regions of the disc ($\lesssim 0.2$ au), due to $\taus$ becoming too large for pebbles to settle. The sudden and sharp increase in $\epsilon_\mathrm{IGM16}$ in the innermost part of the disc is an unintended modelling effect from the accretion impact parameter $B$ becoming smaller than the planet radius $R_\mathrm{pl}$.} \label{fig: eps_in_rsnow}
\end{figure}

The influence of the different pebble radius models within the snowline on the accretion efficiency\footnote{It is important to note that the accretion efficiency needs to be multiplied with the pebble mass flux to find the actual accretion rate. The accretion rate decreases exponentially with time, and is reduced by 50\% within the snowline. So, even though at first glance it seems as if planets grow fastest after 5 Myrs, this is not true. The reason the accretion efficiency is shown instead of the accretion rate, is that the accretion efficiency provides a fair comparison of the PA mechanics, unobscured by the evolution of the pebble mass flux, which is a completely independent model. } is shown in Fig. \ref{fig: eps_in_rsnow}. These results have been calculated for a 10$^{-3}$ M$_\mathrm{E}$ planetesimal, which is a typical mass for planets in the early stages of rapid PA. Fragmentation at the snowline causes a rapid drop in accretion efficiency, especially at early times in the disc. This reduction might be even stronger if aerodynamic deflection of tiny pebbles coupled to the gas is considered \citep{Visser_2016}. This deflection depends on the considered gas flow model around the planet, which is a level of detail that is beyond the scope of this study.

Sublimation, on the other hand, causes an increase in accretion efficiency compared to the non-sublimation model, due to the high Stokes number being slightly reduced, allowing for the pebble to be slowed down slightly faster. Nevertheless, in the innermost regions of the disc, the Stokes number becomes so large that pebbles are no longer efficiently slowed down and do not settle in the planet's gravitational field. This causes the rapid drop in accretion efficiency in both OL18 and IGM16. 

In IGM16, the efficiency reduction for $\taus \gg 1$ is explicitly modelled in the accretion cross-section $B$ through the exponential reduction factor $\kappa$ from \citet{Ormel_and_Kobayashi_2012} (see the parameters of Eq. \ref{eq: eps_IGM16}). However, this approach has a problem of its own. Unlike OL18, the version of IGM16 used in this study does not include expressions for the ballistic regime, which occurs when the accretion impact parameter $B$, which is to say the largest impact parameter for which pebbles can accrete, becomes less than the geometric limit $R_\mathrm{pl}$. In this regime, pebbles no longer accrete because they settle in the gravitational field at a distance $B$ away from the planet, but because their trajectory directly intersects the planet's surface. The planet's surface thus becomes the impact parameter $B$ for accretion. As $\taus$ rapidly increases in the innermost regions of the disc, $\kappa$ drastically decreases, so that $B \lll R_\mathrm{pl}$. In our version of IGM16, the cross-section cannot decrease below $R_\mathrm{pl}$. However, using the geometric limit in the settling efficiency equations leads to a serious overestimation of the accretion efficiency, as can be seen from the sharp increase in $\epsilon_\mathrm{IGM16}$ in Fig. \ref{fig: eps_in_rsnow}. This is because $\epsilon_\mathrm{IGM16}$ scales approximately with $b^3\equiv B^3r^{-3}=R_\mathrm{pl}^3r^{-3}$, the slope of which matches exactly with what is observed in the figure. 

Allowing $B$ to asymptotically decrease to 0 due to $\kappa$ and ignoring the ballistic regime altogether, has less significant consequences on the growth rate of planets in these inner regions of the disc, since, as can be seen from the OL18 results, growth in the ballistic regime is small. 

Either way, this shortcoming of IGM16 has little to no effect on the full SyMBA simulations, since it applies to a region far closer to the star than the planetesimal disc. Planets can only reach these regions through migration, and planets generally reach the pebble isolation mass in a much shorter timescale than they migrate.

The final caveat of the sublimation model is that in the inner regions of the disc, $\taus \gg 1$. As a result, the drift slows down significantly, since 
\begin{equation*}
    t_\mathrm{drift}\ \propto\ \frac{1+\taus^2}{\taus} \approx \taus, \quad \mathrm{for\ } \taus > 1.     
\end{equation*}

As $\taus$ continues to increase, the drift timescale again becomes longer than the growth timescale. This occurs when $\taus$ exceeds \citep{Ida_2016}
\begin{equation}\label{eq: taus_crit2}
\begin{split}
    \tau_\mathrm{s,crit2} &\simeq \left(\frac{3\sqrt{3\pi}}{16\left| 3-0.5\gamma\right|} \frac{\alpha_\mathrm{acc}}{\eta}\frac{\dot{M}_\mathrm{F}}{\dot{M}_*}\right)^{-1/2} \\
    &\simeq 13 \left(\alpha_3^{-1}\dot{M}_{*8}\dot{M}_\mathrm{F4}^{-1} \right)^{1/2}\left(\frac{\eta}{10^{-3}} \right)^{1/2} \quad \propto \quad r^q.
\end{split}
\end{equation}
 In theory, this could result in runaway coagulation \citep{Okuzumi_2012}, since $\taus\ \propto\ R^2$. This means that once $\taus>\tau_\mathrm{s,crit2}$, the condition will always be satisfied and pebble drift will come to a complete halt.

\begin{figure}
    \centering
    {\includegraphics[width=1.0\linewidth]{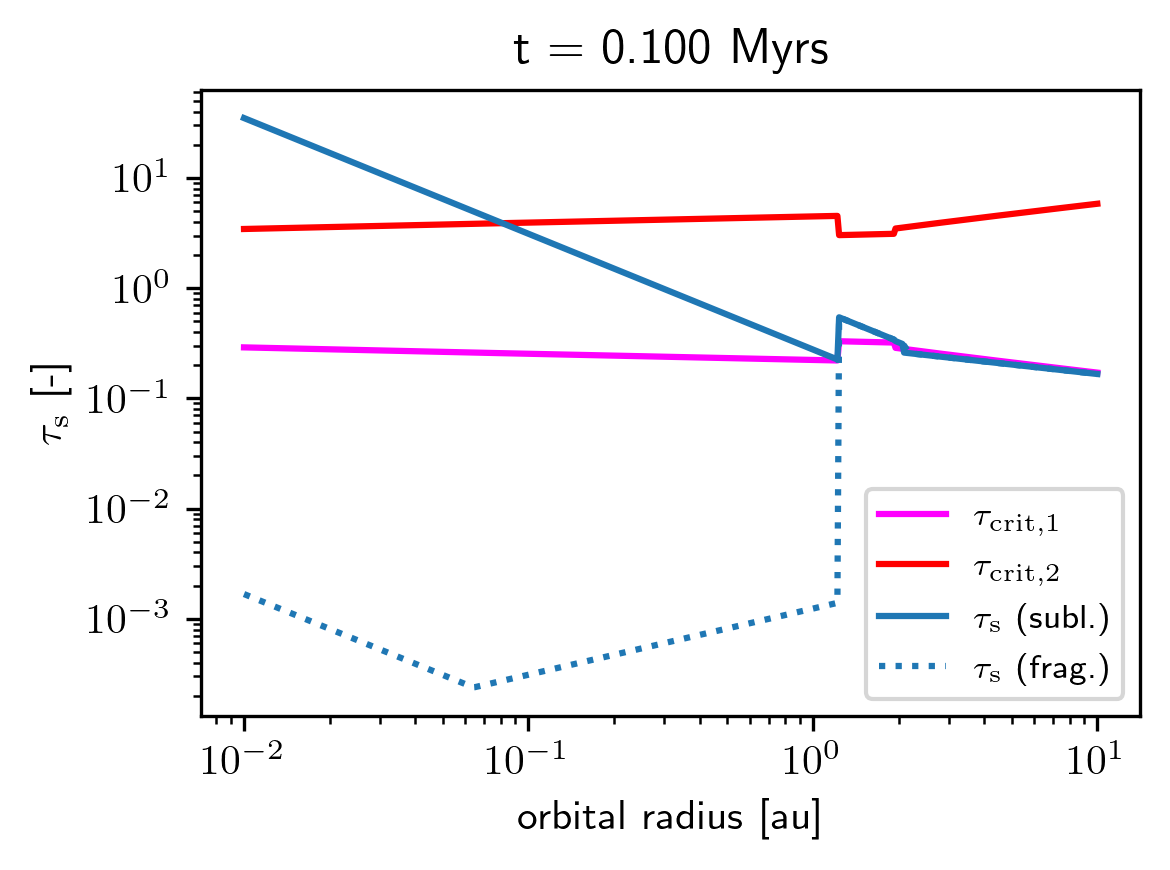}}
        \caption[Comparison between $\taus$ and $\tau_\mathrm{crit}$ for the sublimation and fragmentation model of a 1.00 M$_\Sun$ star.]{Comparison between $\tau_\mathrm{crit}$ and $\taus$ in both the sublimation and fragmentation model for a 1.00 M$_\Sun$ star. In the sublimation model, $\taus>\tau_\mathrm{crit,2}$ for $r\lesssim 0.2$ au, which could lead to runaway coagulation. As the disc evolves, $\taus$ decreases, and remains within the bounds up to smaller orbital radii. In the fragmentation model, $\taus \ll \tau_\mathrm{crit,1}$, which suggests these pebbles no longer drift and should therefore grow in situ as well. The values for $\tau_\mathrm{crit,1}$ and $\tau_\mathrm{crit,2}$ were calculated by numerically solving $t_\mathrm{grow}=t_\mathrm{drift}$.} 
    \label{fig: taus_vs_tauscrit_1.00Msun}
\end{figure}

\begin{figure}
    \centering
    {\includegraphics[width=1.0\linewidth]{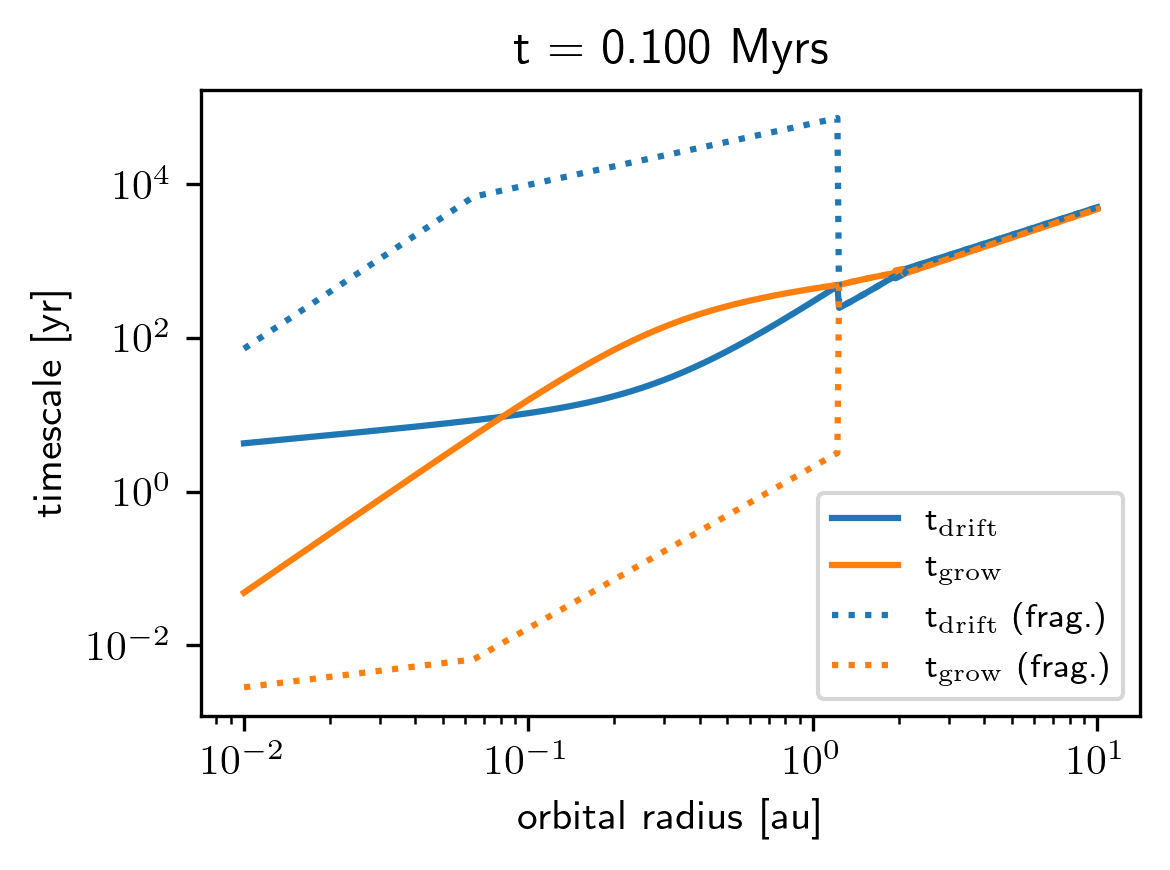}}
        \caption[Comparison between the growth and drift timescale of pebbles in the sublimation and fragmentation model around a 1.00 M$_\Sun$ star.]{Comparison between $t_\mathrm{drift}$ and $t_\mathrm{grow}$ in both the sublimation and fragmentation model for a 1.00 M$_\Sun$ star. For both models, $t_\mathrm{grow}<t_\mathrm{drift}$ in the inner disc, suggesting (runaway) coagulation could occur. However, while the fragmented pebbles remain stuck around the snowline for tens of thousands of years, four orders of magnitude longer than their growth timescale, pebbles in the sublimation model drift into the star in only a few years, giving them significantly less time to grow. It is therefore unlikely that runaway coagulation would be a serious problem in the sublimation model.} 
    \label{fig: tgrow_vs_tdrift_1.00Msun}
\end{figure}

Figure \ref{fig: taus_vs_tauscrit_1.00Msun} shows that for the sublimation model around a solar-mass star, $\taus$ indeed exceeds $\tau_\mathrm{crit,2}$, suggesting the pebbles could enter this phase of runaway coagulation. However, when comparing the growth timescale to the drift timescale, as is done in Fig. \ref{fig: tgrow_vs_tdrift_1.00Msun}, it turns out that in the region where $t_\mathrm{grow}<t_\mathrm{drift}$, $t_\mathrm{drift}$ is of the order of a few years, and $t_\mathrm{grow}$ is only a few orders shorter. Moreover, when the compact silicate pebbles within the snowline come together, they do not perfectly stick to one another \citep{Morbidelli_2015,Ida_2016}, which significantly limits the growth rate, meaning that the growth timescale is underestimated. Therefore, it is probable that pebbles drift into the central star before they have time to significantly grow. For smaller stars, $\taus$ always remains within the confines of $\tau_\mathrm{crit}$.

Some might argue that the fragmentation model should be preferred, since these small pebbles never exceed $\tau_\mathrm{s,crit2}$. However, these small pebbles have Stokes numbers significantly smaller than $\tau_\mathrm{s,crit1}$, meaning they are also prone to rapid growth. In fact, pebbles in the fragmentation model stay at the snowline for tens of thousands of years due to their long drift timescale. Meanwhile, their growth timescale is of the order of a year or less, as can be seen in Fig. \ref{fig: tgrow_vs_tdrift_1.00Msun}. We believe that it is, therefore, far more likely that the mm-sized silicate grains of the fragmentation model would start growing, than that the pebbles in the sublimation model would. Moreover, in the fragmentation model, the growth of the pebbles would be around the snowline, and would therefore significantly impact the growth rate of the planetesimals. The potential runaway growth of pebbles in the sublimation model would occur far closer to the star, where, as argued before, they could only encounter planets that have already reached the pebble isolation mass. 

We therefore favour the sublimation model over the fragmentation model, and use this model for the full SyMBA simulations. However, detailed dust coagulation computations using the disc models of \citet{Ida_2016} are required to further validate the assumptions mentioned above. This is beyond the scope of this study.

\FloatBarrier

\section{Supplementary figures}\label{app:supplement}
This appendix presents additional figures showing more data or alternative visualisations complementary to Sect. \ref{Chapter: Nbody-results}.

\begin{figure}[h]
    \centering
    {\includegraphics[width=1\linewidth]{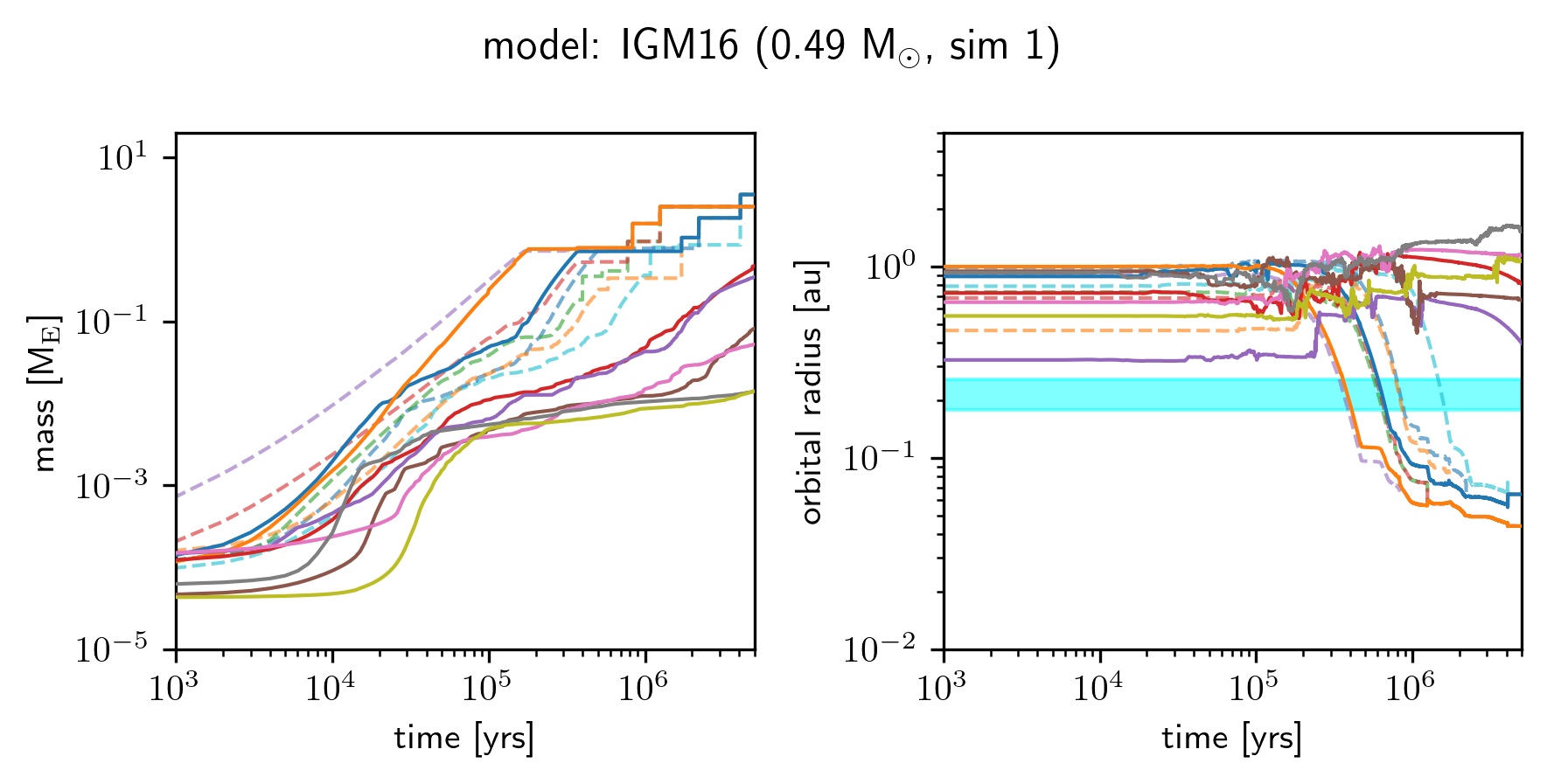}}
        \caption[The dynamical evolution tracks ($M_\mathrm{p}$ and $a$) of all large planets in a single IGM16 simulation for a 0.49 M$_\Sun$ star.]{The dynamical evolution tracks ($M_\mathrm{p}$ and $a$) of all large planets in IGM16 simulation 1 around a 0.49 M$_\Sun$ star. The coloured solid lines represent different planets. The transparent dashed lines represent large planetesimals that merged with the planets. Only planets with masses $> 0.01$ M$_\mathrm{E}$ are included. The cyan shaded region indicates the habitable zone. In the 0.49 M$_\Sun$ simulations, the planets that form at a late stage in the disc ($t\gtrsim1$ Myrs) form outside the HZ and remain too small to migrate into it.}
    \label{fig: 0.49Msun_case_IGM16}
\end{figure}

\begin{figure}
    \centering
    {\includegraphics[width=1\linewidth]{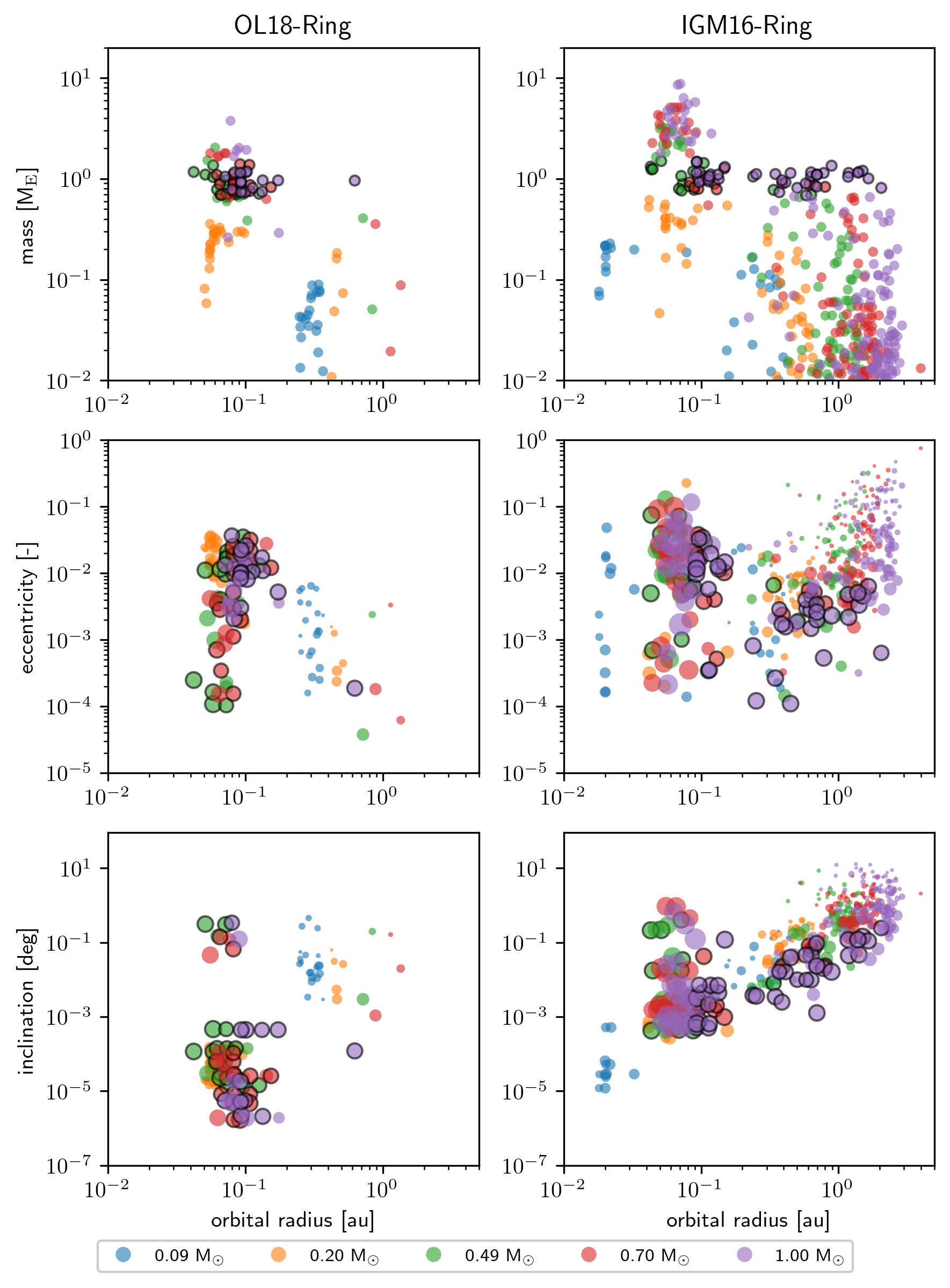}}
        \caption[Mass, eccentricity, and inclination of all planets formed in the OL18-Ring and IGM16-Ring simulations.]{Overview of the mass, eccentricity, and inclination of all planets formed in the OL18-Ring and IGM16-Ring simulations. Earth-like planets have been highlighted using a black edge around the marker. In the $e$ and $i$ plots, the size of the marker is proportional to its mass. Overall, the results are very similar to those of the normal runs, presented in Fig. \ref{fig: M,e,i_vs_a_all_planets_normal}.}
    \label{fig: M,e,i_vs_a_all_planets_Ring}
\end{figure} \end{appendix}

\end{document}